\documentclass[aps,10pt,notitlepage,nofootinbib,superscriptaddress]{revtex4-1}

\usepackage[utf8]{inputenc}
\usepackage{newtxtext,newtxmath}

\usepackage{bbold}
\usepackage{bm}
\usepackage{graphicx}
\usepackage[usenames,dvipsnames]{xcolor}
\usepackage{color}
\usepackage[colorlinks=true,linkcolor=blue,urlcolor=blue,citecolor=blue]{hyperref}
\usepackage{amsmath,amssymb,amsfonts}
\usepackage{slashed}
\usepackage[english]{babel}
\usepackage{dcolumn}
\usepackage{pifont}
\usepackage{dsfont,mathrsfs}
\usepackage{cancel}
\usepackage{bigints}
\usepackage{accents}
\usepackage{soul}
\usepackage{multirow}

\newcommand{\nn}{\nonumber}
\newcommand{\ket}[1]{\left|#1\right\rangle}

\newcommand{\expectationvalue}[3]{\langle#1|#2|#3\rangle}
\newcommand{\ExpectationValue}[3]{\left\langle#1\left|#2\right|#3\right\rangle}
\newcommand{\ensembleaverage}[1]{\left\langle#1\right\rangle}

\newcommand{\MB}[1]{\left|#1\right|}
\newcommand{\FB}[1]{\left(#1\right)}
\newcommand{\fb}[1]{(#1)}
\newcommand{\SB}[1]{\left\{#1\right\}}
\newcommand{\TB}[1]{\left[#1\right]}

\newcommand{\mcT}{\mathcal{T}}
\newcommand{\mcTc}{\mathcal{T}_C}
\newcommand{\mcN}{\mathcal{N}}
\newcommand{\mcI}{\mathcal{I}}
\newcommand{\mcF}{\mathcal{F}}
\newcommand{\mcSh}{\hat{\mathcal{S}}}
\newcommand{\mcZ}{\mathcal{Z}}
\newcommand{\mcW}{\mathcal{W}}
\newcommand{\mcL}{\mathcal{L}}

\newcommand{\scrL}{\mathscr{L}}

\newcommand{\scrD}{\mathscr{D}}

\newcommand{\munu}{{\mu\nu}}

\newcommand{\alphabeta}{{\alpha\beta}}
\newcommand{\IM}{\text{Im}}

\newcommand{\Tr}{\text{Tr}}

\newcommand{\Psibar}{\overline{\Psi}}
\newcommand{\psibar}{\overline{\psi}}

\newcommand{\kpll}{k_\parallel}
\newcommand{\qpll}{q_\parallel}
\newcommand{\ppll}{p_\parallel}

\newcommand{\gpll}{g_\parallel}
\newcommand{\gper}{g_\perp}

\newcommand{\del}{\partial}
\newcommand{\identity}{\mathds{1}}

\begin{document}
\title{Effects of anomalous magnetic moment of quarks on the dilepton production from hot and dense magnetized quark matter using the NJL model}


\author{Snigdha Ghosh}
\email{snigdha.physics@gmail.com, snigdha.ghosh@saha.ac.in}
\affiliation{Saha Institute of Nuclear Physics, 1/AF Bidhannagar, Kolkata - 700064, India}
\author{Nilanjan Chaudhuri}
\email{sovon.nilanjan@gmail.com}
\affiliation{Variable Energy Cyclotron Centre, 1/AF Bidhannagar, Kolkata 700 064, India}
\affiliation{Homi Bhabha National Institute, Training School Complex, Anushaktinagar, Mumbai - 400085, India}
\author{Sourav Sarkar}
\email{sourav@vecc.gov.in}
\affiliation{Variable Energy Cyclotron Centre, 1/AF Bidhannagar, Kolkata 700 064, India}
\affiliation{Homi Bhabha National Institute, Training School Complex, Anushaktinagar, Mumbai - 400085, India}
\author{Pradip Roy}
\email{pradipk.roy@saha.ac.in}
\affiliation{Saha Institute of Nuclear Physics, 1/AF Bidhannagar, Kolkata - 700064, India}
\affiliation{Homi Bhabha National Institute, Training School Complex, Anushaktinagar, Mumbai - 400085, India}

\begin{abstract}
Dilepton production rate (DPR) from hot and dense quark matter is studied in the presence of an arbitrary external magnetic field using the 2-flavour Nambu--Jona-Lasinio (NJL) model. The anomalous magnetic moment (AMM) of the quarks is taken into consideration while calculating the constituent quark mass as well as the DPR from the thermo-magnetic medium. An infinite number of quark Landau levels is incorporated so that no approximations are made on the strength of the background magnetic field. The analytic structure of the two point vector current correlation function in the complex energy plane reveals that, in addition to the usual Unitary cut, a non-trival Landau cut appears in the physical kinematic domains solely due to the external magnetic field. Moreover, these kinematic domains of the Unitary and Landau cuts are found to be significantly modified due to the AMM of the quarks. With finite AMM of the quarks, for certain values of the external magnetic field, the kinematically forbidden gap between the Unitary and Landau cuts are shown to vanish leading to the generation of a continuous spectrum of dilepton emission over the whole invariant mass region not observed earlier.
\end{abstract}

\maketitle
%
\section{INTRODUCTION}\label{sec.intro}

The study of ``strongly'' interacting nuclear matter under extreme conditions of high temperature and/or density along with the strong external electromagnetic field 
has gained intensity in contemporary research over the past few years~\cite{Kharzeev:2012ph}. 
Such non-trivial background is expected to lead to a large number of interesting physical phenomena~\cite{Kharzeev:2012ph,Kharzeev:2007tn,Chernodub:2010qx, Chernodub:2012tf} owing 
to the rich vacuum structure of the underlying Quantum Chromodynamics (QCD), e.g. the Chiral Magnetic Effect (CME)~\cite{Fukushima:2008xe,Kharzeev:2007jp,Kharzeev:2009pj,Bali:2011qj}, 
Magnetic Catalysis (MC)~\cite{Shovkovy:2012zn,Gusynin:1994re,Gusynin:1995nb,Gusynin:1999pq}, Inverse Magnetic Catalysis (IMC)~\cite{Preis:2010cq,Preis:2012fh}, 
Chiral Vortical Effect (CVE), vacuum superconductivity and superfluidity~\cite{Chernodub:2011gs,Chernodub:2011mc} {\it etc}. 
Significant research efforts on these topics is also made in the context of astrophysical and cosmological studies~\cite{Duncan:1992hi,Ferrer:2005vd,Ferrer:2006vw,Ferrer:2007iw,Fukushima:2007fc,Feng:2009vt,Fayazbakhsh:2010gc,Fayazbakhsh:2010bh}.

In recent studies~\cite{Kharzeev:2007jp,Skokov:2009qp}, it is revealed that, in a non-central or asymmetric Heavy Ion Collision (HIC) experiment, 
extremely strong magnetic fields of the order $ \approx 10^{18} $ Gauss or larger are transiently created~\cite{Kharzeev:2007jp}. 
However, high values of the electrical conductivity of the medium could possibly make it sustain longer than thought earlier~\cite{,Gursoy:2014aka}. 
Apart from this, existence of magnetic field $\sim 10^{15}$ Gauss has been conjectured in the interior of certain astrophysical objects called magnetars~\cite{Duncan:1992hi,Thompson:1993hn}. 
Besides these, in the early universe during electroweak phase transition, the magnetic field as high as $ \approx 10^{23}$ Gauss~\cite{Vachaspati:1991nm,Campanelli:2013mea} might have been produced. 
The magnitude of the magnetic field being comparable to the typical QCD energy scale ($eB\sim \Lambda_\text{QCD}^2$), the bulk as well as 
microscopic properties of the QCD matter could be non-trivially modified under such a strong external magnetic field. Thus, apart from 
the theoretical intricacies, there also exists the possibility of an experimental verification.

The large value of the QCD coupling constant in low energy regime puts huge constraints and complexities on the first principle calculations. 
The non-perturbative aspects of QCD at intermediate temperatures (comparable to the QCD scale) and low baryonic density can be best addressed 
in Lattice QCD simulations ~\cite{Forcrand1,Forcrand2,Forcrand3,Bali_lat,Luschevskaya,Aoki,Aoki2}. 
It is also worthwhile to mention that recent investigations on the matter produced in HIC experiments is 
strongly interacting~\cite{Adler:2006yt,Adare:2006nq,Adare:2006ti,Abelev:2006db,Aamodt:2010pa,Aamodt:2010pb,Aamodt:2010jd,Aad:2010bu} which for the above mentioned 
reasons can not be dealt with perturbative-QCD. As an alternative, the effective models carrying some of the essential features of QCD are mathematically 
tractable and thus extensively used for the study of QCD matter in the low energy regime. The Nambu--Jona-Lasinio (NJL) model~\cite{Nambu1,Nambu2} is one the most 
successful effective models (see~\cite{Klevansky,Hatsuda1,Vogl,Buballa} for reviews) which provides a useful scheme to probe the vacuum structure of QCD at finite temperature and density. 
The studies of the non-perturbative properties of the QCD vacuum have been extensively performed using the NJL model as it respects the global symmetries of QCD, 
most importantly the chiral symmetry. The chiral phase transition in the NJL model in presence of external magnetic field is studied 
earlier in Refs~\cite{Klevansky2,Gusynin:1994re,Gusynin:1995nb,Gusynin:1999pq,Mao,Sadooghi1,Ruggieri,Ruggieri2,Andersen,Ayala2,Mao}.

It has been indicated Ref.~\cite{Strickland}, that the Anomalous Magnetic Moment (AMM) of the nucleons increases the level of pressure anisotropies for a system 
of proton and neutrons. Also in Ref.~\cite{Arghya}, the vacuum to nuclear matter phase transition has been studied using the Walecka model in presence of weak 
external magnetic field  including the AMM of the nucleons. Recently in Refs.~\cite{Sadooghi,Chaudhuri:2019lbw}, the effect of the AMM of the quarks are studied in the NJL model. 
In both the Refs.~\cite{Sadooghi,Chaudhuri:2019lbw}, the dependence of the constituent quark mass on temperature ($T$), chemical potential ($\mu_B$), magnetic field ($eB$) and AMM ($\kappa$) of the quarks were studied and the complete phase potrait of the model in the parameter space $\{T,\mu_B,eB,\kappa\}$ was explored. Moreover IMC was reported when the AMM of the quarks are taken into 
consideration.

The medium formed in a HIC experiment is a `transient' state that exists for a very short time ($\sim$ few fm/c) and this can not be observed directly. There are various indirect probes and observables~\cite{Wong:1995jf} that are used to extract the microscopic as well as bulk properties of QCD matter such as electromagnetic probes (photon and dileptons)~\cite{Arnold:2001ms, Aurenche:1999tq,Aurenche:2000gf,Chatterjee:2009rs,Alam:1996fd,Alam:1999sc,Kajantie:1986dh,McLerran:1984ay,Rapp:1999ej,Weldon:1990iw}, heavy quarks~\cite{Rapp:2009my}, collective flow~\cite{Poskanzer:1998yz,Voloshin:2008dg,Ollitrault:1992bk,Kolb:2003dz,Adams:2005dq,Aamodt:2010pa}, quarkonia~\cite{Matsui:1986dk}, jets~\cite{Wang:1991xy} and so on. The study of different $n$-point current-current correlation functions or the in-medium spectral functions of local currents could be an important theoretical tool to probe the microscopic properties of the medium. One such is the electromagnetic spectral function, which is obtained from the vector-vector current correlator. The vector current correlator can also be related to the Dilepton Production Rate (DPR) from the hot and dense magnetized medium. Unlike the `strong' probes, the dileptons are emitted from the entire space-time volume of the medium evolution. Since the dileptons interact only through the electromagnetic interaction owing to their larger mean free paths than the system size, they come out of the thermal medium soon after their production  without suffering more collisions. Therefore, the dileptons carry the information of the exact thermodynamic state of the medium where they are produced.

The DPR in the presence of external magnetic field has been studied quite extensively in~\cite{Tuchin:2012mf,Tuchin:2013bda,Sadooghi:2016jyf,Mamo:2013efa,Bandyopadhyay:2016fyd,Bandyopadhyay:2017raf,Ghosh:2018xhh,Islam:2018sog,Das:2019nzv}. In Refs.~\cite{Tuchin:2012mf,Tuchin:2013bda}, the authors have obtained the DPR from hot magnetized Quark Gluon Plasma (QGP) in a phenomenological way including the effects of synchrotron radiation as well as quark-antiquark annihilation. In Ref.~\cite{Sadooghi:2016jyf}, the Ritus formalism has been used to calculate the photon polarization tensor and DPR under external magnetic field. The DPR in strong as well as in weak magnetic field approximations has been reported in Refs.~\cite{Bandyopadhyay:2016fyd,Bandyopadhyay:2017raf}. In Ref.~\cite{Ghosh:2018xhh}, the DPR in presence of arbitrary external magnetic field has been calculated employing the Effective fugacity Quasi-Particle Model (EQPM) in which the effect of strong interactions are captured in the temperature dependent fugacities of the partons. In Ref.~\cite{Islam:2018sog}, the DPR has been obtained in the lowest landau level (LLL) approximation using the 2-flavour NJL and Polyakov-NJL model. In Ref.~\cite{Das:2019nzv}, the authors have used Hard Thermal Loop (HTL) resumed thermo-magnetic quark propagator to estimate the DPR from weakly magnetized hot QCD medium and found marginal modification in the DPR due to the weak magnetic field.

In this work, we aim to calculate the DPR using the 2-flavour NJL model in presence of arbitrary external magnetic field at finite temperature and baryon density incorporating the AMM of the quarks. The Real Time Formalism (RTF) of finite temperature field theory and the full Schwinger proper time propagator (with the AMM of the charged fermions) including all the Landau levels are used to calculate the current-current correlator in the vector channel in a thermo-magnetic dense medium. The temperature, density and magnetic field dependent effective quark mass calculated from the NJL model is used in the expression of the DPR. Moreover, no approximation like strong or weak magnetic field has been made in the analysis as usually done in most of the works in the literature. 
We have shown that, the DPR at non-zero magnetic field have contributions both from the Unitary as well as the Landau cuts corresponding to the physical time-like and positive energy dilepton production. The Unitary cuts (corresponding to the decay/formation processes) are also present at zero temperature and zero external magnetic field whereas the non-trivial Landau cuts (corresponding to the scattering processes) in the time-like domain appear only at finite temperature and due to the non-zero external magnetic field. In particular, the  Landau cuts lead to the physical processes as time-like positive energy photon emission or absorption by a quark/antiquark in magnetized hot quark matter which is forbidden in zero magnetic field case due to kinematic constraint. Moreover, upon analyzing the analytic structure of the two-point vector current correlator, we have shown that the thresholds of the Unitary and Landau cuts have non-trivial dependence on the magnetic field as well as on the AMM of the quarks. The DPR obtained is found to be largely enhanced in the low invariant mass region due to the appearance of the Landau cuts. Also the introduction of the AMM of the quarks leads a significant enhancement of DPR as compared to the zero AMM case.

The paper is organized as follows. In Sec.~\ref{sec.DPR} along with its two subsections, the DPR is obtained at both zero and non-zero external magnetic field case. Sec.~\ref{sec.NJL} is devoted to estimate the constituent quark mass using the NJL model. In Sec.~\ref{sec.results} we show our numerical results and finally summarize and conclude in Sec.~\ref{sec.summary}.


\section{DILEPTON PRODUCTION RATE} \label{sec.DPR}
\begin{figure}[h]
	\begin{center}
		\includegraphics[angle=0,scale=0.5]{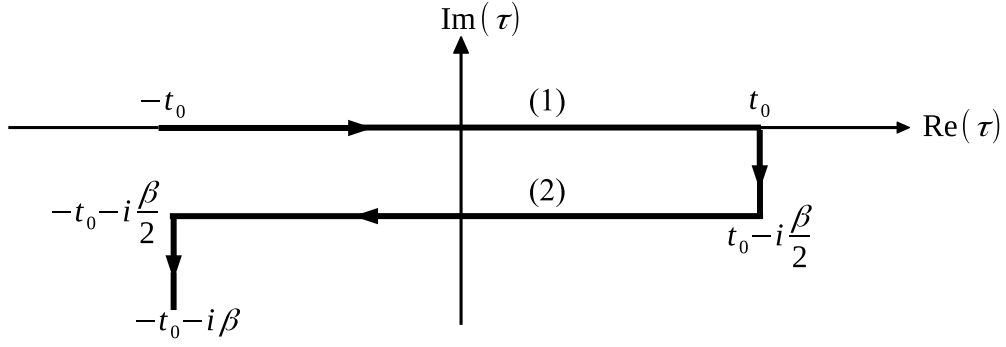}
	\end{center}
	\caption{The symmetric Schwinger-Keldysh contour $C$ in the complex time plane used in the Real Time Formalism with $t_0\to\infty$ and $\beta = 1/T$. The labels `(1)' and `(2)' refer to the two horizontal segments of the contour. }
	\label{fig.contour}
\end{figure}

In order to calculate the dilepton production rate (DPR) from hot and dense magnetized QCD medium, we use the standard prescriptions as given in Refs.~\cite{Mallik:2016anp, Bandyopadhyay:2017raf, Ghosh:2018xhh} from which, for completeness we briefly sketch the important steps here. Let us consider an initial state of quark/antiquark $\ket{i}=\ket{\mcI(p_\mcI)}$ containing a quark/antiquark with momentum $p_\mcI$ going to a final state $\ket{f}=\ket{\mcF(p_\mcF),l^+(p_+)l^-(p_-)}$ which contains a quark/antiquark of momentum $p_\mcF$ plus a pair of leptons of momenta $p_+$ and $p_-$ respectively. The probability amplitude for the transition $\ket{i}\to\ket{f}$ is $|\expectationvalue{f}{\mcSh}{i}|^2$ where $\mcSh$ is the scattering matrix given by
\begin{eqnarray}
\mcSh = \mcT \TB{ \exp \SB{ i\int \scrL_\text{int}(x)d^4x } }
\end{eqnarray}
in which $\mcT$ is the time-ordering symbol and 
\begin{eqnarray}
\scrL_\text{int}(x) = j^\mu(x)A_\mu(x)+J^\mu(x)A_\mu(x)
\end{eqnarray}
is the local interaction Lagrangian (density). We will be using metric tensor with signature $g^\munu=\text{diag}(1,-1,-1,-1)$. In the above equation $j^\mu(x)$ and $J^\mu(x)$ are respectively the conserved vector currents corresponding to leptons and quarks which couple to the photon field $A^\mu(x)$. An expansion of $|\expectationvalue{f}{\mcSh}{i}|^2$ up to second order in perturbation series, after some simplifications, yields~\cite{Ghosh:2018xhh}
\begin{eqnarray}
|\expectationvalue{f}{\mcSh}{i}|^2 = \int\int d^4x' d^4x e^{i(p_++p_-)\cdot x'}\frac{1}{(p_++p_-)^4}
\ExpectationValue{l^+(p_+)l^-(p_-)}{j^\mu(0)}{0} \ExpectationValue{0}{j^{\nu\dagger}(0)}{l^+(p_+)l^-(p_-)} \nn \\
\ExpectationValue{\mcF(p_\mcF)}{J_\mu(x')}{\mcI(p_\mcI)} \ExpectationValue{\mcI(p_\mcI)}{J_\nu^\dagger(0)}{\mcF(p_\mcF)}. \label{eq.sfi}
\end{eqnarray}
The dilepton multiplicity is then obtained from the following expression
\begin{eqnarray}
N = \frac{1}{\mcZ} \sum_{\text{spins}}^{}\int\frac{d^3p_+}{(2\pi)^32p_+^0}\int\frac{d^3p_-}{(2\pi)^32p_-^0} \sum_{\mcI,\mcF}^{}
\exp\FB{-p_\mcI^0/T} |\expectationvalue{f}{\mcSh}{i}|^2 \label{eq.DM}
\end{eqnarray}
where $\mcZ$ is the partition function of the system and the sum refers to sum over all leptonic spin configuration. Substituting Eq.~\eqref{eq.sfi} into Eq.~\eqref{eq.DM}, we get, after some calculations,
\begin{eqnarray}
N = \int d^4x\int \frac{d^4q}{(2\pi)^4} e^{-q^0/T}\frac{1}{q^4} \mcW_{+\munu}(q) \mcL_+^\munu(q) 
\label{eq.dilepton.mult}
\end{eqnarray}
where,
\begin{eqnarray}
\mcW_+^\munu(q) &=& \int d^4x e^{iq\cdot x} \ensembleaverage{J^\mu(x)J^{\nu\dagger}(0)}, \label{eq.W+}\\
\mcL_+^\munu(q) &=& \int d^4x e^{iq\cdot x} \ExpectationValue{0}{j^{\nu\dagger}(x)j^{\mu}(0)}{0} \label{eq.L+}
\end{eqnarray}
in which $\ensembleaverage{...}$ represents the ensemble average. Therefore the DPR is obtained from Eq.~\eqref{eq.dilepton.mult} as
\begin{eqnarray}
\text{DPR} = \frac{dN}{d^4xd^4q} = \frac{1}{(2\pi)^4} \frac{e^{-q^0/T}}{q^4}  \mcW_{+\munu}(q) \mcL_+^\munu(q). \label{eq.DPR.1}
\end{eqnarray}
Note that, both the $\mcW_+^\munu(q)$ and $\mcL_+^\munu(q)$ in Eqs.~\eqref{eq.W+} and \eqref{eq.L+} contain the Fourier transform of the two-point current-current correlation functions. However, it is more useful to express them in terms of the time-ordered correlators for which Eqs.~\eqref{eq.W+} and \eqref{eq.L+} can be rewritten as
\begin{eqnarray}
\mcW_+^\munu(q) &=& \FB{\frac{1}{1+e^{-q^0/T}}} 2 \IM \mcW_{11}^\munu(q)~, \label{eq.W+.2}\\
\mcL_+^\munu(q) &=& 2 \IM \mcL^\munu(q) \label{eq.L+.2}
\end{eqnarray}
where,
\begin{eqnarray}
\mcW_{11}^\munu(q) &=& i\int d^4x e^{iq\cdot x} \ensembleaverage{\mcTc J^\mu(\tau,\vec{x})J^{\nu\dagger}(0,\vec{0})}_{11}~, \label{eq.W11}\\
\mcL^\munu(q) &=& i\int d^4x e^{iq\cdot x} \ExpectationValue{0}{\mcT j^{\nu\dagger}(x)j^{\mu}(0)}{0} \label{eq.L}
\end{eqnarray}
in which, $\mcTc$ refers to the time ordering with respect to the symmetric Schwinger-Keldysh complex time contour used in the RTF of finite temperature field theory as shown in Fig.~\ref{fig.contour}: the subscript $11$ in Eq.~\eqref{eq.W11} implies that the two points are on the real horizontal segment `(1)' of the contour $C$ in which case $\mcTc$ reduces to the ordinary time ordering denoted by $\mcT$.
The quantities $\mcW_{11}^\munu(q)$ and $\mcL^\munu(q)$ are termed as matter and lepton tensors respectively. Substituting Eqs.~\eqref{eq.W+.2} and \eqref{eq.L+.2} in Eq.~\eqref{eq.DPR.1}, one obtains
\begin{eqnarray}
\text{DPR} = \frac{dN}{d^4xd^4q} = \frac{1}{4\pi^4q^4} \FB{\frac{1}{e^{q^0/T}+1}}\IM \mcW_{11}^\munu(q)\IM \mcL_\munu(q). \label{eq.DPR.2}
\end{eqnarray}
Our next task is to calculate the quantities $\mcW_{11}^\munu(q)$ and $\mcL^\munu(q)$ for which we now require the explicit form of the currents $J^\mu(x)$ and $j^\mu(x)$. They are given by
\begin{eqnarray}
J^\mu(x) &=& e\Psibar(x)\hat{Q}\gamma^\mu\Psi(x)~, \label{eq.J} \\
j^\mu(x) &=& -e\psibar(x)\gamma^\mu\psi(x) \label{eq.j}
\end{eqnarray}
where, $\Psi=\begin{pmatrix} u \\ d \end{pmatrix}$ is the 2-flavour quark isospin doublet with $u$ and $d$ being respectively the up and down quark fields, $\hat{Q}=\frac{1}{3} \begin{pmatrix} 2 & 0 \\ 0 & -1 \end{pmatrix}$, $\psi$ is the lepton field and $e$ is the electric charge of a proton. Substituting Eqs.~\eqref{eq.J} and \eqref{eq.j} into Eqs.~\eqref{eq.W11} and \eqref{eq.L} followed by using Wick's theorem, we get, after some simplifications,
\begin{eqnarray}
\mcW_{11}^\munu(q) &=& i\int\frac{d^4k}{(2\pi)^4} \Tr_\text{d,f,c} 
\TB{\gamma^\mu \hat{Q} S_{11}(p=q+k)\gamma^\nu \hat{Q} S_{11}(k)}~, \label{eq.W11.2}\\
\mcL^\munu(q) &=& i e^2\int\frac{d^4k}{(2\pi)^4} \Tr_\text{d} \TB{\gamma^\nu S(p=q+k)\gamma^\mu S(k)} \label{eq.L.2}
\end{eqnarray}
where the subscript `d', `f', and `c' in the trace correspond to trace over Dirac, flavour and color spaces respectively; $S_{11}(p)$ is the 11-component of the real time quark propagator and $S(k)$ is the vacuum lepton Feynman propagator given by 
\begin{eqnarray}
S(p) = \frac{-(\cancel{p}+m_L)}{p^2-m_L^2+i\varepsilon} \label{eq.lepton.propagator}
\end{eqnarray}
with $m_L$ being the mass of the lepton. It is to be noted that, both the matter tensor $\mcW_{11}^\munu(q)$ and the leptonic tensor $\mcL^\munu(q)$ are transverse to the momentum $q^\mu$ i.e.
\begin{eqnarray}
q_\mu \mcW_{11}^\munu(q) = q_\mu \mcL^\munu(q) = 0 \label{eq.transversality}
\end{eqnarray}
which is a consequence of the conservation of the currents $J^\mu(x)$ and $j^\mu(x)$: $\del_\mu J^\mu(x)=\del_\mu j^\mu(x)=0$. We now consider the two separate cases: (i) zero external magnetic field $(B=0)$ and (ii) non-zero external magnetic field $(B\ne 0)$ in order to calculate the DPR in the following subsections.
\subsection{DPR AT $B=0$} \label{subsec.DPR.1}
\begin{figure}[h]
	\begin{center}
		\includegraphics[angle=0,scale=0.65]{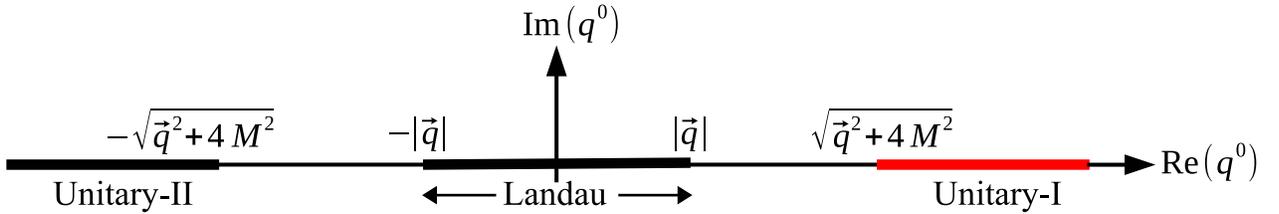}
	\end{center}
	\caption{(Color Online) The branch cuts of $\mcW_{11}^\munu(q)$ in the complex $q^0$ plane for a given $\vec{q}$. The red region corresponds to 
		kinematic domain for the physical dileptons production defined in terms of $q^0>0$ and $q^2>0$. }
	\label{fig.analytic0}
\end{figure}

At $B=0$, the transversility condition $q_\mu \mcL^\munu(q)$ of Eq.~\eqref{eq.transversality} implies that the Lorentz structure of $\mcL^\munu(q)$ must be of the form:
\begin{eqnarray}
\mcL^\munu(q) = \FB{g^\munu-\frac{q^\mu q^\nu}{q^2}}\FB{\frac{1}{3}g_\alphabeta \mcL^\alphabeta}. \label{eq.L.0}
\end{eqnarray}
Substituting the above equation in Eq.~\eqref{eq.DPR.2}, and making use of $q_\mu \mcW_{11}^\munu(q)=0$ we get the DPR at $B=0$ as
\begin{eqnarray}
\text{DPR}_{B=0} = \FB{\frac{dN}{d^4xd^4q}}_{B=0} = \frac{1}{12\pi^4q^4} \FB{\frac{1}{e^{q^0/T}+1}}
g_\munu\IM \mcW_{11}^\munu(q) g_\alphabeta \IM \mcL^\alphabeta(q). \label{eq.DPR.3}
\end{eqnarray}
The calculation of $g_\alphabeta \IM \mcL^\alphabeta(q)$ is straightforward for which we substitute Eq.~\eqref{eq.lepton.propagator} into Eq.~\eqref{eq.L.2} and get, after some algebra,
\begin{eqnarray}
g_\alphabeta \IM \mcL^\alphabeta(q) = \frac{-e^2}{4\pi}q^2\FB{1+\frac{2m_L^2}{q^2}}\FB{1-\frac{4m_L^2}{q^2}}^{1/2}\Theta\FB{q^2-4m_L^2}.
\label{eq.L.3}
\end{eqnarray}
On the other hand, in order to calculate $g_\munu\IM \mcW_{11}^\munu(q)$, we note that the 11-component of the real time thermal quark propagator is 
\begin{eqnarray}
S_{11}(p,M) = \FB{\cancel{p}+M}\TB{\frac{-1}{p^2-M^2+i\varepsilon}-2\pi i \eta(p\cdot u)\delta(p^2-M^2)}
\otimes\identity_\text{Flavour}\otimes\identity_\text{Colour} \label{eq.S11.T}
\end{eqnarray}
where, $M$ is the \textit{constituent quark mass}, $u^\mu$ is the four-velocity of the thermal bath, $\eta(x)=\Theta(x)f^+(x)+\Theta(-x)f^-(-x)$, 
\begin{eqnarray}
f^\pm(x) = \TB{\exp\FB{\frac{x\mp\mu_B/3}{T}}+1}^{-1}
\end{eqnarray}
and $\mu_B$ is the baryon chemical potential. In the local rest frame (LRF) of the medium, $u^\mu_\text{LRF}\equiv(1,\vec{0})$. Substitution of Eq.~\eqref{eq.S11.T} into Eq.~\eqref{eq.W11.2} yields, after some simplifications,
\begin{eqnarray}
g_\munu\IM \mcW_{11}^\munu(q) = N_c\sum_{f\in\{u,d\}} e_f^2 \pi \int\frac{d^3k}{(2\pi)^3}&&\frac{1}{4\omega_k\omega_p}\big[
\SB{1-f^-(\omega_k)-f^+(\omega_p)+2f^-(\omega_k)f^+(\omega_p)}\mcN(k^0=-\omega_k)\delta(q^0-\omega_k-\omega_p) \nn \\
&&+ \SB{1-f^+(\omega_k)-f^-(\omega_p)+2f^+(\omega_k)f^-(\omega_p)}\mcN(k^0=\omega_k)\delta(q^0+\omega_k+\omega_p) \nn \\
&&+ \SB{-f^-(\omega_k)-f^-(\omega_p)+2f^-(\omega_k)f^-(\omega_p)}\mcN(k^0=-\omega_k)\delta(q^0-\omega_k+\omega_p) \nn \\
&&+ \SB{-f^+(\omega_k)-f^+(\omega_p)+2f^+(\omega_k)f^+(\omega_p)}\mcN(k^0=\omega_k)\delta(q^0+\omega_k-\omega_p) \big]
\label{eq.W11.7}
\end{eqnarray}
where, $e_u=2e/3$, $e_d=-e/3$, $\omega_k=\sqrt{\vec{k}^2+M^2}$, $\omega_p=\sqrt{\vec{p}^2+M^2}=\sqrt{(\vec{q}+\vec{k})^2+M^2}$ and 
$\mcN (q,k)= 8(k^2+q\cdot k - 2M^2)$. The terms with the four different Dirac delta functions appearing in the above equation 
correspond to different physical processes giving rise to the branch cuts of $\mcW_{11}^\munu(q)$ in the complex $q^0$ plane. 
The first delta function is the Unitary-I cut which corresponds to the contribution from quark-antiquark 
annihilation to a time-like virtual photon produced with positive energy (and the time reversed process as the decay of the photon to the quark-antiquark pair). 
The term with the second delta function is the Unitary-II cut which corresponds to the quark-antiquark-photon annihilation to vacuum 
(and the time reversed process as the vacuum to quark-antiquark-photon transition) in which the virtual photon is time-like but carries negative energy. 
Similarly, the terms with the last two delta functions (called the Landau cuts) correspond to the scattering processes where 
a quark/antiquark absorbs a space-like virtual photon (and the corresponding time reversed process as the emission of the photon by the quark/antiquark).  
Each of the Dirac delta functions contributes in their respective kinematic domains as shown in Fig.~\ref{fig.analytic0}. 
The kinematic domains for the Unitary-I and Unitary-II cuts are respectively $\sqrt{\vec{q}^2+4M^2}<q^0<\infty$ and 
$-\infty<q^0<-\sqrt{\vec{q}^2+4M^2}$ whereas the same for the Landau cuts is $|q^0|<|\vec{q}|$. 
We restrict ourselves to the physical dileptons of positive total energy and time-like four-momentum i.e. $q^0>0$ and $q^2>0$ 
for which only the Unitary-I cut contributes as shown by the red region in Fig.~\ref{fig.analytic0}.

Evaluating the angular integrals of Eq.~\eqref{eq.W11.7}, using the first Dirac delta function, we get for $q^0>0$ and $q^2>0$:
\begin{eqnarray}
g_\munu\IM \mcW_{11}^\munu(q) = \Theta\FB{q^2-4M^2}\sum_{f\in\{u,d\}} \frac{N_c e_f^2}{16\pi|\vec{q}|} \int_{\omega_-}^{\omega_+}d\omega_k
\SB{1-f^-(\omega_k)-f^+(\omega_p)+2f^-(\omega_k)f^+(\omega_p)}\mcN(k^0=-\omega_k)\Big|_{\theta=\theta_0} \label{eq.W11.3}
\end{eqnarray}
where, $\omega_\pm = \frac{1}{2q^2}\TB{q^0q^2\pm|\vec{q}|\lambda^{1/2}(q^2,M^2,M^2)}$, $\theta$ is the angle between 
$\vec{q}$ and $\vec{k}$ and $\theta_0 = \cos^{-1}\FB{\frac{q^2-2q^0\omega_k}{2|\vec{q}||\vec{k}|}}$ with $\lambda(x,y,z)$ being the 
K\"all\'en function. Evaluating the remaining $d\omega_k$ integral of Eq.~\eqref{eq.W11.3} we arrive at
\begin{eqnarray}
g_\munu\IM \mcW_{11}^\munu(q) = -\Theta\FB{q^2-4M^2}N_c \FB{e_u^2+e_d^2} \frac{Tq^2}{4\pi|\vec{q}|} \coth\FB{\frac{q^0}{2T}}
\FB{1+\frac{2M^2}{q^2}} \ln \TB{\SB{\frac{e^{(q^0+q_-)/T}+1}{e^{(q^0+q_+)/T}+1}}\FB{\frac{e^{q_+/T}+1}{e^{q_-/T}+1}}} \label{eq.W11.4}
\end{eqnarray}
where, $q_\pm = -\frac{1}{2}\TB{q^0 \pm |\vec{q}|\sqrt{1-\frac{4M^2}{q^2}}} + \mu_B/3$. Finally substituting Eqs.~\eqref{eq.L.3} and 
\eqref{eq.W11.4} into Eq.~\eqref{eq.DPR.3} we arrive at the following analytical expression for the DPR
\begin{eqnarray}
\text{DPR}_{B=0} = \FB{\frac{dN}{d^4xd^4q}}_{B=0} = \Theta\FB{q^2-4m_L^2} \Theta\FB{q^2-4M^2}  N_c\frac{ e^2 (e_u^2+e_d^2)}{ 192\pi^6} \frac{T}{|\vec{q}|} \FB{\frac{1}{e^{q^0/T}-1}} \nn \\ \FB{1+\frac{2m_L^2}{q^2}}\FB{1-\frac{4m_L^2}{q^2}}^{1/2} 
\FB{1+\frac{2M^2}{q^2}} \ln \TB{\SB{\frac{e^{(q^0+q_-)/T}+1}{e^{(q^0+q_+)/T}+1}}\FB{\frac{e^{q_+/T}+1}{e^{q_-/T}+1}}}
 \label{eq.DPR.4}
\end{eqnarray}
which agrees with the expression for the Born rate of dilepton production of Ref.~\cite{Greiner:2010zg}. It is to be noted that, the presence of the 
step functions in the above equation restricts the production of dileptons with invariant mass $q^2<4m_L^2$ and  $q^2<4M^2$.


\subsection{DPR AT $B\ne0$} \label{subsec.DPR.2}
Let us now consider a constant magnetic field $B$ in the positive z-direction. The presence of external magnetic 
field will break the rotational symmetry and thus any four vector $a^\mu$ can be decomposed as $a^\mu=(a_\parallel^\mu+a_\perp^\mu)$ 
where $a_\parallel^\mu = \gpll^\munu a_\nu$ and $a_\perp^\mu = \gper^\munu a_\nu$; the corresponding decomposition of the metric 
tensor reads $g^\munu=(\gpll^\munu+\gper^\munu)$ with $\gpll^\munu=\text{diag}(1,0,0,-1)$ and $\gper^\munu=\text{diag}(0,-1,-1,0)$.

For simplicity in the analytical calculations, we take the resultant transverse momentum of the dileptons to be zero i.e. $q_\perp=0$. 
Therefore, the transversility condition $\qpll^\mu \mcL_\munu(\qpll)=0$ of Eq.~\eqref{eq.transversality} enforces that the 
Lorentz structure of $\mcL^\munu(\qpll)$ has to be of the form:
\begin{eqnarray}
\mcL^\munu(\qpll) = \FB{\gpll^\munu-\frac{\qpll^\mu \qpll^\nu}{\qpll^2}}\FB{\gpll^\alphabeta \mcL_\alphabeta}
+ \gper^\munu\FB{\frac{1}{2}\gper^\alphabeta \mcL_\alphabeta}.
\end{eqnarray}
Substituting the above equation in Eq.~\eqref{eq.DPR.2}, and making use of $q_{\parallel\mu} \mcW_{11}^\munu(\qpll)=0$ we get 
the DPR at $B\ne0$ as 
\begin{eqnarray}
\text{DPR}_{B\ne0} = \FB{\frac{dN}{d^4xd^4q}}_{B\ne0} = \frac{1}{4\pi^4\qpll^4} \FB{\frac{1}{e^{q^0/T}+1}}
\TB{ g_\parallel^\munu\IM \mcW^{11}_\munu(\qpll) ~ g_\parallel^\alphabeta \IM \mcL_\alphabeta(\qpll)
	+ \frac{1}{2}g_\perp^\munu\IM \mcW^{11}_\munu(\qpll) ~ g_\perp^\alphabeta \IM \mcL_\alphabeta(\qpll)}. 
\label{eq.DPR.5}
\end{eqnarray}
Let us now proceed to calculate the four quantities $g_{\parallel,\perp}^\munu\IM \mcW^{11}_\munu(\qpll)$ and 
$g_{\parallel,\perp}^\munu\IM \mcL_\munu(\qpll)$ which appear within the third bracket of the above equation. 
First we note that, the 11-component of the thermo-magnetic quark propagator becomes 
\begin{eqnarray}
S_{11}(p) = \FB{\begin{array}{cc}
	S_{11}^u(p) & 0 \\ 
	0 & S_{11}^d(p)
	\end{array}} \otimes\identity_\text{Colour}.
\end{eqnarray}
in which, unlike Eq.~\eqref{eq.S11.T}, the diagonal elements (corresponding to up and down quark) have become 
different due to the presence of the external magnetic field. In the above equation, 
\begin{eqnarray}
S_{11}^f(p) = \sum_{s\in\{\pm1\}}^{}\sum_{n=0}^{\infty} \scrD^f_{ns}(p)\TB{\frac{-1}{\ppll^2-(M_n^f-s\kappa_fB)^2+i\varepsilon}
-2\pi i \eta(p\cdot u)\delta\FB{\ppll^2-(M_n^f-s\kappa_fB)^2}} \label{eq.S11.TB}
\end{eqnarray}
where $M_n^f = \sqrt{M^2+2n|e_fB|}$, $\kappa_f$ is the anomalous magnetic moment of quark flavour $f\in \{u,d\}$ and
\begin{eqnarray}
\scrD^f_{ns}(p) = (-1)^ne^{-\alpha_p^f}\frac{1}{2M_n^f}(1-\delta_n^0\delta_s^{-1})
\Big[(M_n^f+sM)(\cancel{p}_\parallel-\kappa_fB+sM_n^f) \FB{\identity+\text{sign}(e_f)i\gamma^1\gamma^2}L_n(2\alpha_p^f) \nn \\
-(M_n^f-sM)(\cancel{p}_\parallel-\kappa_fB-sM_n^f)\FB{\identity-\text{sign}(e_f)i\gamma^1\gamma^2}L_{n-1}(2\alpha_p^f) \nn \\
-4s \FB{ \cancel{p}_\parallel - \text{sign}(e_f)i\gamma^1\gamma^2 (\kappa_fB-sM_n^f)} \text{sign}(e_f)i\gamma^1\gamma^2 \cancel{p}_\perp
L_{n-1}^1(2\alpha_p^f)   \Big]
\end{eqnarray}
in which $\alpha_p^f=-p_\perp^2/|e_fB|$.

Few comments on the AMM of the quarks are in order here. The AMM of a particle may have different origin apart from having an internal structure. For example, it is well known from Quantum Electrodynamics (QED) that, the electron, although being an elementary particle having no internal structure, possesses AMM due to the quantum corrections. The Land\'e g-factor of the electron comes out to be $2+\alpha/\pi$ upto one-loop in QED where $\alpha$ is the fine structure constant. A simpler way of understanding the existence of the AMM of electron with charge $e$ is as follows. When the electrons are coupled to the photons via the minimal coupling, the ordinary derivatives ($\del^\mu$) are modified to the covariant derivatives 
\begin{eqnarray}
\del^\mu \to D^\mu = \del^\mu + ieA^\mu  
\end{eqnarray}
where $A^\mu$ is the photon field. In that case, the Dirac equation can be recast as~\cite{Schwartz:2013pla,Peskin:1995ev}
\begin{eqnarray}
\TB{D_\mu D^\mu + g\frac{e}{4} F^{\mu\nu}\sigma_{\mu\nu} + m^2}\psi = 0
\end{eqnarray}
where, $F^\munu=(\del^\mu A^\nu-\del^\nu A^\mu)$ is the electromagnetic field strength tensor and $ \sigma_{\mu\nu}= i[\gamma_\mu, \gamma_\nu]/2 $. In the above equation, the second term within the square bracket corresponds to a magnetic dipole moment and the Dirac equation predicts that $g=2$. However, there can be an anomalous contribution to $g$, which comes from the quantum fluctuations viz. higher order loop correction to the QED vertex that has the same effect as an additional $ F^{\mu\nu}\sigma_{\mu\nu} $ term.

Quarks, being charged fermions, similar loop corrections lead to the AMM of quarks. But, the non-perturbative nature of Quantum Chromodynamics (QCD) forbids one to perform a first principle analytical calculation to extract the anomalous contribution to $g$. As an alternative, in this work, we have taken the values of the AMM of quarks calculated using the constituent quark model (CQM)~\cite{Halzen:1984mc,Sadooghi}. In CQM, the experimental values of the nucleons AMM are used to extract the AMM of the quarks.

As the quarks carry non-zero AMM, it is justified to consider the AMM of the quarks while calculating the matrix elements which will lead to an explicit AMM dependence in the DPR (since the quark propagator now has explicit AMM dependence). Moreover, to capture the effect of `strong' interaction in the DPR from the hot and dense magnetized medium, we will be using the NJL model which gives the effective quark mass $M$ as a function of temperature ($T$), baryon chemical potential ($\mu_B$), magnetic field ($eB$) as well as AMM ($\kappa$) of the quarks naively incorporating the non-perturbative aspects of QCD. As the $M=M(T,\mu_B,eB,\kappa)$ will go as an input in the expression of the DPR, an implicit AMM dependence in DPR will also come through the $\kappa$ dependence of $M$.

Substituting Eq.~\eqref{eq.S11.TB} into Eq.~\eqref{eq.W11.2}, 
we get after some simplifications
\begin{eqnarray}
g^\munu_{\parallel,\perp}\IM \mcW^{11}_\munu(\qpll) = N_c\sum_{f\in\{u,d\}} e_f^2 &&\sum_{s_k\in\{\pm1\}} \sum_{s_p\in\{\pm1\}}
\sum_{l=0}^{\infty}\sum_{n=0}^{\infty} \pi \int\frac{d^3k}{(2\pi)^3}\frac{1}{4\omega_k^{lf}\omega_p^{nf}} \nn \\
&&\times \Big[ \SB{1-f^-(\omega_k^{lf})-f^+(\omega_p^{nf})+2f^-(\omega_k^{lf})f^+(\omega_p^{nf})}
\mcN_{\parallel,\perp}^{fln}(k^0=-\omega_k^{lf})\delta(q^0-\omega_k^{lf}-\omega_p^{nf}) \nn \\
&&+ \SB{1-f^+(\omega_k^{lf})-f^-(\omega_p^{nf})+2f^+(\omega_k^{lf})f^-(\omega_p^{nf})}
\mcN_{\parallel,\perp}^{fln}(k^0=\omega_k^{lf})\delta(q^0+\omega_k^{lf}+\omega_p^{nf}) \nn \\
&&+\SB{-f^-(\omega_k^{lf})-f^-(\omega_p^{nf})+2f^-(\omega_k^{lf})f^-(\omega_p^{nf})}
\mcN_{\parallel,\perp}^{fln}(k^0=-\omega_k^{lf})\delta(q^0-\omega_k^{lf}+\omega_p^{nf}) \nn \\
&&+\SB{-f^+(\omega_k^{lf})-f^+(\omega_p^{nf})+2f^+(\omega_k^{lf})f^+(\omega_p^{nf})}
\mcN_{\parallel,\perp}^{fln}(k^0=\omega_k^{lf})\delta(q^0+\omega_k^{lf}-\omega_p^{nf}) \Big] \label{eq.W11.5}
\end{eqnarray}
where, $\omega_k^{lf} = \sqrt{k_z^2+(M_l-s_k\kappa_fB)^2}$, $\omega_p^{nf} = \sqrt{p_z^2+(M_n-s_p\kappa_fB)^2}$ and
\begin{eqnarray}
\mcN_{\parallel,\perp}^{fln}(\qpll,\kpll) = g^\munu_{\parallel,\perp}\Tr_\text{d}\TB{\gamma^\mu\scrD_{ns_p}^f(p=q+k)\gamma^\nu\scrD_{ls_k}^f(k)}. \label{eq.N.1}
\end{eqnarray}
\begin{figure}[h]
	\begin{center}
		\includegraphics[angle=0,scale=0.65]{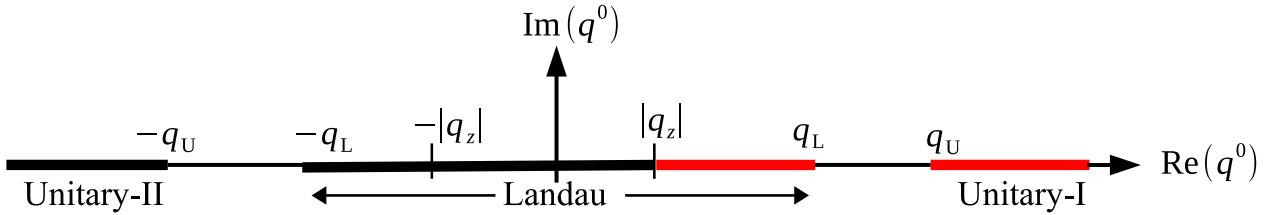}
	\end{center}
	\caption{(Color Online) The branch cuts of $\mcW_{11}^\munu(\qpll)$ in the complex $q^0$ plane for a given $q_z$ and $B$. The points  $q_\text{U}$ and $q_\text{L}$ refer to 
		$q_\text{U} = \sqrt{q_z^2+4\FB{M-\kappa_uB}^2}$ and $q_\text{L} = \sqrt{q_z^2+\FB{\sqrt{M^2+2|e_uB|}+\kappa_uB-\MB{M-\kappa_uB}}^2}$. 
		The red region corresponds to the physical dileptons defined in terms of $q^0>0$ and $\qpll^2>0$.}
	\label{fig.analytic1}
\end{figure}
Alike Eq.~\eqref{eq.W11.7}, Eq.~\eqref{eq.W11.5} also contains four Dirac delta functions corresponding to Unitary and Landau cuts. 
However, because of the external magnetic field and the non-zero AMM of the quarks, the kinematic domains for the cuts are largely 
modified as discussed in Refs.~\cite{Ghosh:2017rjo,Ghosh:2019fet}. Analyzing Eq.~\eqref{eq.W11.5}, we find that, the kinematic domains for the Unitary-I and 
Unitary-II cuts are respectively $\sqrt{q_z^2+4\FB{M-\kappa_uB}^2}<q^0<\infty$ and $-\infty<q^0<-\sqrt{q_z^2+4\FB{M-\kappa_uB}^2}$. 
The kinematic domain for the Landau cuts comes out to be 
\begin{eqnarray}
|q^0|<\sqrt{q_z^2+\FB{\sqrt{M^2+2|e_uB|}+\kappa_uB-\MB{M-\kappa_uB}}^2}. \label{eq.LandauCut}
\end{eqnarray}
The analytic structure of $\mcW_{11}^\munu(\qpll)$ in presence of external magnetic field is shown in Fig.~\ref{fig.analytic1}. 
Considering the physical dileptons with positive total energy and time-like resultant momentum defined in terms of $q^0>0$ and $\qpll^2>0$ 
(shown as red region in Fig.~\ref{fig.analytic1}), we find that the Unitary-II cut does not contribute. 
Moreover, unlike the $B=0$ case, it is interesting to notice that a portion of the Landau cut also contributes to the physical region 
$q^0>0$ and $\qpll^2>0$ . Physically it corresponds to the emission/absorption of time-like virtual photon with positive energy by 
a quark/antiquark which changes its Landau level by unity (i.e. the quark in the Landau level $l$ goes to the Landau level $(l\pm1)$ after 
absorbing/emitting the photon). The appearance of the Landau cut in the physical region will have consequences and will lead to the enhancement 
of dilepton yields in the low invariant mass region.

Next, we analytically evaluate the $d^2k_\perp$ integral of Eq.~\eqref{eq.W11.5} by using the orthogonality of the 
Laguerre polynomials appearing in Eq.~\eqref{eq.N.1} followed by the evaluation of the remaining $dk_z$ integral using the 
Dirac delta functions. This leads to the following analytic expression
\begin{eqnarray}
g^\munu_{\parallel,\perp}\IM \mcW^{11}_\munu(\qpll) &=& N_c\sum_{f\in\{u,d\}} e_f^2 \sum_{s_k\in\{\pm1\}} \sum_{s_p\in\{\pm1\}}
\sum_{l=0}^{\infty}\sum_{n=(l-1)}^{(l+1)} \sum_{k_z\in\{k_z^\pm\}} 
\frac{1}{4}\lambda^{-\frac{1}{2}}\FB{\qpll^2,M_{lfs_k}^2,M_{nfs_p}^2} \nn \\
&& \times  \Big[ \SB{1-f^-(\omega_k^{lf})-f^+(\omega_p^{nf})+2f^-(\omega_k^{lf})f^+(\omega_p^{nf})}
\tilde{\mcN}_{\parallel,\perp}^{fln}(k^0=-\omega_k^{lf}) \Theta \FB{q^0-\sqrt{q_z^2+(M_{lfs_k}+M_{nfs_p})^2}} \nn \\
&& + \SB{1-f^+(\omega_k^{lf})-f^-(\omega_p^{nf})+2f^+(\omega_k^{lf})f^-(\omega_p^{nf})}
\tilde{\mcN}_{\parallel,\perp}^{fln}(k^0=\omega_k^{lf})\Theta \FB{-q^0-\sqrt{q_z^2+(M_{lfs_k}+M_{nfs_p})^2}} \nn \\
&&+\SB{-f^-(\omega_k^{lf})-f^-(\omega_p^{nf})+2f^-(\omega_k^{lf})f^-(\omega_p^{nf})}
\tilde{\mcN}_{\parallel,\perp}^{fln}(k^0=-\omega_k^{lf})\Theta\FB{-q^0-q_\text{min}} \Theta\FB{q^0+q_\text{max}}\nn \\
&& +\SB{-f^+(\omega_k^{lf})-f^+(\omega_p^{nf})+2f^+(\omega_k^{lf})f^+(\omega_p^{nf})}
\tilde{\mcN}_{\parallel,\perp}^{fln}(k^0=\omega_k^{lf})\Theta\FB{q^0-q_\text{min}} \Theta\FB{-q^0+q_\text{max}} \Big]
 \label{eq.W11.6}
\end{eqnarray}
where 
\begin{eqnarray}
\tilde{\mcN}_{\parallel}^{fln}(\qpll,\kpll) &=& \frac{|e_fB|}{2\pi}\frac{1}{M_l^fM_n^f}
(1-\delta_l^0\delta_{s_k}^{-1})(1-\delta_n^0\delta_{s_p}^{-1})(M_l^f-s_k\kappa_fB)(M_n^f-s_p\kappa_fB) \nn \\
&& \times \TB{-4|e_fB|n\delta_{l-1}^{n-1} -\delta_{l-1}^{n-1}(M-s_kM_l^f)(M-s_pM_n^f)-\delta_l^n(M+s_kM_l^f)(M+s_pM_n^f)}, \\
\tilde{\mcN}_{\perp}^{fln}(\qpll,\kpll) &=& -\frac{|e_fB|}{2\pi}\frac{1}{M_l^fM_n^f}
(1-\delta_l^0\delta_{s_k}^{-1})(1-\delta_n^0\delta_{s_p}^{-1}) 
\TB{ s_ks_p(\kpll^2+\qpll\cdot\kpll)+ (M_l^f-s_k\kappa_fB)(M_n^f-s_p\kappa_fB)} \nn \\
&& \times \TB{ \delta_{l-1}^{n}(M-s_kM_l^f)(M+s_pM_n^f)-\delta_{l}^{n-1}(M+s_kM_l^f)(M-s_pM_n^f)}
\end{eqnarray}
in which 
\begin{eqnarray}
M_{lfs_k}=|M_l^f-s_k\kappa_fB| 
\end{eqnarray}
is the AMM dependent effective quark mass and 
\begin{eqnarray}
k_z^\pm &=& \frac{1}{2\qpll^2}\TB{-q_z(\qpll^2+M_{lfs_k}^2-M_{nfs_p}^2) \pm |q^0|\lambda^{\frac{1}{2}}\fb{\qpll^2,M_{lfs_k}^2,M_{nfs_p}^2} }, \\
q_\text{min} &=& \text{min}\FB{q_z,\frac{M_{lfs_k}-M_{nfs_p}}{|M_{lfs_k}\pm M_{nfs_p}|}\sqrt{q_z^2+(M_{lfs_k}\pm M_{nfs_p})^2}},\\
q_\text{max} &=& \text{max}\FB{q_z,\frac{M_{lfs_k}-M_{nfs_p}}{|M_{lfs_k}\pm M_{nfs_p}|}\sqrt{q_z^2+(M_{lfs_k}\pm M_{nfs_p})^2}}.
\end{eqnarray}
The presence of the step functions in Eq.~\eqref{eq.W11.6} dictates the kinematic domains in which the quantity 
$g^\munu_{\parallel,\perp}\IM \mcW^{11}_\munu(\qpll)$ is non-zero. It can now clearly be seen that, for $q^0>0$ and $\qpll^2>0$, 
the second term within the third bracket (the Unitary-II cut) of Eq.~\eqref{eq.W11.6} does not contribute.


It is now straightforward to substitute Eqs.~\eqref{eq.L.0}, \eqref{eq.L.3} and \eqref{eq.W11.6} into Eq.~\eqref{eq.DPR.5} to get the DPR in 
presence of arbitrary external magnetic field. It is to be noted that, in the calculation of DPR, one requires the value 
of the constituent quark mass $M$ which will essentially depend on the external parameters like temperature $T$, 
chemical potential ($\mu_B$) and/or external magnetic field ($B$). In the next section, we will use the 2-flavour NJL model 
for the estimation of $M=M(T,\mu_B,B)$. 


\section{THE CONSTITUENT QUARK MASS IN THE NJL MODEL} \label{sec.NJL}
The Lagrangian for the 2-flavour NJL model in presence of an external electromagnetic field characterized by the classical four-potential $A_\mu^\text{ext}$ or the field 
strength tensor $F_\munu^\text{ext}=\del_\mu A_\nu^\text{ext}-\del_\nu A_\mu^\text{ext}$ is given by 
\begin{eqnarray}
\scrL_\text{NJL} = \Psibar\FB{i\gamma^\mu\del_\mu - e\hat{Q}\gamma^\mu (A_\mu + A_\mu^\text{ext}) +\frac{1}{2}\hat{\kappa}\sigma^\munu F_\munu^\text{ext}-m}\Psi 
+ G \SB{(\Psibar\Psi)^2+(\Psibar i\gamma^5\vec{\tau}\Psi)^2} \label{eq.lagrangian}
\end{eqnarray}
where, $\Psi=\begin{pmatrix} u \\ d \end{pmatrix}$ is the quark flavour doublet, 
$\hat{\kappa}=\begin{pmatrix} \kappa_u & 0 \\ 0 & \kappa_d \end{pmatrix}$ is the AMM flavour matrix, $G$ is the coupling in the scalar channel and $m$ is the 
current quark mass. In the Mean Field Approximation (MFA), the constituent quark mass $M$ can be obtained by solving the gap equations. At $B=0$, 
the gap equation reads~\cite{Chaudhuri:2019lbw},
\begin{eqnarray}
M = m + 2G N_c\sum_{f\in\{u,d\}} \int\frac{d^3k}{(2\pi)^3}\frac{2M}{\omega_k}\TB{\Theta(\Lambda-|\vec{k}|)-f_+(\omega_k)-f_-(\omega_k)}
\end{eqnarray}
where, $\Lambda$ is the three-momentum cutoff that will regulate the Ultra-Violet (UV) divergences. 
For a constant magnetic field along the $\hat{z}$ direction $\vec{B}=B\hat{z}$, the gap equation becomes~\cite{Klevansky,Chaudhuri:2019lbw}
\begin{eqnarray}
M = m + 2G N_c \sum_{f\in\{u,d\}}&&\frac{|e_fB|}{2\pi}\sum_{l=0}^{\infty}\sum_{s\in\{\pm1\}}(1-\delta_l^0\delta_s^{-1})
\FB{1-\frac{s\kappa_fB}{M_l^f}} \nn \\
&& \times\int_{0}^{\infty}\frac{dk_z}{2\pi}\frac{2M}{\omega_k^{lf}}
\TB{\Theta\FB{\Lambda^2-k_z^2-\vec{k}_{\perp nfs}^2}\Theta\FB{\vec{k}_{\perp nfs}^2}-f^+(\omega_k^{lf})-f^-(\omega_k^{lf})}
\end{eqnarray}
where, $\vec{k}_{\perp nfs}^2 = 2n|e_fB|+\kappa_fB(\kappa_fB-2sM_n^f)$ is the AMM dependent effective transverse momentum of the 
quarks. The two step functions in the above equation respectively correspond to the UV and AMM blocking as discussed in 
Ref.~\cite{Chaudhuri:2019lbw}. Follwoing Refs.~\cite{Zhang}, the values of the parameters of the NJL model are chosen as $\Lambda=587.9$ MeV, $G=2.44\Lambda^2$ and $m = 5.6$ MeV. 
The values of the AMM of the quarks are taken from Ref.~\cite{Sadooghi} as $\kappa_u=0.29016\times e_u$ GeV$^{-1}$ and $\kappa_d=0.35986\times e_d$ GeV$^{-1}$.
\begin{figure}[h]
	\begin{center}
		\includegraphics[angle=-90,scale=0.33]{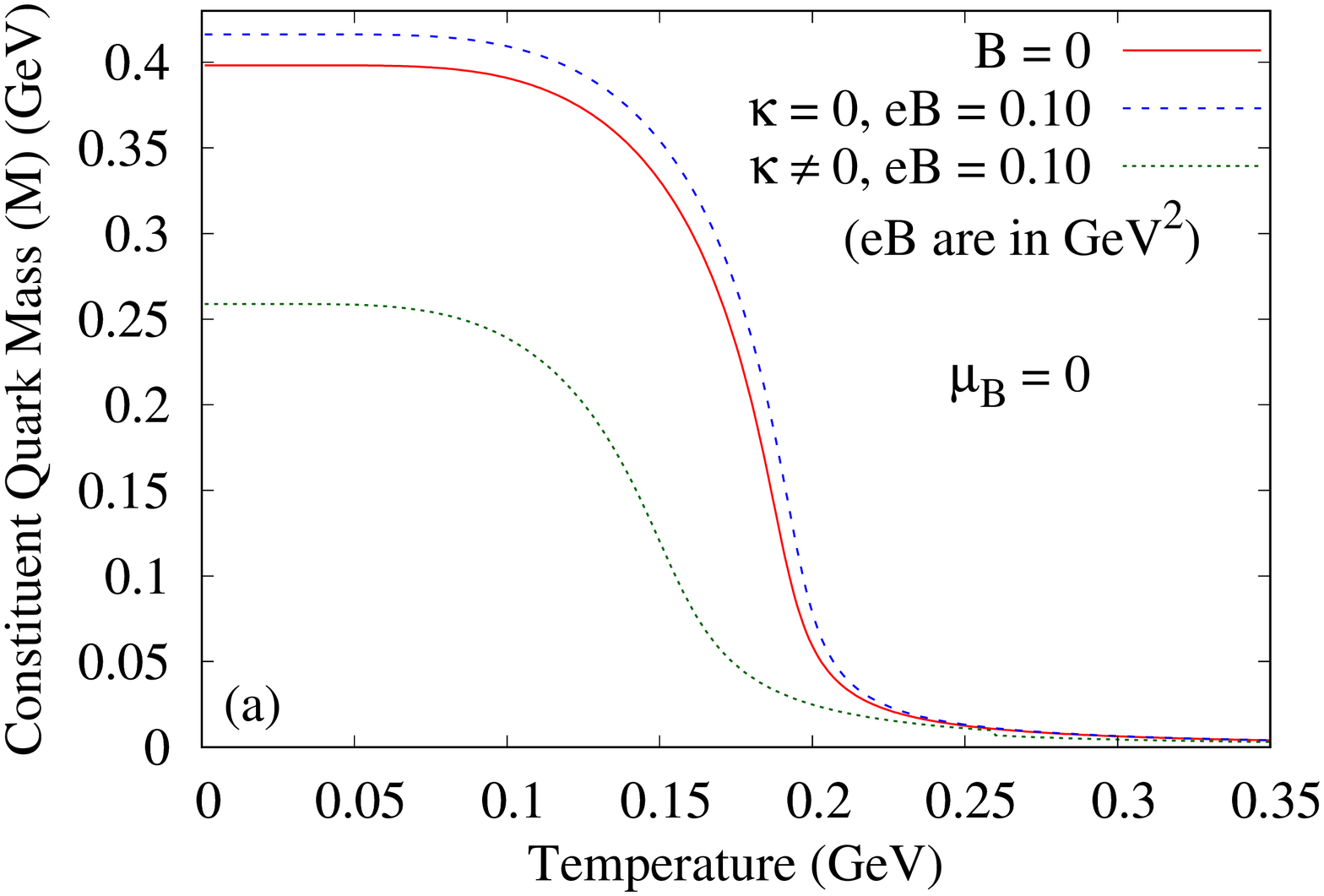}  \includegraphics[angle=-90,scale=0.33]{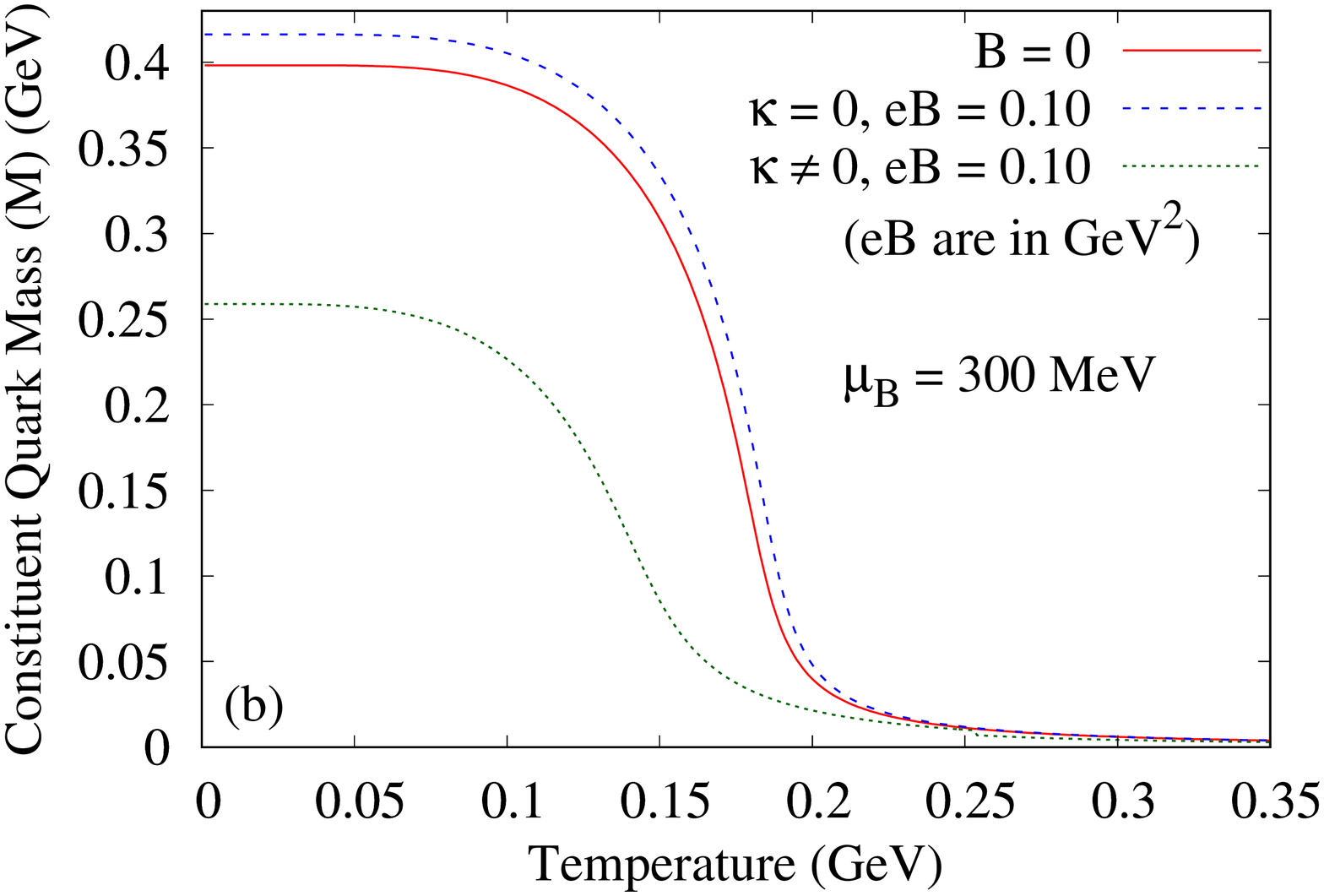}
	\end{center}
	\caption{(Color Online) Variation of the constituent quark mass ($M$) as a function of temperature for different values of external magnetic field and AMM of the quarks at (a) $\mu_B=0$ and  (b) $\mu_B=300$ MeV.}
	\label{fig.M}
\end{figure}

\section{NUMERICAL RESULTS}\label{sec.results}
We begin this section by showing the variation of the constituent quark mass as a function of temperature for different values of magnetic field, AMM 
and baryon chemical potential in Fig.~\ref{fig.M}. For all the numerical calculations in this work, we have considered upto 2000 Landau levels. 
As can be noticed from the figure, in all the cases, $M$ remains almost constant in the lower temperature 
region ($T\precsim 100$ MeV) and drops sharply at a particular temperature signifying the phase transition from chiral symmetry broken to restored phase. 
The effect of increase of $\mu_B$ is found to decrease the transition temperature (though by a small amount) as one expects in a typical QCD phase diagram. 
Since, we have taken $m\ne0$, the chiral symmetry is only partially restored. Moreover, when the AMM of the quarks is switched off (on), the graph of 
$M$ at non-zero $B$ remains always above (below) the corresponding $B=0$ graph. The constituent quark mass at $B\ne0$ has a very strong dependence on the AMM of 
the quarks; for example, at $T\simeq0$ and $eB=0.10$ GeV$^2$, the constituent quark mass decreases about 40\% when the AMM of the quarks is switched on. 
This strong dependence of $M$ on the AMM of the quarks can be understood from Eq.~\eqref{eq.lagrangian} in which the term proportional to $\kappa$ 
is $i\gamma^1\gamma^2$ for constant magnetic field along $\hat{z}$ direction. Considering the Dirac as well as Weyl representation of the gamma matrices, 
the term $i\gamma^1\gamma^2$ is a diagonal matrix in Dirac space, which in turn effectively adds to $m \times \identity_\text{Dirac}$ in the Lagrangian. 
As it is well known, the solution of the gap equation is highly sensitive to the value of $m$ considered, which in turn leads to the strong AMM 
dependence of the constituent quark mass. 
\begin{figure}[h]
	\begin{center}
		\includegraphics[angle=-90,scale=0.33]{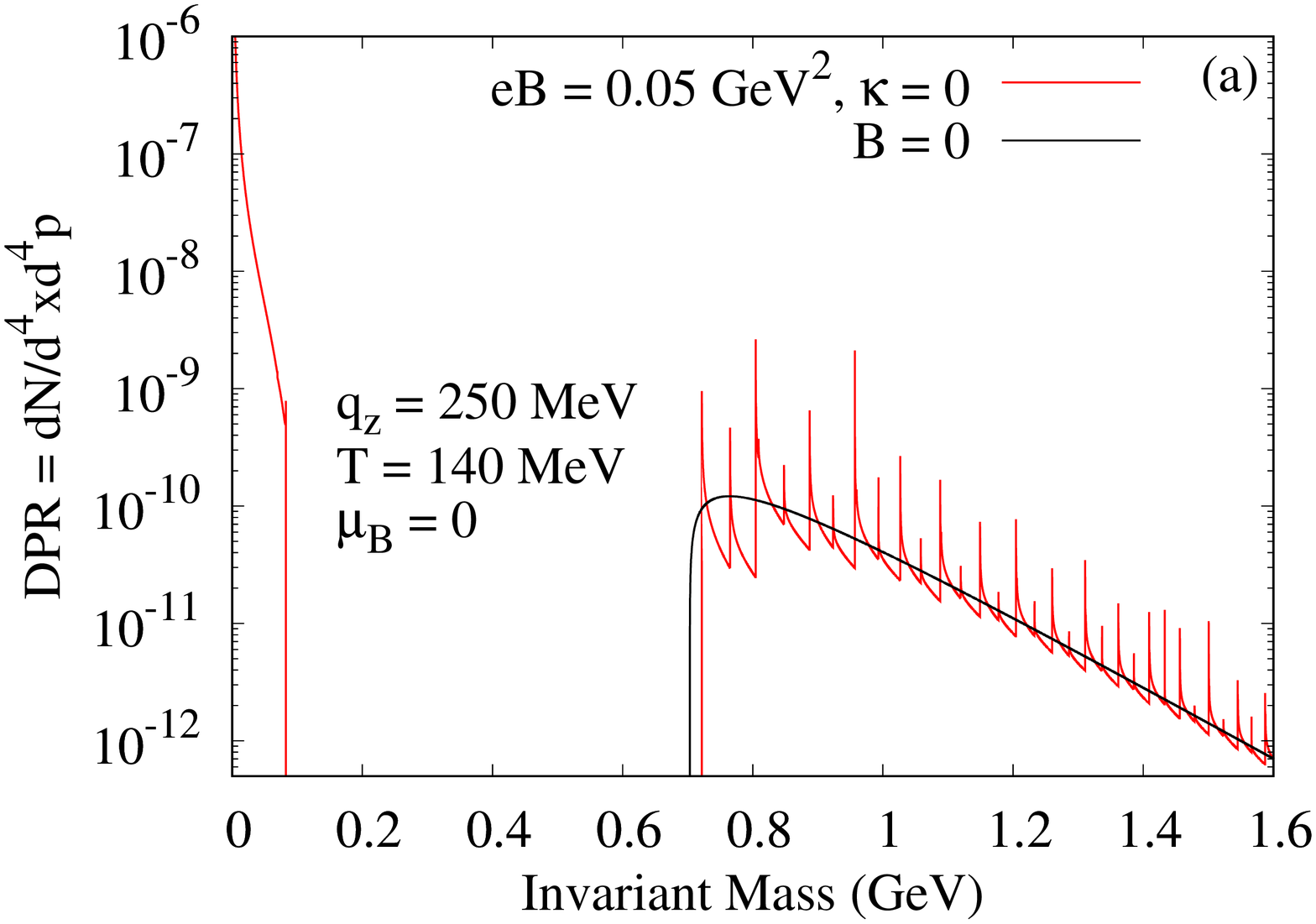}  \includegraphics[angle=-90,scale=0.33]{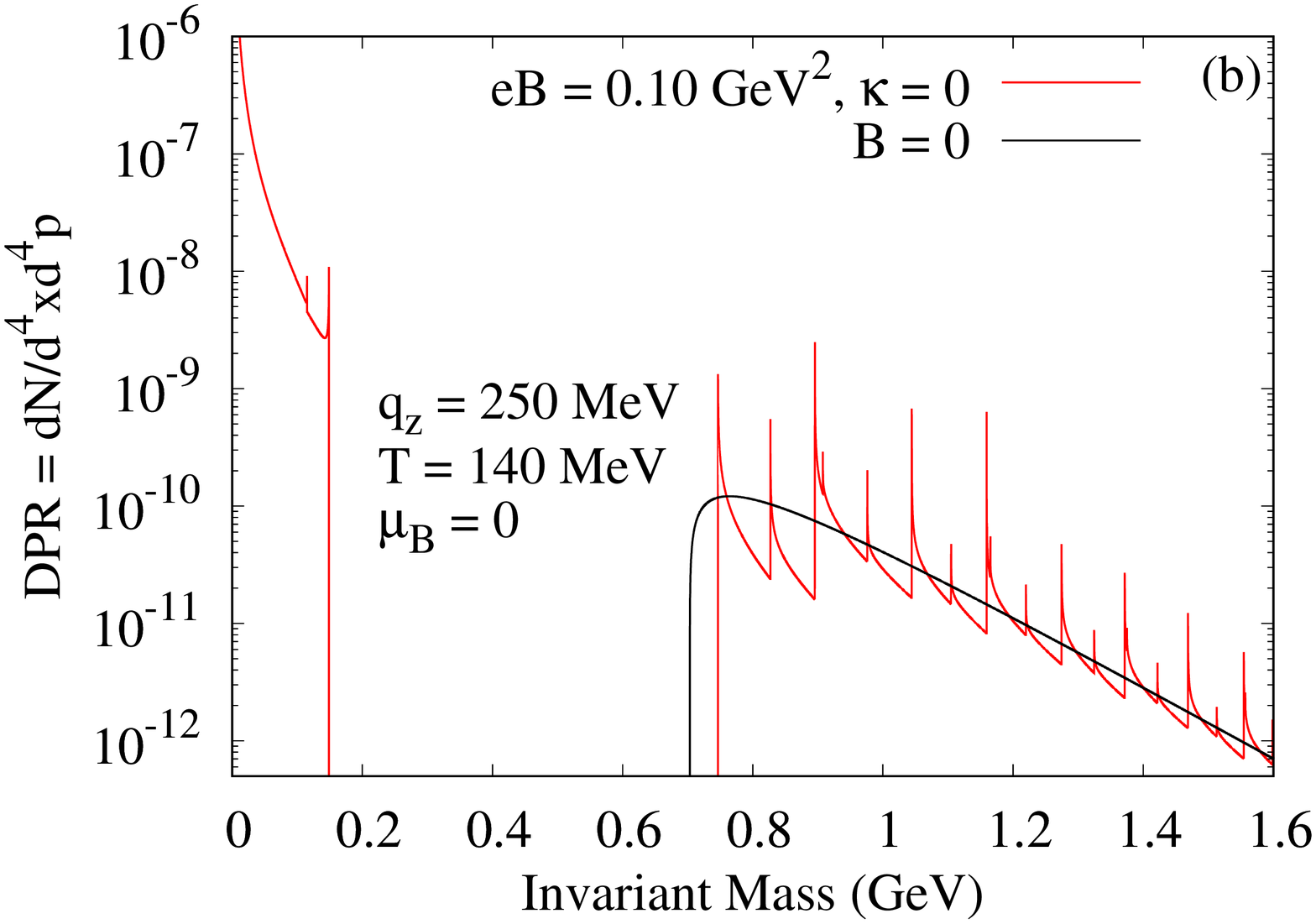}
		\includegraphics[angle=-90,scale=0.33]{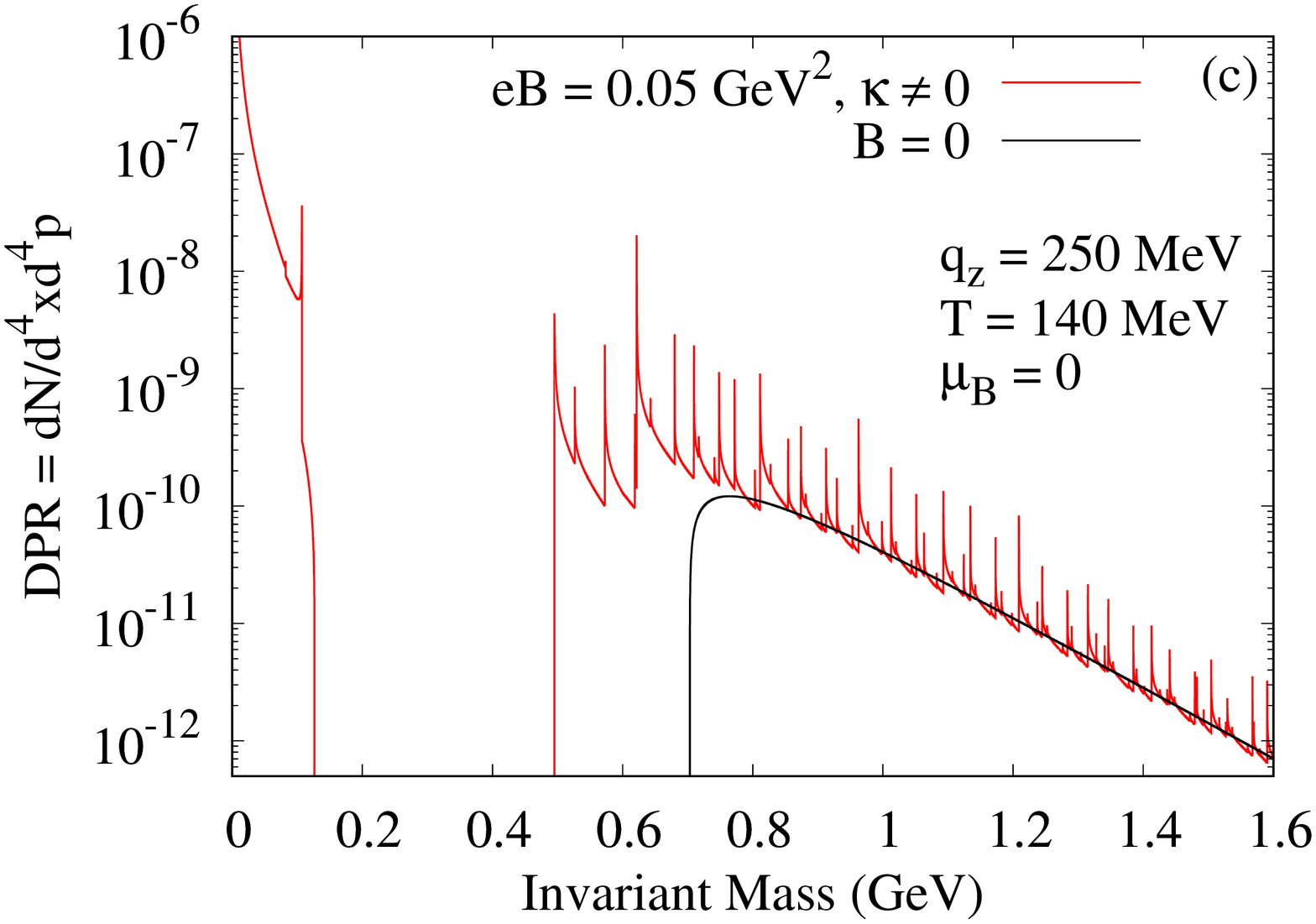}  \includegraphics[angle=-90,scale=0.33]{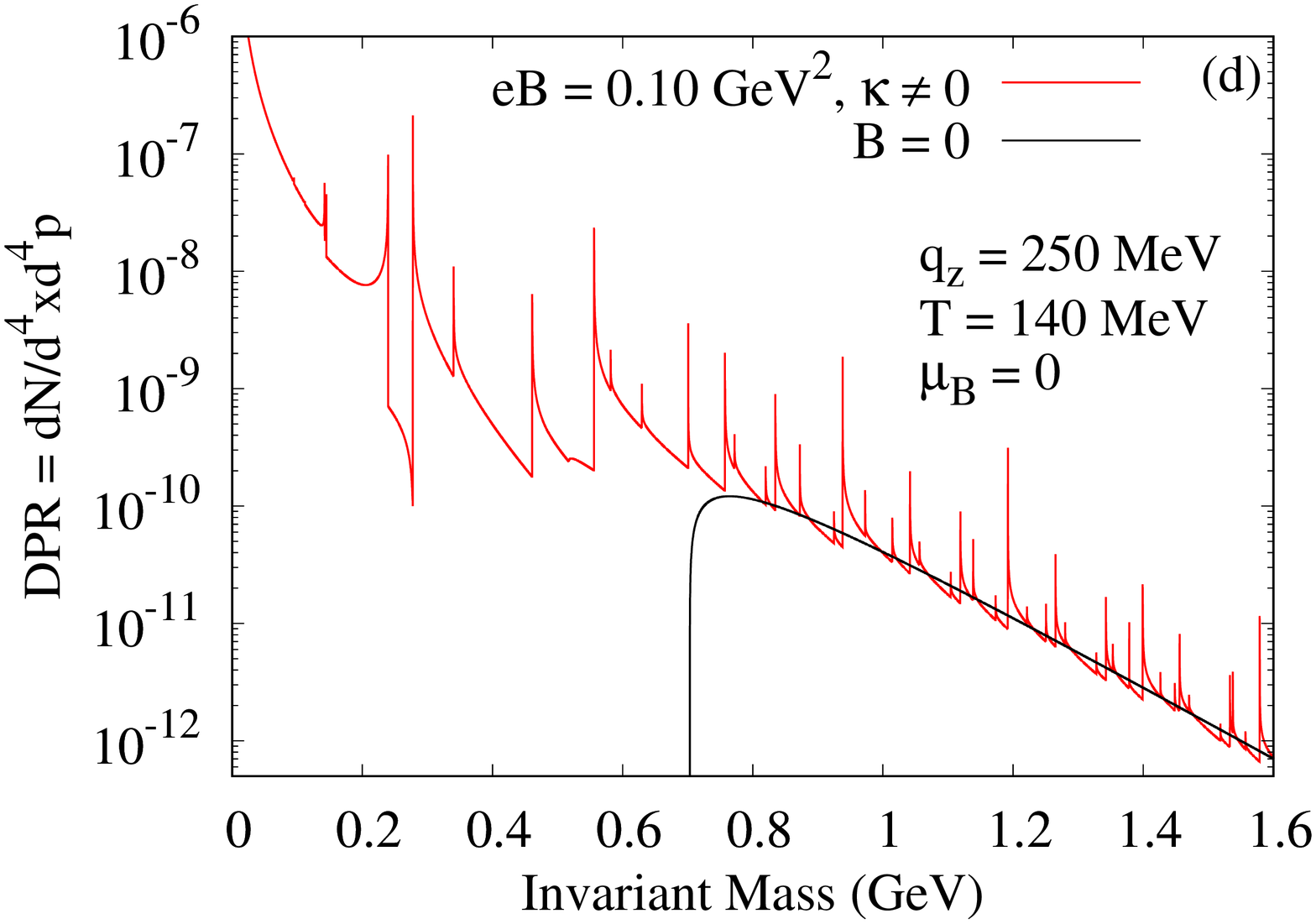}
	\end{center}
	\caption{(Color Online) Dilepton production rate 
		at $q_z=250$ MeV, $T=140$ MeV, $\mu_B=0$ with different values of external magnetic field and AMM of the quarks as (a) $eB=0.05$ GeV$^2$, $\kappa=0$, (b) $eB=0.10$ GeV$^2$, $\kappa=0$, 
		(c) $eB=0.05$ GeV$^2$, $\kappa\ne0$ and (d) $eB=0.10$ GeV$^2$, $\kappa\ne0$. The DPR at $B=0$ is also shown for comparison.}
	\label{fig.dpr1}
\end{figure}

\begin{figure}[h]
	\begin{center}
		\includegraphics[angle=-90,scale=0.33]{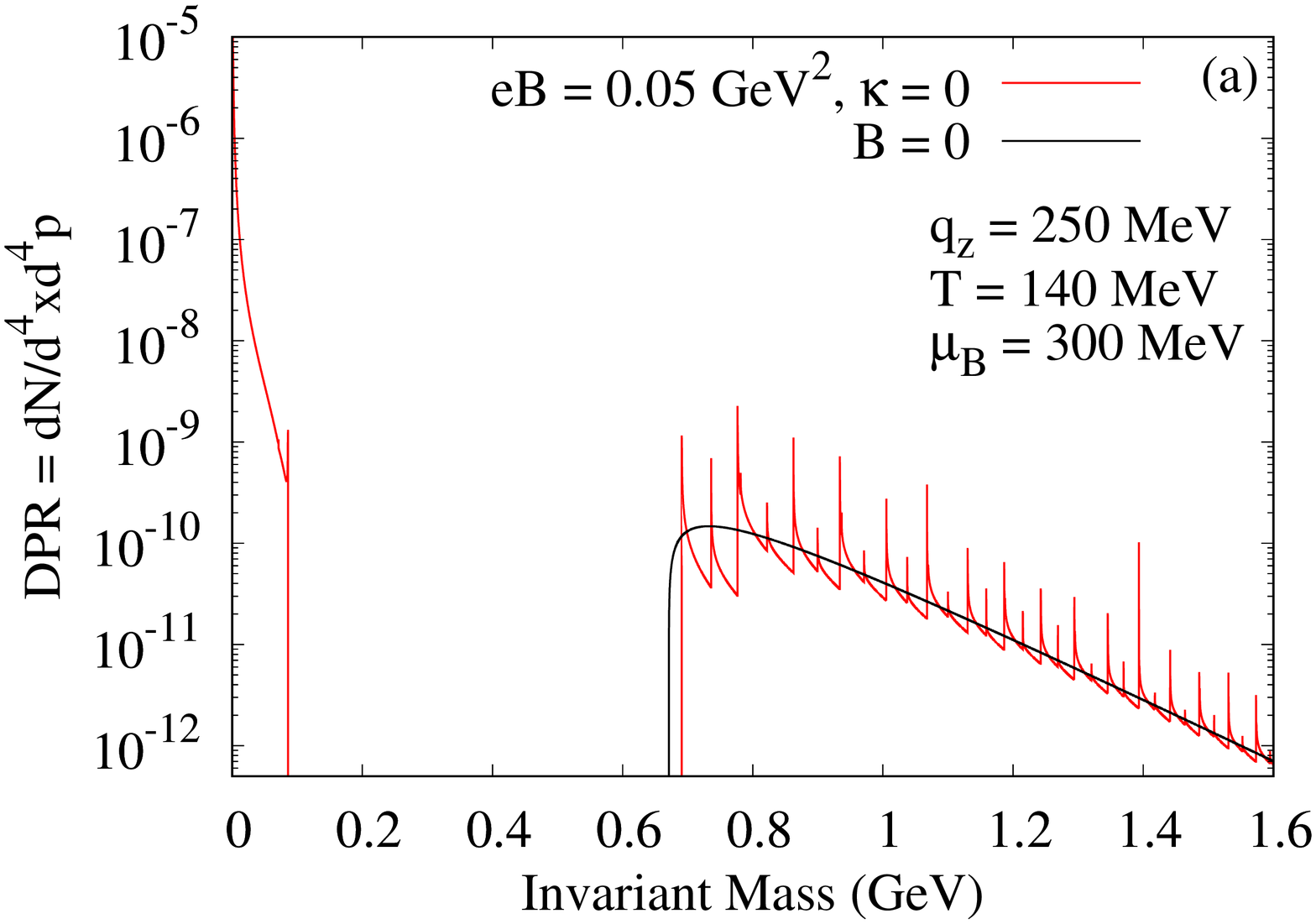}  \includegraphics[angle=-90,scale=0.33]{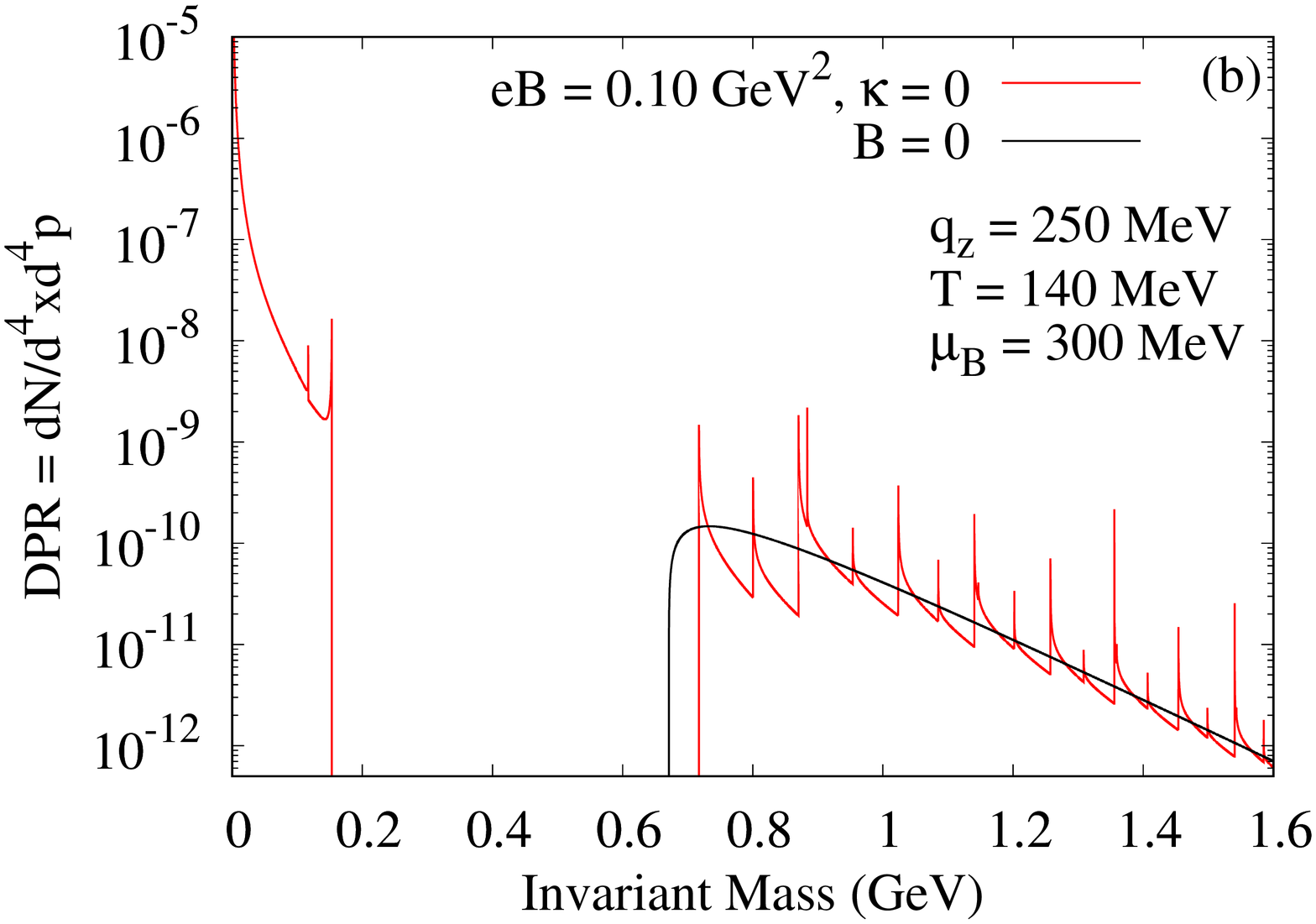}
		\includegraphics[angle=-90,scale=0.33]{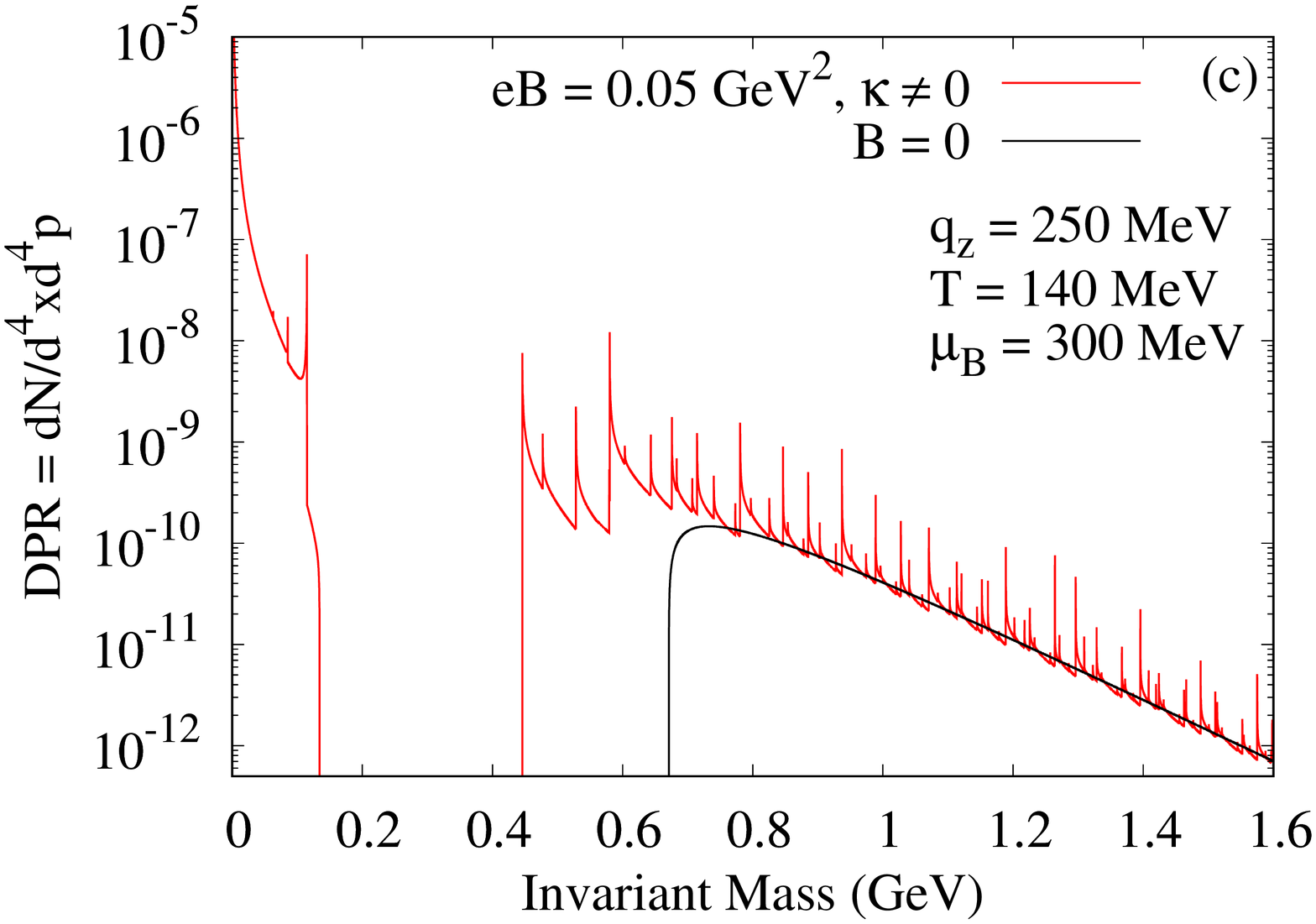}  \includegraphics[angle=-90,scale=0.33]{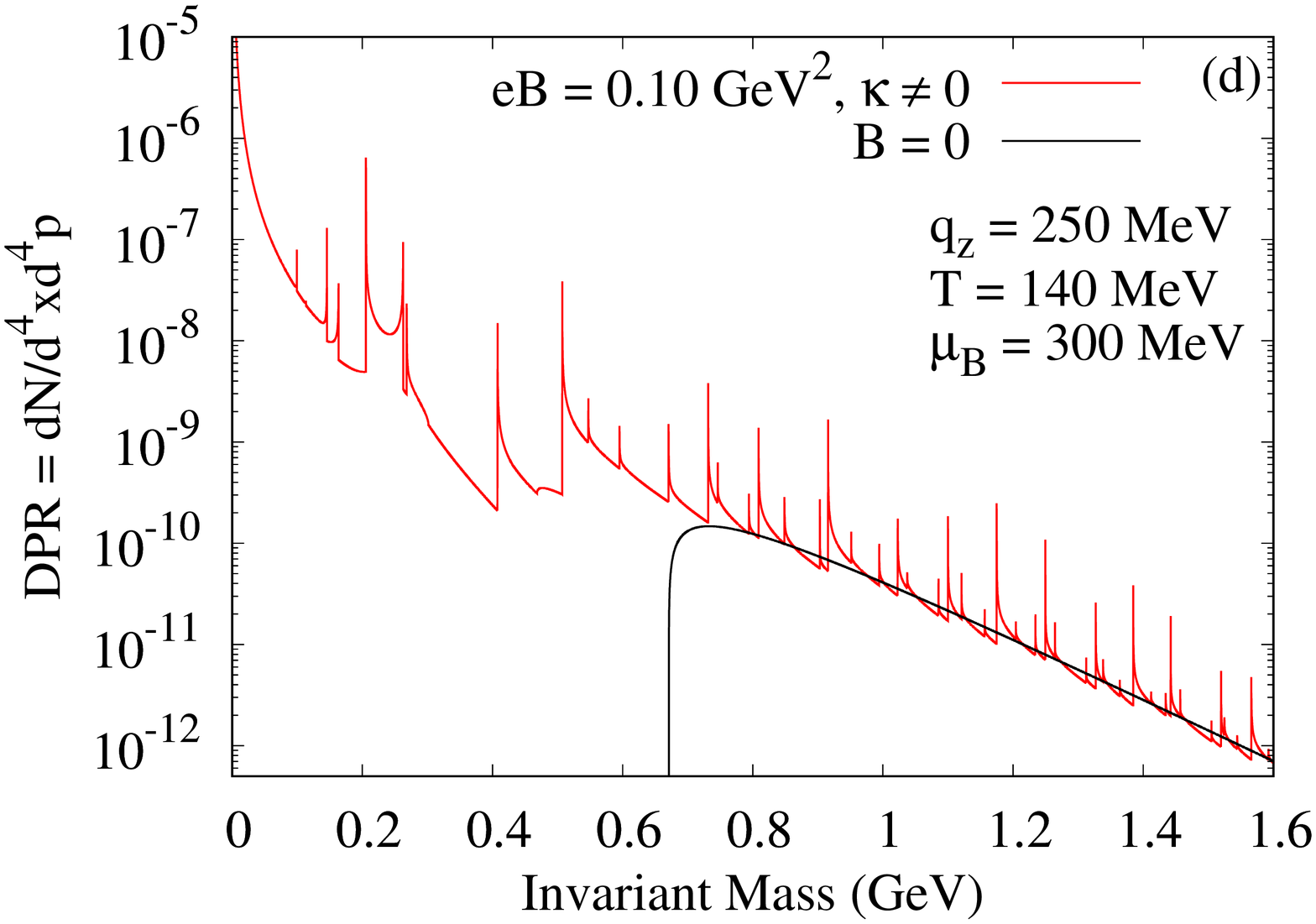}
	\end{center}
	\caption{(Color Online) Dilepton production rate 
		at $q_z=250$ MeV, $T=140$ MeV, $\mu_B=300$ MeV with different values of external magnetic field and AMM of the quarks as (a) $eB=0.05$ GeV$^2$, $\kappa=0$, (b) $eB=0.10$ GeV$^2$, $\kappa=0$, 
		(c) $eB=0.05$ GeV$^2$, $\kappa\ne0$ and (d) $eB=0.10$ GeV$^2$, $\kappa\ne0$. The DPR at $B=0$ is also shown for comparison.}
	\label{fig.dpr3}
\end{figure}
\begin{figure}[h]
	\begin{center}
		\includegraphics[angle=-90,scale=0.33]{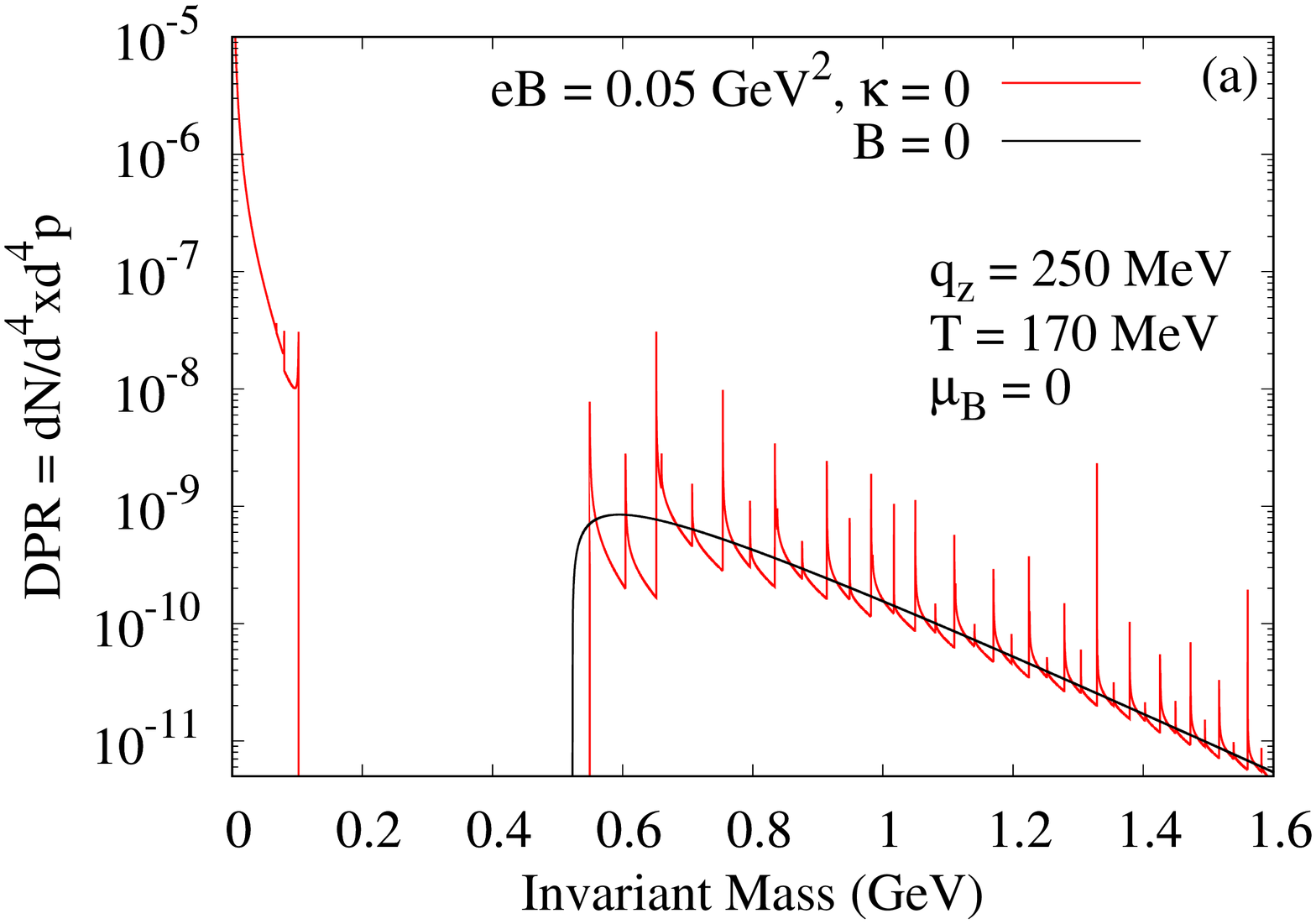}  \includegraphics[angle=-90,scale=0.33]{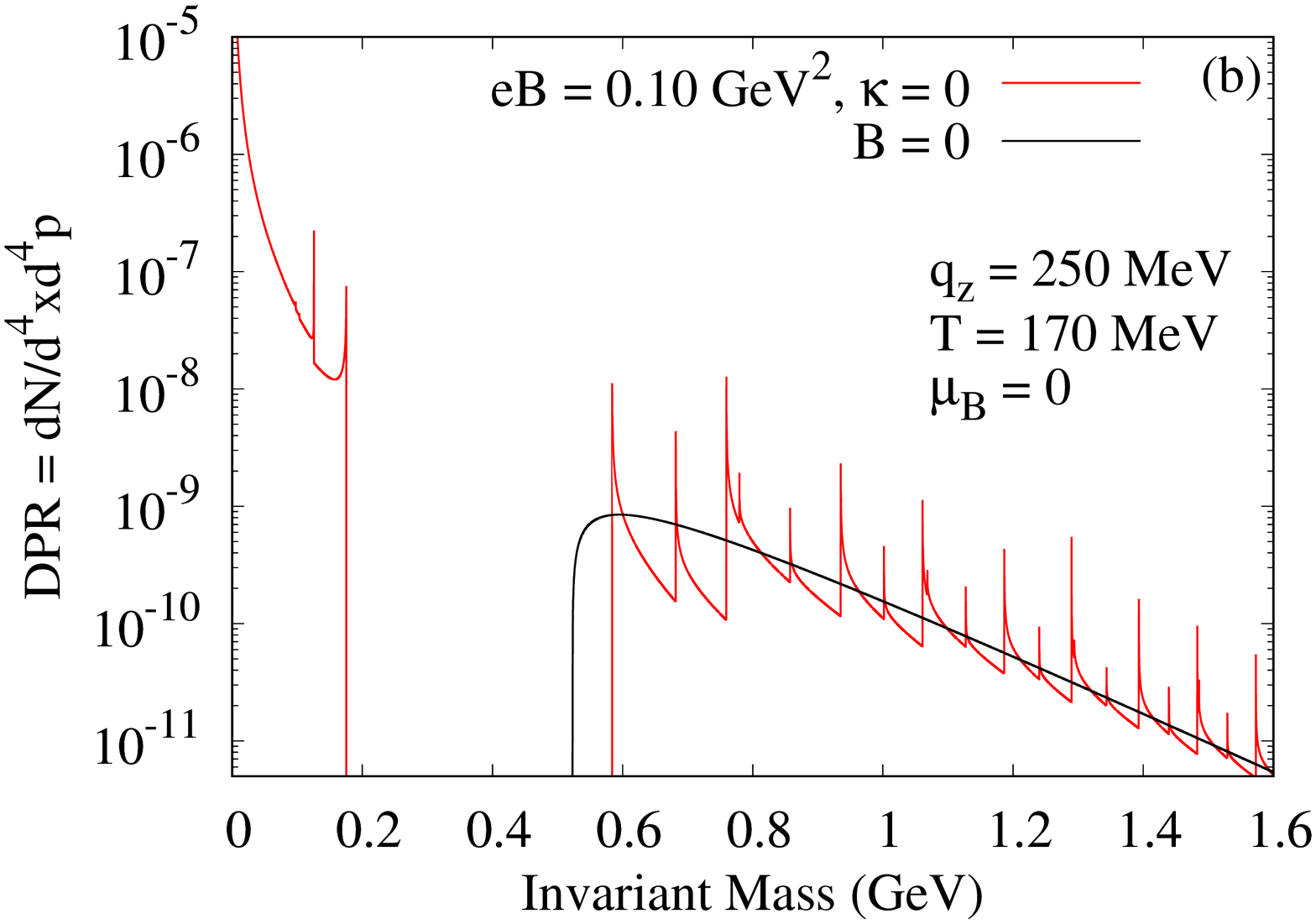}
		\includegraphics[angle=-90,scale=0.33]{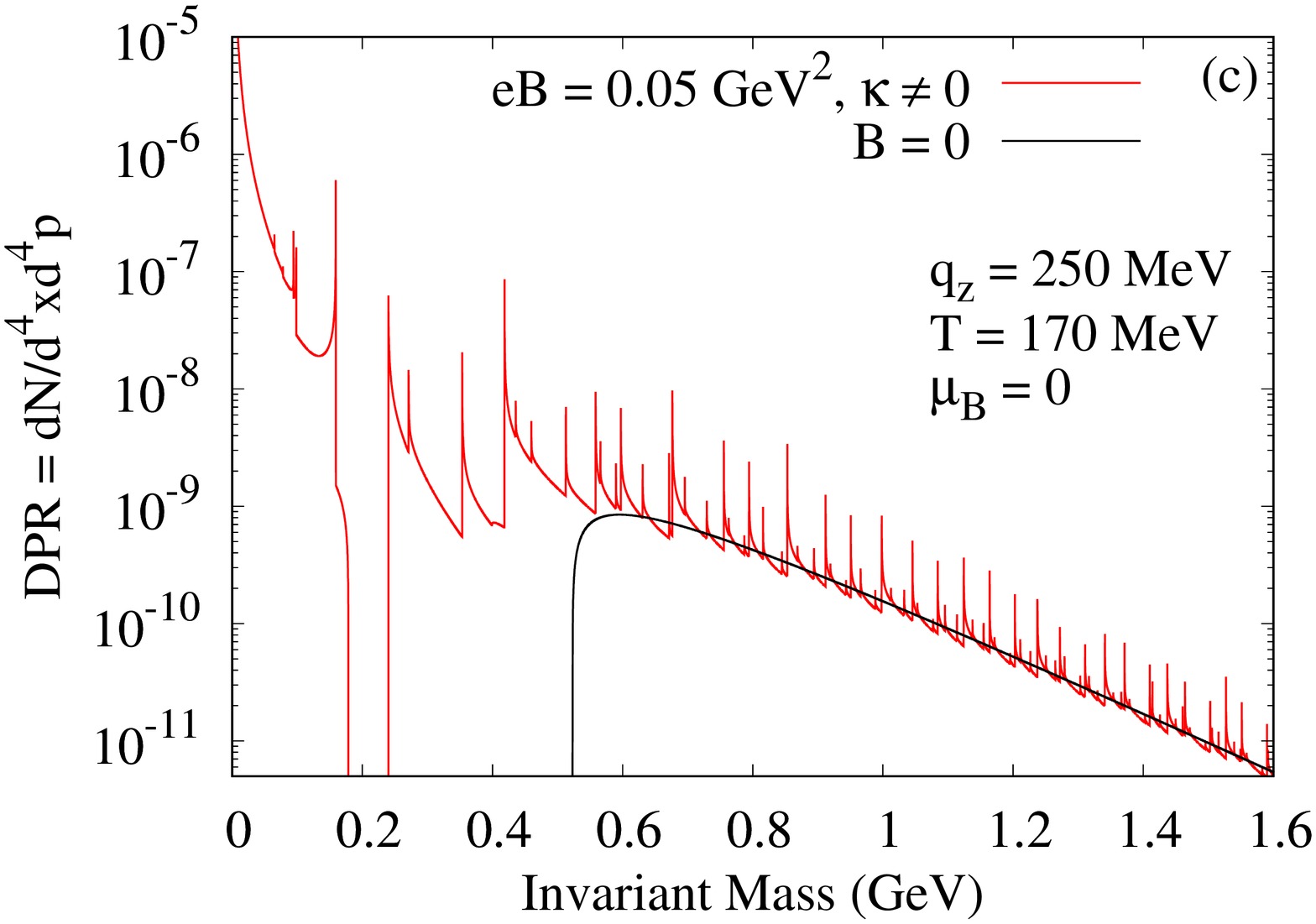}  \includegraphics[angle=-90,scale=0.33]{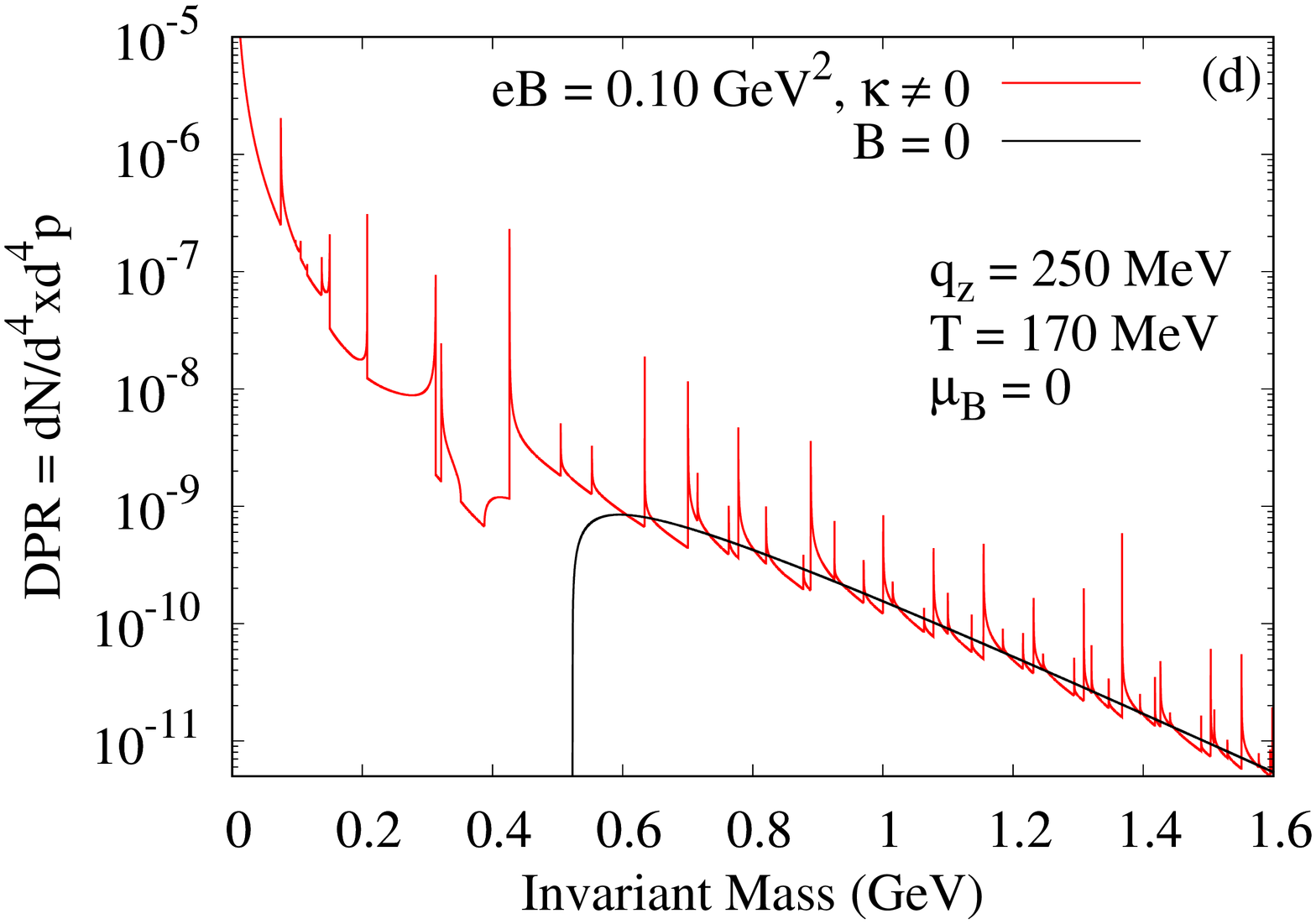}
	\end{center}
	\caption{(Color Online) Dilepton production rate 
		at $q_z=250$ MeV, $T=170$ MeV, $\mu_B=0$ with different values of external magnetic field and AMM of the quarks as (a) $eB=0.05$ GeV$^2$, $\kappa=0$, (b) $eB=0.10$ GeV$^2$, $\kappa=0$, 
		(c) $eB=0.05$ GeV$^2$, $\kappa\ne0$ and (d) $eB=0.10$ GeV$^2$, $\kappa\ne0$. The DPR at $B=0$ is also shown for comparison.}
	\label{fig.dpr4}
\end{figure}
\begin{figure}[h]
	\begin{center}
		\includegraphics[angle=-90,scale=0.33]{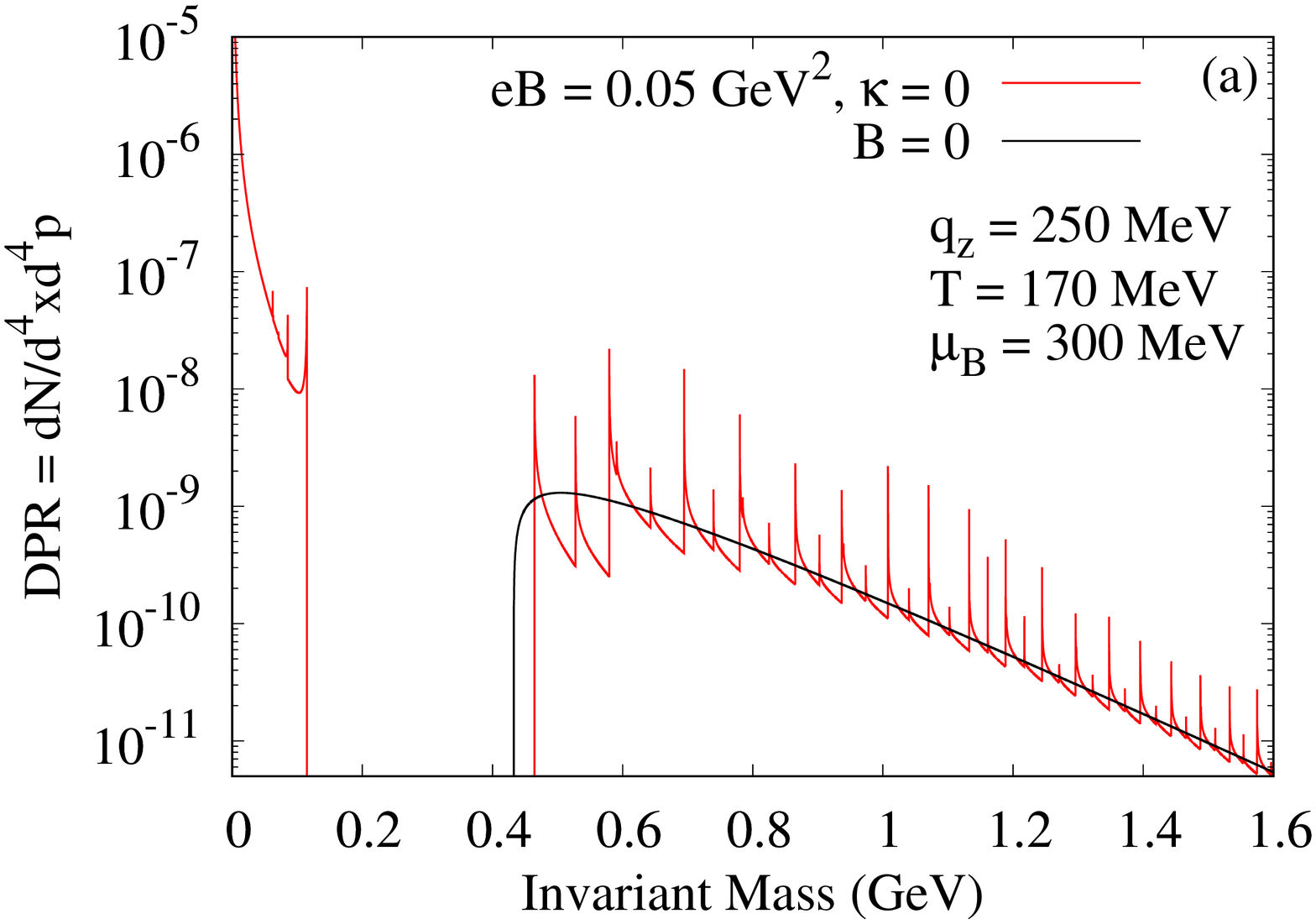}  \includegraphics[angle=-90,scale=0.33]{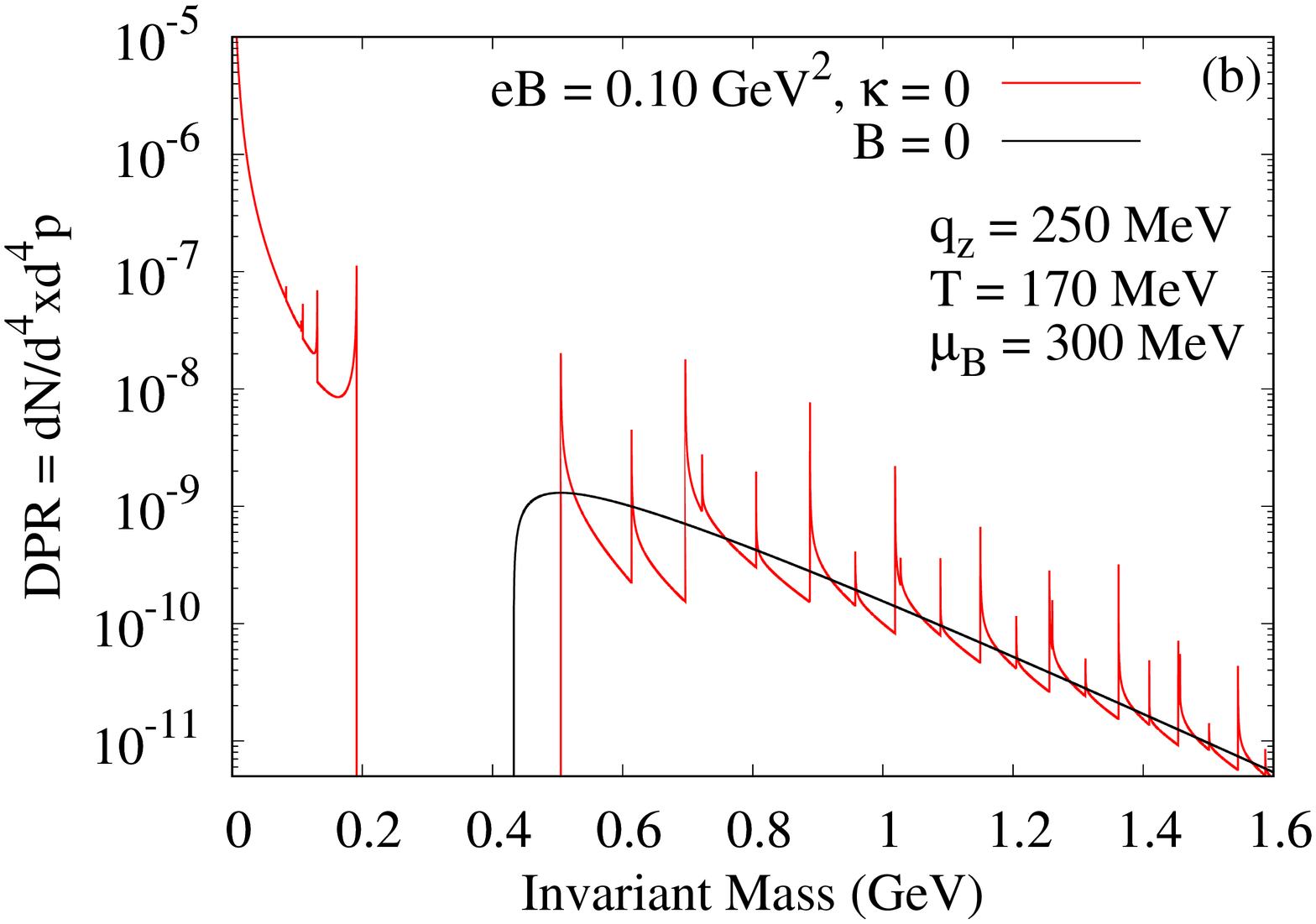}
		\includegraphics[angle=-90,scale=0.33]{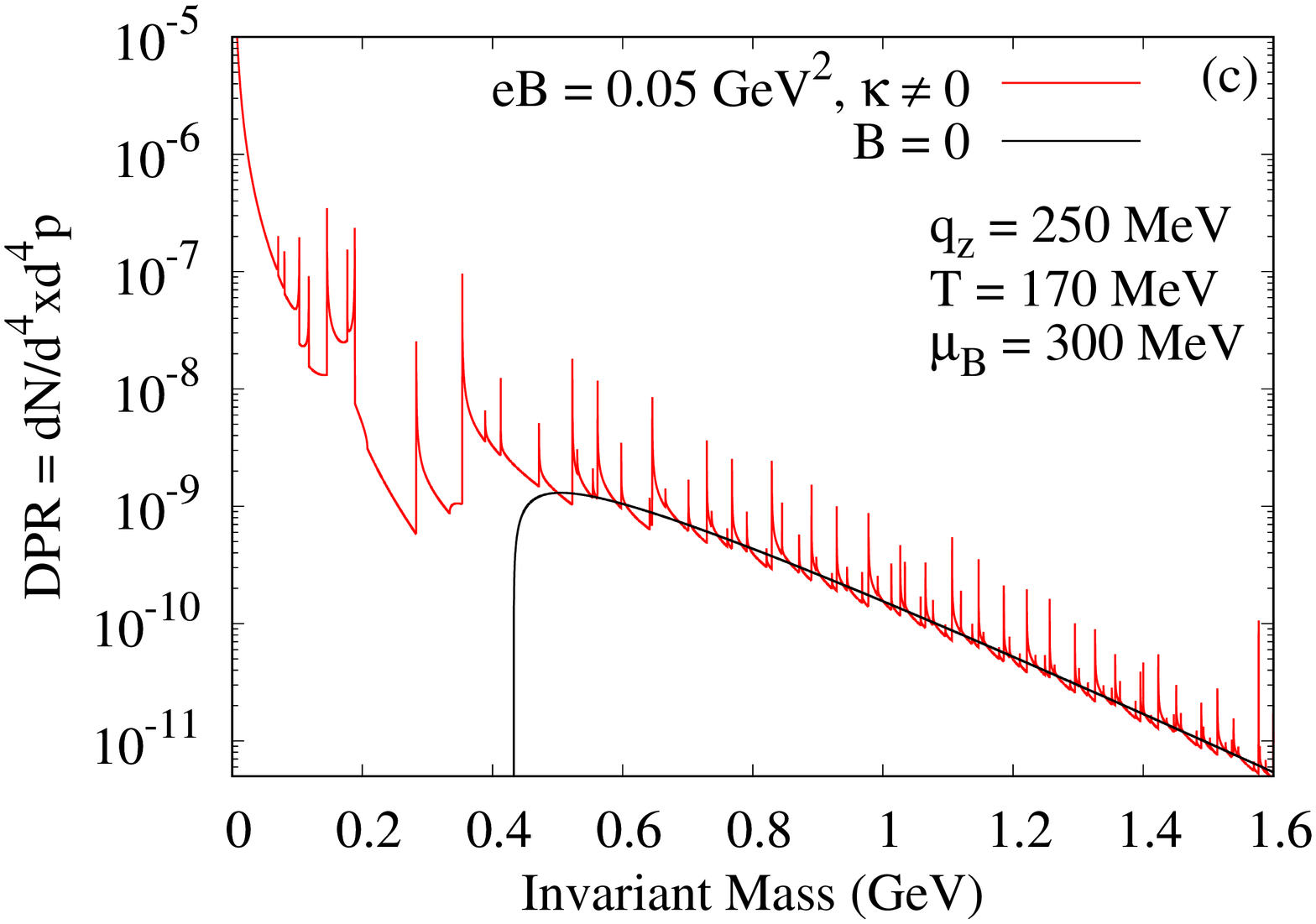}  \includegraphics[angle=-90,scale=0.33]{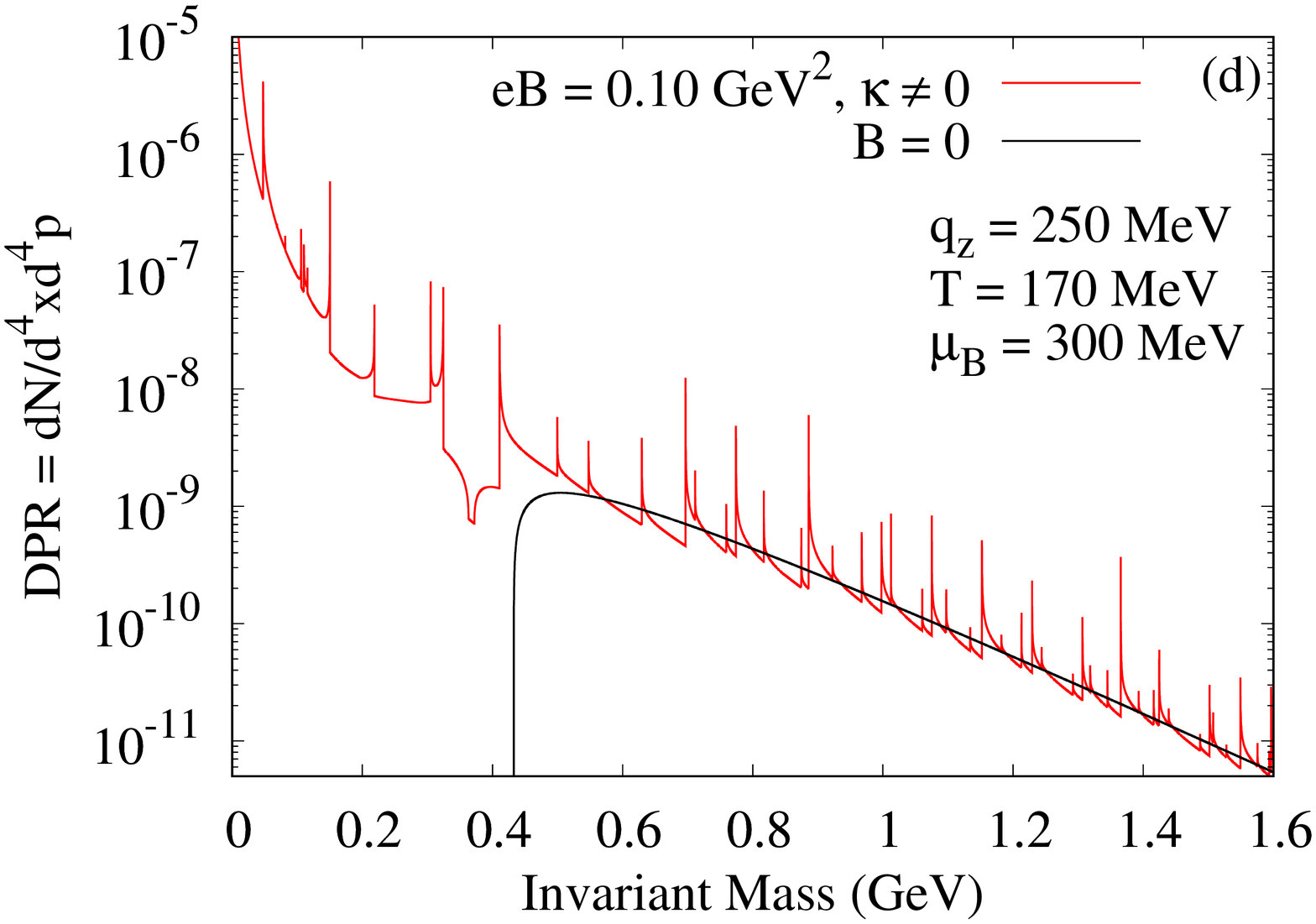}
	\end{center}
	\caption{(Color Online) Dilepton production rate at $q_z=250$ MeV, $T=170$ MeV, $\mu_B=300$ MeV with different values of external magnetic field and AMM of the quarks as (a) $eB=0.05$ GeV$^2$, $\kappa=0$, (b) $eB=0.10$ GeV$^2$, $\kappa=0$, 
		(c) $eB=0.05$ GeV$^2$, $\kappa\ne0$ and (d) $eB=0.10$ GeV$^2$, $\kappa\ne0$. The DPR at $B=0$ is also shown for comparison.}
	\label{fig.dpr5}
\end{figure}
\begin{figure}[h]
	\begin{center}
		\includegraphics[angle=-90,scale=0.33]{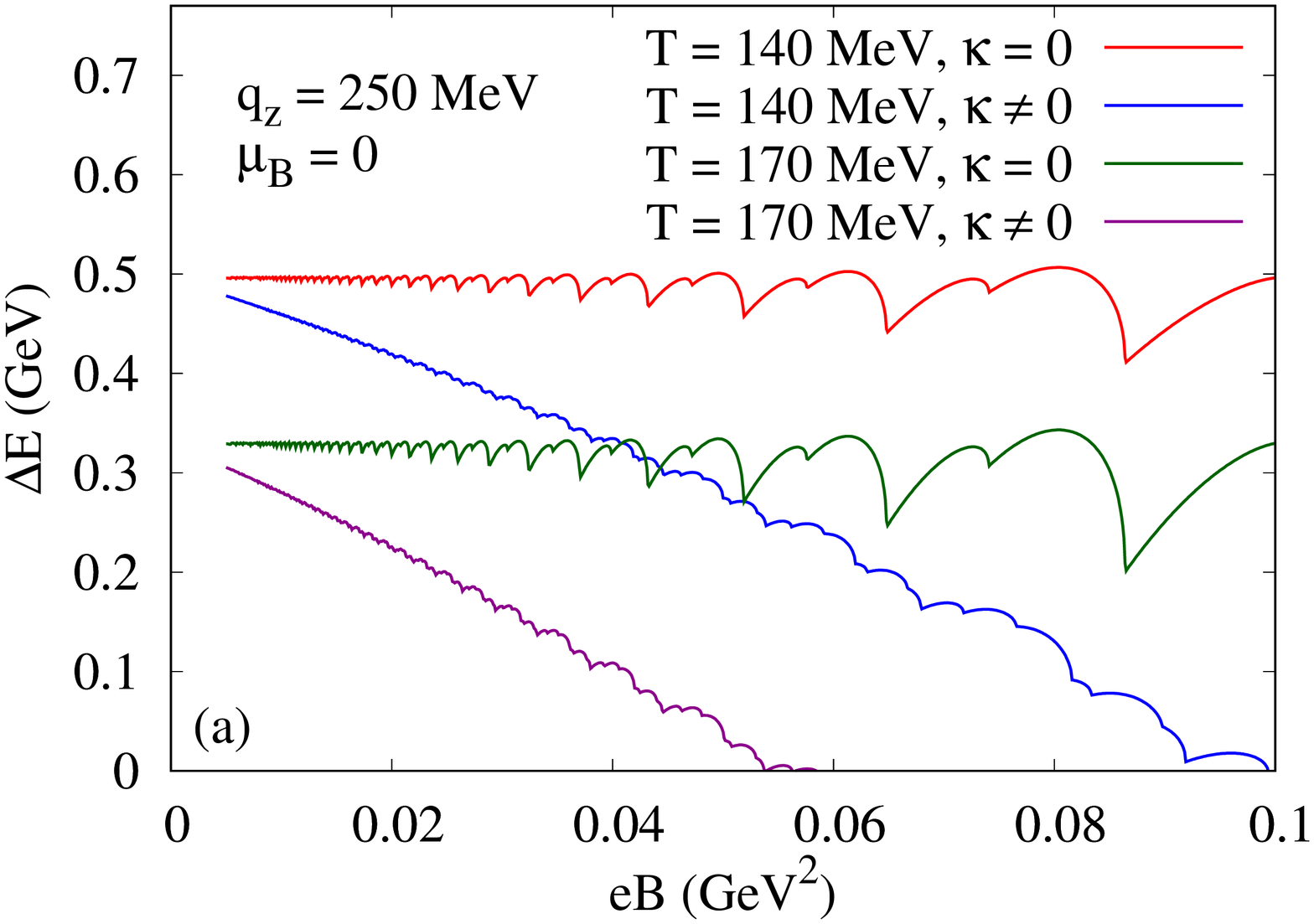}  \includegraphics[angle=-90,scale=0.33]{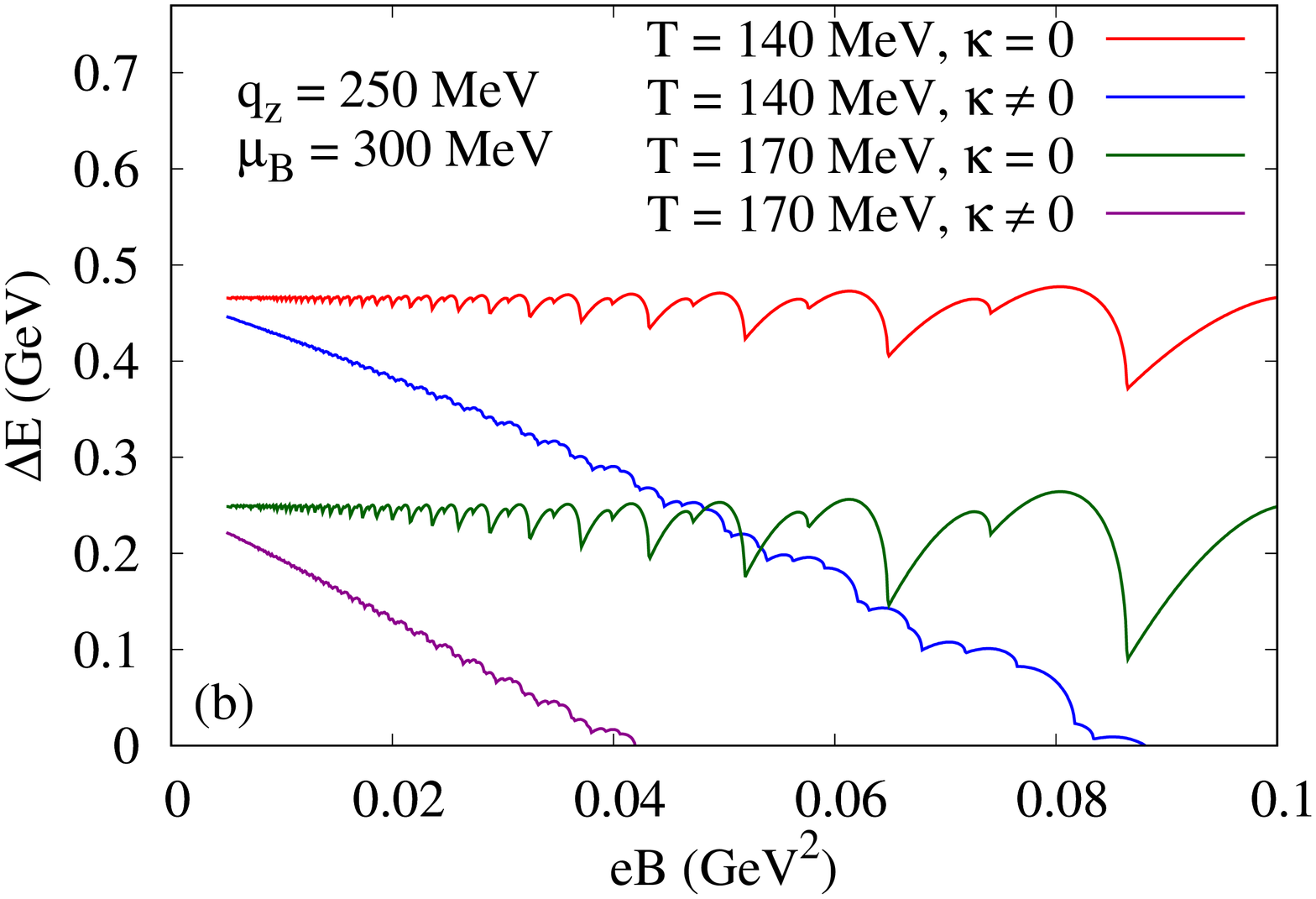}
	\end{center}
	\caption{(Color Online) Variation of the kinematic gap $\Delta E$ as a function of external magnetic 
		field at $q_z=250$ MeV and at different values of temperature and AMM of the quarks for baryon chemical potential (a) $\mu_B=0$ and (b) $\mu_B=300$ MeV.}
	\label{fig.dE}
\end{figure}

Following Refs.~\cite{Shovkovy:2012zn,Gusynin:1994re}, we give a qualitative argument of how the external magnetic field enhances the chiral condensate, which is opposite when compared to its role in the case of superconductivity. 
The chiral condensate is made of overall neutral fermion-antifermion pairs, in contrats to the charged Cooper pairs. 
In a Cooper pair, the two electrons have opposite spins and therefore opposite magnetic moments. 
Now, in the presence of a magnetic field only one of them can orient along the direction of the field leaving the other one in a frustrating position. This results in a energy stress and tends to break the Cooper pair. 
However, in the case of chiral condensate, the magnetic moments of the fermion (with a fixed charge and spin) and the antifermion (with the opposite charge and spin) can comfortably align along the direction of the magnetic field without producing any frustration in the pair. 
This makes the chiral condensate `stronger' and leads to the Magnetic Catalysis (MC) of chiral condensate. 
A similar type of physical argument for the Inverse Magnetic Catalysis (IMC) in presence of AMM of the quarks is difficult to construct. 
The possible reason could be an energy mismatch (could be due to the repulsive interaction in presence of AMM) between the Left-handed (L) and the Righ-handed (R) quark-antiquark pair (note that $ \langle \overline{\psi}\psi \rangle = \langle \overline{\psi}_L\psi_R \rangle + \langle \overline{\psi}_R\psi_L \rangle $), which will tend to dissolve the chiral condensate as happens in the case of color superconductivity where the Bardeen--Cooper--Schrieffer (BCS) pair breaks down due to Fermi level mismatch arising out of isospin asymmetry~\cite{Shovkovy:2003uu}.
%

Let us now proceed to show the numerical results for the DPR in the presence of external magnetic field. For our numerical estimates, we have taken the lepton mass to be zero i.e. $m_L=0$. 
In Fig.~\ref{fig.dpr1}, the DPRs 
at longitudinal net dilepton momentum $q_z=250$ MeV, temperature $T=140$ MeV and baryon chemical potential $\mu_B=0$ with different values of external magnetic field and AMM of the quarks have been depicted as a function of the dilepton invariant mass $\sqrt{q^2}$. 
Since the lepton mass $m_L$ is taken to be zero, from Eq.~\eqref{eq.DPR.4}, one obtains the threshold for the DPR production at $B=0$ as 
the Unitary-I cut threshold of $\IM\mcW_{11}^\munu(q)$ which is $\sqrt{q^2}>2M(T,\mu_B)$ (see Fig.~\ref{fig.analytic0}) where $M(T,\mu_B)$ is 
the constituent quark mass. In Fig.~\ref{fig.dpr1}, corresponding to $T=140$ MeV and $\mu_B=0$, the value of the constituent quark mass 
at $B=0$ is $M\simeq350$ MeV in turn making the DPR threshold $\sqrt{q^2} \gtrsim 750$ MeV as can be clearly observed in the figure.

As already discussed, at $B\ne0$ the DPR will have contributions from both the Landau as well as Unitary-I cuts (see Fig.~\ref{fig.analytic1}). For $\kappa=0$, the Unitary-I 
and Landau cut thresholds respectively become $\sqrt{\qpll^2} > 2M$ and $\sqrt{\qpll^2}<\FB{\sqrt{M^2+2|e_uB|}-M}$. Both these thresholds 
increase with the increase in magnetic field which can be clearly seen by comparing Figs.~\ref{fig.dpr1}(a) and (b). 
Moreover, the threshold of the Unitary-I cut for $B\ne0$ and $\kappa=0$ is more as compared to that of $B=0$ because of the fact that the value of the 
constituent quark mass is larger at non-zero magnetic field when the AMM of the quarks is switched off as can be seen in Fig.~\ref{fig.M}. 
At $\kappa\ne0$, the Landau cut threshold is more than that of at $\kappa=0$ as 
can be obtained from Eq.~\eqref{eq.LandauCut} whereas, the Unitary-I cut threshold $\sqrt{\qpll^2} > 2\MB{M-\kappa_uB}$ becomes 
smaller due to the introduction of non-zero values of the AMM of the quarks. Together these two threshold modifications due to the AMM of the quarks, the gap 
between the Landau and Unitary-I cuts becomes smaller with the increase in $B$ at $\kappa\ne0$. This behaviour can be seen if one compares 
Figs.~\ref{fig.dpr1}(c) and (d) with (a) and (b). More interestingly, at sufficiently high values of the magnetic field the Landau and Unitary-I cuts 
merge with each other as depicted in Figs.~\ref{fig.dpr1}(d) at $eB=0.10$ GeV$^2$.


For all the cases, in the $B\ne0$ graphs, we can observe spike like structures spreading over all the allowed invariant mass regions. 
The appearance of the spike like structure in any momentum space two point correlation function at $B\ne0$ is an well known phenomena. 
These spikes are basically ``threshold singularities'' that appears at each Landau levels as can be understood from Eq.~\eqref{eq.W11.6}, 
in which the K\"all\'en functions in the denominators vanish at each threshold defined in terms of the step functions in 
those equations. Irrespective of the spikes, the overall/average magnitude of the DPR at $B\ne0$ is of same order as $B=0$ and at higher invariant mass 
region the graphs of $B\ne0$ oscillate about the $B=0$ graphs. 

In Fig.~\ref{fig.dpr3}, we have introduced non-zero baryon density taking $T=140$ MeV and $\mu_B=300$ MeV. The plots in Fig.~\ref{fig.dpr3} are 
qualitatively similar (in terms of the thresholds) to that of the plots in Fig.~\ref{fig.dpr1}. This is due the fact that, 
the constituent quark mass does not change significantly at $\mu_B\simeq300$ MeV as can be seen from Fig.~\ref{fig.M}. However, because of the increase in density, the phase space 
availability increases owing to an overall increase in the magnitude of the DPR as can be noticed on comparing  Fig.~\ref{fig.dpr3} with Fig.~\ref{fig.dpr1}.

The DPR at $T=170$ MeV with $\mu_B=0$ and 300 MeV are shown in Figs.~\ref{fig.dpr4} and \ref{fig.dpr5}. The only difference from Figs.~\ref{fig.dpr1} and \ref{fig.dpr3} is the 
increase in the temperature and baryon chemical potential for which the constituent quark mass $M$ decreases (see Fig.~\ref{fig.M}) in all the cases. This in turn decreases the 
threshold of the Unitary-I cut and increase the same of the Landau cut. It can be observed that, at $T=170$ MeV and $\mu_B=300$ MeV, the Landau and Unitary-I cuts have merged even at $eB=0.05$ GeV$^2$ when the AMM of the quarks is considered. Moreover, with the increase in temperature, the overall magnitudes of the DPR have increased as compared to the $T=140$ MeV because of the enhancement of the thermal phase space.

In this work, we have considered the constituent quark mass ($M$) obtained from the gap equation in NJL model to be constant in momentum space while, in principle $M$ should have have momentum dependence. The momentum dependent quark mass $M=M(q)$ has been obtained earlier using Dyson-Schwinger approach to QCD~\cite{Roberts:2000aa,Roberts:2007ji,Bhagwat:2003vw,Siringo:2016jrc} as well as in instanton-dyon model calculations~\cite{Liu:2015jsa,Shuryak:2018fjr} where $M(q)$ is shown have a monotonically decreasing trend with the increase in $q$. 
The qualitative behavior of $M(q)$ is also confirmed in the lattice QCD simulations in Refs.~\cite{Bowman:2002kn,Bowman:2005vx}. So in a more realistic situation where $M$ has additional momentum dependence, we expect that the DPR would have enhanced in the high invariant mass region with respect to the case with constant $M$ for a particular temperature.

Few more comments about the DPR obtained in this work are in order here. The non-trivial Landau cuts have appeared in the physical kinematic region solely due to the external magnetic field and finite temperature. It can be noticed from Figs.~\ref{fig.analytic1} that, there exists a kinematically forbidden gap $\Delta E$ between the Landau and Unitary-I cuts where the DPR is zero and 
$\Delta E$ is given by 
\begin{eqnarray}
\Delta E = \sqrt{q_z^2+4\FB{M-\kappa_uB}^2} - \sqrt{q_z^2+\FB{\sqrt{M^2+2|e_uB|}+\kappa_uB-\MB{M-\kappa_uB}}^2}. \label{eq.lugap}
\end{eqnarray}
At vanising AMM of the quarks, the forbidden gap becomes $\Delta E_{\kappa=0} = \sqrt{q_z^2+4M^2} - \sqrt{q_z^2+\FB{\sqrt{M^2+2|e_uB|}-M}^2}$, which is always non-zero for any value of the external magnetic field. However, for $\kappa\ne0$, at a certain value of the magnetic field, the gap $\Delta E$ could be zero as can be obtained from Eq.~\eqref{eq.lugap}. In Fig.~\ref{fig.dE}, the quantity $\Delta E$ has been plotted as a function of external magnetic field with different values of temperature, baryon chemical potential and AMM of the quarks for $q_z=250$ MeV. As can be seen from the figure, irrespective of $\kappa$, $\Delta E$ shows oscillatory behaviour with external magnetic field which is due to the fact that the constituent quark mass $M$ is itself an oscillatory quantity in presence of external magnetic field (see Refs.~\cite{Chaudhuri:2019lbw,Fayazbakhsh:2010gc,Fayazbakhsh:2010bh,Sadooghi1,Sadooghi,Ebert:1998gx,Inagaki:2004ih}) and can be attributed to the well-known de Haas-van Alphen (dHvA) effect~\cite{Landau:1980mil}. Also, when the AMM of the quarks is not considered, the overall magnitude of $\Delta E$ remains almost constant and finite with the increase in external magnetic field maintaining the finite gap between the Landau and Unitary-I cut. However, switching on the AMM of the quarks leads to a monotonic decrease of $\Delta E$ with magnetic field and at some value of $eB$, $\Delta E$ becomes zero leading to a continuous spectrum of dilepton yield over the whole invariant mass region.  Moreover the value of the magnetic field $B_C$ at which the gap becomes zero decreases with the increase in temperature as can be noticed by comparing the blue and purple curves of Fig.~\ref{fig.dE} which is due to the decrease in the constituent quark mass with temperature. Finally comparing Fig.~\ref{fig.dE}(a) with Fig.~\ref{fig.dE}(b), we also notice that, $B_C$ also decreases with the increase in baryon chemical potential which is due to the decrease in the constituent quark mass with density.

\section{SUMMARY \& CONCLUSIONS} \label{sec.summary}
In summary, we have calculated the DPR using the 2-flavour NJL model in presence of arbitrary external magnetic field at finite temperature and baryon density with AMM of the quarks. 
Using the RTF of finite temperature field theory and the full Schwinger proper-time propagator, we have evaluated the current-current correlator in the vector channel 
in a thermo-magnetic dense medium. 
The NJL model is used to obtain the temperature, density, magnetic field and AMM dependent effective quark mass $M$ which in turn is used in the expression of the DPR.

We find that, the DPR at non-zero magnetic field has contributions both from the Unitary as well as the Landau cuts corresponding to the physical time-like and positive energy dilepton production. The Unitary cuts (corresponding to the decay/formation processes) are also present at zero temperature and zero external magnetic field whereas the non-trivial Landau cuts (corresponding to the scattering processes) in the time-like domain appear only at finite temperature and due to the non-zero external magnetic field. In particular, the Landau cuts lead to the physical processes as time-like positive energy photon emission or absorption by a quark/antiquark in magnetized hot quark matter which is forbidden in zero magnetic field case due to kinematic constraint. Moreover, on analyzing the analytic structure of the two-point correlator, we have seen that the thresholds of the Unitary and Landau cuts have non-trivial dependence on the magnetic field as well as the AMM of the quarks. 
The DPR obtained is found to be largely enhanced in the low invariant mass region due to the appearance of the Landau cuts. 
Finally, we have shown that, there always exists a kinematic region between the Landau and Unitary-I cut where the DPR is zero if the AMM of the quarks is switched off irrespective of the value 
of the external magnetic field. However, considering finite AMM, of the quarks, this forbidden gap monotonically decreases to zero with the increase in external magnetic field. 
This results in a continuous spectrum of dilepton emission over the whole range of invariant mass which is a novel finding not observed in earlier 
calculations~\cite{Tuchin:2012mf,Tuchin:2013bda,Sadooghi:2016jyf,Mamo:2013efa,Bandyopadhyay:2016fyd,Bandyopadhyay:2017raf,Ghosh:2018xhh,Islam:2018sog,Das:2019nzv}.

\section*{Acknowledgments}
The authors were funded by the Department of Atomic Energy (DAE), Government of India.

\bibliography{snigdha}

\begin{thebibliography}{109}%
\makeatletter
\providecommand \@ifxundefined [1]{%
 \@ifx{#1\undefined}
}%
\providecommand \@ifnum [1]{%
 \ifnum #1\expandafter \@firstoftwo
 \else \expandafter \@secondoftwo
 \fi
}%
\providecommand \@ifx [1]{%
 \ifx #1\expandafter \@firstoftwo
 \else \expandafter \@secondoftwo
 \fi
}%
\providecommand \natexlab [1]{#1}%
\providecommand \enquote  [1]{``#1''}%
\providecommand \bibnamefont  [1]{#1}%
\providecommand \bibfnamefont [1]{#1}%
\providecommand \citenamefont [1]{#1}%
\providecommand \href@noop [0]{\@secondoftwo}%
\providecommand \href [0]{\begingroup \@sanitize@url \@href}%
\providecommand \@href[1]{\@@startlink{#1}\@@href}%
\providecommand \@@href[1]{\endgroup#1\@@endlink}%
\providecommand \@sanitize@url [0]{\catcode `\\12\catcode `\$12\catcode
  `\&12\catcode `\#12\catcode `\^12\catcode `\_12\catcode `\%12\relax}%
\providecommand \@@startlink[1]{}%
\providecommand \@@endlink[0]{}%
\providecommand \url  [0]{\begingroup\@sanitize@url \@url }%
\providecommand \@url [1]{\endgroup\@href {#1}{\urlprefix }}%
\providecommand \urlprefix  [0]{URL }%
\providecommand \Eprint [0]{\href }%
\providecommand \doibase [0]{http://dx.doi.org/}%
\providecommand \selectlanguage [0]{\@gobble}%
\providecommand \bibinfo  [0]{\@secondoftwo}%
\providecommand \bibfield  [0]{\@secondoftwo}%
\providecommand \translation [1]{[#1]}%
\providecommand \BibitemOpen [0]{}%
\providecommand \bibitemStop [0]{}%
\providecommand \bibitemNoStop [0]{.\EOS\space}%
\providecommand \EOS [0]{\spacefactor3000\relax}%
\providecommand \BibitemShut  [1]{\csname bibitem#1\endcsname}%
\let\auto@bib@innerbib\@empty
\bibitem [{\citenamefont {Kharzeev}\ \emph {et~al.}(2013)\citenamefont
  {Kharzeev}, \citenamefont {Landsteiner}, \citenamefont {Schmitt},\ and\
  \citenamefont {Yee}}]{Kharzeev:2012ph}%
  \BibitemOpen
  \bibfield  {author} {\bibinfo {author} {\bibfnamefont {D.~E.}\ \bibnamefont
  {Kharzeev}}, \bibinfo {author} {\bibfnamefont {K.}~\bibnamefont
  {Landsteiner}}, \bibinfo {author} {\bibfnamefont {A.}~\bibnamefont
  {Schmitt}}, \ and\ \bibinfo {author} {\bibfnamefont {H.-U.}\ \bibnamefont
  {Yee}},\ }\href {\doibase 10.1007/978-3-642-37305-3_1} {\bibfield  {journal}
  {\bibinfo  {journal} {Lect. Notes Phys.}\ }\textbf {\bibinfo {volume}
  {871}},\ \bibinfo {pages} {1} (\bibinfo {year} {2013})},\ \Eprint
  {http://arxiv.org/abs/1211.6245} {arXiv:1211.6245 [hep-ph]} \BibitemShut
  {NoStop}%
\bibitem [{\citenamefont {Kharzeev}\ and\ \citenamefont
  {Zhitnitsky}(2007)}]{Kharzeev:2007tn}%
  \BibitemOpen
  \bibfield  {author} {\bibinfo {author} {\bibfnamefont {D.}~\bibnamefont
  {Kharzeev}}\ and\ \bibinfo {author} {\bibfnamefont {A.}~\bibnamefont
  {Zhitnitsky}},\ }\href {\doibase 10.1016/j.nuclphysa.2007.10.001} {\bibfield
  {journal} {\bibinfo  {journal} {Nucl. Phys.}\ }\textbf {\bibinfo {volume}
  {A797}},\ \bibinfo {pages} {67} (\bibinfo {year} {2007})},\ \Eprint
  {http://arxiv.org/abs/0706.1026} {arXiv:0706.1026 [hep-ph]} \BibitemShut
  {NoStop}%
\bibitem [{\citenamefont {Chernodub}(2010)}]{Chernodub:2010qx}%
  \BibitemOpen
  \bibfield  {author} {\bibinfo {author} {\bibfnamefont {M.~N.}\ \bibnamefont
  {Chernodub}},\ }\href {\doibase 10.1103/PhysRevD.82.085011} {\bibfield
  {journal} {\bibinfo  {journal} {Phys. Rev.}\ }\textbf {\bibinfo {volume}
  {D82}},\ \bibinfo {pages} {085011} (\bibinfo {year} {2010})},\ \Eprint
  {http://arxiv.org/abs/1008.1055} {arXiv:1008.1055 [hep-ph]} \BibitemShut
  {NoStop}%
\bibitem [{\citenamefont {Chernodub}(2013)}]{Chernodub:2012tf}%
  \BibitemOpen
  \bibfield  {author} {\bibinfo {author} {\bibfnamefont {M.~N.}\ \bibnamefont
  {Chernodub}},\ }\href {\doibase 10.1007/978-3-642-37305-3_6} {\bibfield
  {journal} {\bibinfo  {journal} {Lect. Notes Phys.}\ }\textbf {\bibinfo
  {volume} {871}},\ \bibinfo {pages} {143} (\bibinfo {year} {2013})},\ \Eprint
  {http://arxiv.org/abs/1208.5025} {arXiv:1208.5025 [hep-ph]} \BibitemShut
  {NoStop}%
\bibitem [{\citenamefont {Fukushima}\ \emph {et~al.}(2008)\citenamefont
  {Fukushima}, \citenamefont {Kharzeev},\ and\ \citenamefont
  {Warringa}}]{Fukushima:2008xe}%
  \BibitemOpen
  \bibfield  {author} {\bibinfo {author} {\bibfnamefont {K.}~\bibnamefont
  {Fukushima}}, \bibinfo {author} {\bibfnamefont {D.~E.}\ \bibnamefont
  {Kharzeev}}, \ and\ \bibinfo {author} {\bibfnamefont {H.~J.}\ \bibnamefont
  {Warringa}},\ }\href {\doibase 10.1103/PhysRevD.78.074033} {\bibfield
  {journal} {\bibinfo  {journal} {Phys. Rev.}\ }\textbf {\bibinfo {volume}
  {D78}},\ \bibinfo {pages} {074033} (\bibinfo {year} {2008})},\ \Eprint
  {http://arxiv.org/abs/0808.3382} {arXiv:0808.3382 [hep-ph]} \BibitemShut
  {NoStop}%
\bibitem [{\citenamefont {Kharzeev}\ \emph {et~al.}(2008)\citenamefont
  {Kharzeev}, \citenamefont {McLerran},\ and\ \citenamefont
  {Warringa}}]{Kharzeev:2007jp}%
  \BibitemOpen
  \bibfield  {author} {\bibinfo {author} {\bibfnamefont {D.~E.}\ \bibnamefont
  {Kharzeev}}, \bibinfo {author} {\bibfnamefont {L.~D.}\ \bibnamefont
  {McLerran}}, \ and\ \bibinfo {author} {\bibfnamefont {H.~J.}\ \bibnamefont
  {Warringa}},\ }\href {\doibase 10.1016/j.nuclphysa.2008.02.298} {\bibfield
  {journal} {\bibinfo  {journal} {Nucl. Phys.}\ }\textbf {\bibinfo {volume}
  {A803}},\ \bibinfo {pages} {227} (\bibinfo {year} {2008})},\ \Eprint
  {http://arxiv.org/abs/0711.0950} {arXiv:0711.0950 [hep-ph]} \BibitemShut
  {NoStop}%
\bibitem [{\citenamefont {Kharzeev}\ and\ \citenamefont
  {Warringa}(2009)}]{Kharzeev:2009pj}%
  \BibitemOpen
  \bibfield  {author} {\bibinfo {author} {\bibfnamefont {D.~E.}\ \bibnamefont
  {Kharzeev}}\ and\ \bibinfo {author} {\bibfnamefont {H.~J.}\ \bibnamefont
  {Warringa}},\ }\href {\doibase 10.1103/PhysRevD.80.034028} {\bibfield
  {journal} {\bibinfo  {journal} {Phys. Rev.}\ }\textbf {\bibinfo {volume}
  {D80}},\ \bibinfo {pages} {034028} (\bibinfo {year} {2009})},\ \Eprint
  {http://arxiv.org/abs/0907.5007} {arXiv:0907.5007 [hep-ph]} \BibitemShut
  {NoStop}%
\bibitem [{\citenamefont {Bali}\ \emph {et~al.}(2012)\citenamefont {Bali},
  \citenamefont {Bruckmann}, \citenamefont {Endrodi}, \citenamefont {Fodor},
  \citenamefont {Katz}, \citenamefont {Krieg}, \citenamefont {Schafer},\ and\
  \citenamefont {Szabo}}]{Bali:2011qj}%
  \BibitemOpen
  \bibfield  {author} {\bibinfo {author} {\bibfnamefont {G.~S.}\ \bibnamefont
  {Bali}}, \bibinfo {author} {\bibfnamefont {F.}~\bibnamefont {Bruckmann}},
  \bibinfo {author} {\bibfnamefont {G.}~\bibnamefont {Endrodi}}, \bibinfo
  {author} {\bibfnamefont {Z.}~\bibnamefont {Fodor}}, \bibinfo {author}
  {\bibfnamefont {S.~D.}\ \bibnamefont {Katz}}, \bibinfo {author}
  {\bibfnamefont {S.}~\bibnamefont {Krieg}}, \bibinfo {author} {\bibfnamefont
  {A.}~\bibnamefont {Schafer}}, \ and\ \bibinfo {author} {\bibfnamefont
  {K.~K.}\ \bibnamefont {Szabo}},\ }\href {\doibase 10.1007/JHEP02(2012)044}
  {\bibfield  {journal} {\bibinfo  {journal} {JHEP}\ }\textbf {\bibinfo
  {volume} {02}},\ \bibinfo {pages} {044} (\bibinfo {year} {2012})},\ \Eprint
  {http://arxiv.org/abs/1111.4956} {arXiv:1111.4956 [hep-lat]} \BibitemShut
  {NoStop}%
\bibitem [{\citenamefont {Shovkovy}(2013)}]{Shovkovy:2012zn}%
  \BibitemOpen
  \bibfield  {author} {\bibinfo {author} {\bibfnamefont {I.~A.}\ \bibnamefont
  {Shovkovy}},\ }\href {\doibase 10.1007/978-3-642-37305-3_2} {\bibfield
  {journal} {\bibinfo  {journal} {Lect. Notes Phys.}\ }\textbf {\bibinfo
  {volume} {871}},\ \bibinfo {pages} {13} (\bibinfo {year} {2013})},\ \Eprint
  {http://arxiv.org/abs/1207.5081} {arXiv:1207.5081 [hep-ph]} \BibitemShut
  {NoStop}%
\bibitem [{\citenamefont {Gusynin}\ \emph {et~al.}(1994)\citenamefont
  {Gusynin}, \citenamefont {Miransky},\ and\ \citenamefont
  {Shovkovy}}]{Gusynin:1994re}%
  \BibitemOpen
  \bibfield  {author} {\bibinfo {author} {\bibfnamefont {V.~P.}\ \bibnamefont
  {Gusynin}}, \bibinfo {author} {\bibfnamefont {V.~A.}\ \bibnamefont
  {Miransky}}, \ and\ \bibinfo {author} {\bibfnamefont {I.~A.}\ \bibnamefont
  {Shovkovy}},\ }\href {\doibase 10.1103/PhysRevLett.76.1005,
  10.1103/PhysRevLett.73.3499} {\bibfield  {journal} {\bibinfo  {journal}
  {Phys. Rev. Lett.}\ }\textbf {\bibinfo {volume} {73}},\ \bibinfo {pages}
  {3499} (\bibinfo {year} {1994})},\ \bibinfo {note} {[Erratum: Phys. Rev.
  Lett.76,1005(1996)]},\ \Eprint {http://arxiv.org/abs/hep-ph/9405262}
  {arXiv:hep-ph/9405262 [hep-ph]} \BibitemShut {NoStop}%
\bibitem [{\citenamefont {Gusynin}\ \emph {et~al.}(1996)\citenamefont
  {Gusynin}, \citenamefont {Miransky},\ and\ \citenamefont
  {Shovkovy}}]{Gusynin:1995nb}%
  \BibitemOpen
  \bibfield  {author} {\bibinfo {author} {\bibfnamefont {V.~P.}\ \bibnamefont
  {Gusynin}}, \bibinfo {author} {\bibfnamefont {V.~A.}\ \bibnamefont
  {Miransky}}, \ and\ \bibinfo {author} {\bibfnamefont {I.~A.}\ \bibnamefont
  {Shovkovy}},\ }\href {\doibase 10.1016/0550-3213(96)00021-1} {\bibfield
  {journal} {\bibinfo  {journal} {Nucl. Phys.}\ }\textbf {\bibinfo {volume}
  {B462}},\ \bibinfo {pages} {249} (\bibinfo {year} {1996})},\ \Eprint
  {http://arxiv.org/abs/hep-ph/9509320} {arXiv:hep-ph/9509320 [hep-ph]}
  \BibitemShut {NoStop}%
\bibitem [{\citenamefont {Gusynin}\ \emph {et~al.}(1999)\citenamefont
  {Gusynin}, \citenamefont {Miransky},\ and\ \citenamefont
  {Shovkovy}}]{Gusynin:1999pq}%
  \BibitemOpen
  \bibfield  {author} {\bibinfo {author} {\bibfnamefont {V.~P.}\ \bibnamefont
  {Gusynin}}, \bibinfo {author} {\bibfnamefont {V.~A.}\ \bibnamefont
  {Miransky}}, \ and\ \bibinfo {author} {\bibfnamefont {I.~A.}\ \bibnamefont
  {Shovkovy}},\ }\href {\doibase 10.1016/S0550-3213(99)00573-8} {\bibfield
  {journal} {\bibinfo  {journal} {Nucl. Phys.}\ }\textbf {\bibinfo {volume}
  {B563}},\ \bibinfo {pages} {361} (\bibinfo {year} {1999})},\ \Eprint
  {http://arxiv.org/abs/hep-ph/9908320} {arXiv:hep-ph/9908320 [hep-ph]}
  \BibitemShut {NoStop}%
\bibitem [{\citenamefont {Preis}\ \emph {et~al.}(2011)\citenamefont {Preis},
  \citenamefont {Rebhan},\ and\ \citenamefont {Schmitt}}]{Preis:2010cq}%
  \BibitemOpen
  \bibfield  {author} {\bibinfo {author} {\bibfnamefont {F.}~\bibnamefont
  {Preis}}, \bibinfo {author} {\bibfnamefont {A.}~\bibnamefont {Rebhan}}, \
  and\ \bibinfo {author} {\bibfnamefont {A.}~\bibnamefont {Schmitt}},\ }\href
  {\doibase 10.1007/JHEP03(2011)033} {\bibfield  {journal} {\bibinfo  {journal}
  {JHEP}\ }\textbf {\bibinfo {volume} {03}},\ \bibinfo {pages} {033} (\bibinfo
  {year} {2011})},\ \Eprint {http://arxiv.org/abs/1012.4785} {arXiv:1012.4785
  [hep-th]} \BibitemShut {NoStop}%
\bibitem [{\citenamefont {Preis}\ \emph {et~al.}(2013)\citenamefont {Preis},
  \citenamefont {Rebhan},\ and\ \citenamefont {Schmitt}}]{Preis:2012fh}%
  \BibitemOpen
  \bibfield  {author} {\bibinfo {author} {\bibfnamefont {F.}~\bibnamefont
  {Preis}}, \bibinfo {author} {\bibfnamefont {A.}~\bibnamefont {Rebhan}}, \
  and\ \bibinfo {author} {\bibfnamefont {A.}~\bibnamefont {Schmitt}},\ }\href
  {\doibase 10.1007/978-3-642-37305-3_3} {\bibfield  {journal} {\bibinfo
  {journal} {Lect. Notes Phys.}\ }\textbf {\bibinfo {volume} {871}},\ \bibinfo
  {pages} {51} (\bibinfo {year} {2013})},\ \Eprint
  {http://arxiv.org/abs/1208.0536} {arXiv:1208.0536 [hep-ph]} \BibitemShut
  {NoStop}%
\bibitem [{\citenamefont {Chernodub}\ \emph {et~al.}(2012)\citenamefont
  {Chernodub}, \citenamefont {Van~Doorsselaere},\ and\ \citenamefont
  {Verschelde}}]{Chernodub:2011gs}%
  \BibitemOpen
  \bibfield  {author} {\bibinfo {author} {\bibfnamefont {M.~N.}\ \bibnamefont
  {Chernodub}}, \bibinfo {author} {\bibfnamefont {J.}~\bibnamefont
  {Van~Doorsselaere}}, \ and\ \bibinfo {author} {\bibfnamefont
  {H.}~\bibnamefont {Verschelde}},\ }\href {\doibase
  10.1103/PhysRevD.85.045002} {\bibfield  {journal} {\bibinfo  {journal} {Phys.
  Rev.}\ }\textbf {\bibinfo {volume} {D85}},\ \bibinfo {pages} {045002}
  (\bibinfo {year} {2012})},\ \Eprint {http://arxiv.org/abs/1111.4401}
  {arXiv:1111.4401 [hep-ph]} \BibitemShut {NoStop}%
\bibitem [{\citenamefont {Chernodub}(2011)}]{Chernodub:2011mc}%
  \BibitemOpen
  \bibfield  {author} {\bibinfo {author} {\bibfnamefont {M.~N.}\ \bibnamefont
  {Chernodub}},\ }\href {\doibase 10.1103/PhysRevLett.106.142003} {\bibfield
  {journal} {\bibinfo  {journal} {Phys. Rev. Lett.}\ }\textbf {\bibinfo
  {volume} {106}},\ \bibinfo {pages} {142003} (\bibinfo {year} {2011})},\
  \Eprint {http://arxiv.org/abs/1101.0117} {arXiv:1101.0117 [hep-ph]}
  \BibitemShut {NoStop}%
\bibitem [{\citenamefont {Duncan}\ and\ \citenamefont
  {Thompson}(1992)}]{Duncan:1992hi}%
  \BibitemOpen
  \bibfield  {author} {\bibinfo {author} {\bibfnamefont {R.~C.}\ \bibnamefont
  {Duncan}}\ and\ \bibinfo {author} {\bibfnamefont {C.}~\bibnamefont
  {Thompson}},\ }\href {\doibase 10.1086/186413} {\bibfield  {journal}
  {\bibinfo  {journal} {Astrophys. J.}\ }\textbf {\bibinfo {volume} {392}},\
  \bibinfo {pages} {L9} (\bibinfo {year} {1992})}\BibitemShut {NoStop}%
\bibitem [{\citenamefont {Ferrer}\ \emph {et~al.}(2005)\citenamefont {Ferrer},
  \citenamefont {de~la Incera},\ and\ \citenamefont {Manuel}}]{Ferrer:2005vd}%
  \BibitemOpen
  \bibfield  {author} {\bibinfo {author} {\bibfnamefont {E.~J.}\ \bibnamefont
  {Ferrer}}, \bibinfo {author} {\bibfnamefont {V.}~\bibnamefont {de~la
  Incera}}, \ and\ \bibinfo {author} {\bibfnamefont {C.}~\bibnamefont
  {Manuel}},\ }\href {\doibase 10.1103/PhysRevLett.95.152002} {\bibfield
  {journal} {\bibinfo  {journal} {Phys. Rev. Lett.}\ }\textbf {\bibinfo
  {volume} {95}},\ \bibinfo {pages} {152002} (\bibinfo {year} {2005})},\
  \Eprint {http://arxiv.org/abs/hep-ph/0503162} {arXiv:hep-ph/0503162 [hep-ph]}
  \BibitemShut {NoStop}%
\bibitem [{\citenamefont {Ferrer}\ \emph {et~al.}(2006)\citenamefont {Ferrer},
  \citenamefont {de~la Incera},\ and\ \citenamefont {Manuel}}]{Ferrer:2006vw}%
  \BibitemOpen
  \bibfield  {author} {\bibinfo {author} {\bibfnamefont {E.~J.}\ \bibnamefont
  {Ferrer}}, \bibinfo {author} {\bibfnamefont {V.}~\bibnamefont {de~la
  Incera}}, \ and\ \bibinfo {author} {\bibfnamefont {C.}~\bibnamefont
  {Manuel}},\ }\href {\doibase 10.1016/j.nuclphysb.2006.04.013} {\bibfield
  {journal} {\bibinfo  {journal} {Nucl. Phys.}\ }\textbf {\bibinfo {volume}
  {B747}},\ \bibinfo {pages} {88} (\bibinfo {year} {2006})},\ \Eprint
  {http://arxiv.org/abs/hep-ph/0603233} {arXiv:hep-ph/0603233 [hep-ph]}
  \BibitemShut {NoStop}%
\bibitem [{\citenamefont {Ferrer}\ and\ \citenamefont {de~la
  Incera}(2007)}]{Ferrer:2007iw}%
  \BibitemOpen
  \bibfield  {author} {\bibinfo {author} {\bibfnamefont {E.~J.}\ \bibnamefont
  {Ferrer}}\ and\ \bibinfo {author} {\bibfnamefont {V.}~\bibnamefont {de~la
  Incera}},\ }\href {\doibase 10.1103/PhysRevD.76.045011} {\bibfield  {journal}
  {\bibinfo  {journal} {Phys. Rev.}\ }\textbf {\bibinfo {volume} {D76}},\
  \bibinfo {pages} {045011} (\bibinfo {year} {2007})},\ \Eprint
  {http://arxiv.org/abs/nucl-th/0703034} {arXiv:nucl-th/0703034 [NUCL-TH]}
  \BibitemShut {NoStop}%
\bibitem [{\citenamefont {Fukushima}\ and\ \citenamefont
  {Warringa}(2008)}]{Fukushima:2007fc}%
  \BibitemOpen
  \bibfield  {author} {\bibinfo {author} {\bibfnamefont {K.}~\bibnamefont
  {Fukushima}}\ and\ \bibinfo {author} {\bibfnamefont {H.~J.}\ \bibnamefont
  {Warringa}},\ }\href {\doibase 10.1103/PhysRevLett.100.032007} {\bibfield
  {journal} {\bibinfo  {journal} {Phys. Rev. Lett.}\ }\textbf {\bibinfo
  {volume} {100}},\ \bibinfo {pages} {032007} (\bibinfo {year} {2008})},\
  \Eprint {http://arxiv.org/abs/0707.3785} {arXiv:0707.3785 [hep-ph]}
  \BibitemShut {NoStop}%
\bibitem [{\citenamefont {Feng}\ \emph {et~al.}(2010)\citenamefont {Feng},
  \citenamefont {Hou}, \citenamefont {Ren},\ and\ \citenamefont
  {Wu}}]{Feng:2009vt}%
  \BibitemOpen
  \bibfield  {author} {\bibinfo {author} {\bibfnamefont {B.}~\bibnamefont
  {Feng}}, \bibinfo {author} {\bibfnamefont {D.}~\bibnamefont {Hou}}, \bibinfo
  {author} {\bibfnamefont {H.-c.}\ \bibnamefont {Ren}}, \ and\ \bibinfo
  {author} {\bibfnamefont {P.-p.}\ \bibnamefont {Wu}},\ }\href {\doibase
  10.1103/PhysRevLett.105.042001} {\bibfield  {journal} {\bibinfo  {journal}
  {Phys. Rev. Lett.}\ }\textbf {\bibinfo {volume} {105}},\ \bibinfo {pages}
  {042001} (\bibinfo {year} {2010})},\ \Eprint {http://arxiv.org/abs/0911.4997}
  {arXiv:0911.4997 [hep-ph]} \BibitemShut {NoStop}%
\bibitem [{\citenamefont {Fayazbakhsh}\ and\ \citenamefont
  {Sadooghi}(2010)}]{Fayazbakhsh:2010gc}%
  \BibitemOpen
  \bibfield  {author} {\bibinfo {author} {\bibfnamefont {S.}~\bibnamefont
  {Fayazbakhsh}}\ and\ \bibinfo {author} {\bibfnamefont {N.}~\bibnamefont
  {Sadooghi}},\ }\href {\doibase 10.1103/PhysRevD.82.045010} {\bibfield
  {journal} {\bibinfo  {journal} {Phys. Rev.}\ }\textbf {\bibinfo {volume}
  {D82}},\ \bibinfo {pages} {045010} (\bibinfo {year} {2010})},\ \Eprint
  {http://arxiv.org/abs/1005.5022} {arXiv:1005.5022 [hep-ph]} \BibitemShut
  {NoStop}%
\bibitem [{\citenamefont {Fayazbakhsh}\ and\ \citenamefont
  {Sadooghi}(2011)}]{Fayazbakhsh:2010bh}%
  \BibitemOpen
  \bibfield  {author} {\bibinfo {author} {\bibfnamefont {S.}~\bibnamefont
  {Fayazbakhsh}}\ and\ \bibinfo {author} {\bibfnamefont {N.}~\bibnamefont
  {Sadooghi}},\ }\href {\doibase 10.1103/PhysRevD.83.025026} {\bibfield
  {journal} {\bibinfo  {journal} {Phys. Rev.}\ }\textbf {\bibinfo {volume}
  {D83}},\ \bibinfo {pages} {025026} (\bibinfo {year} {2011})},\ \Eprint
  {http://arxiv.org/abs/1009.6125} {arXiv:1009.6125 [hep-ph]} \BibitemShut
  {NoStop}%
\bibitem [{\citenamefont {Skokov}\ \emph {et~al.}(2009)\citenamefont {Skokov},
  \citenamefont {Illarionov},\ and\ \citenamefont {Toneev}}]{Skokov:2009qp}%
  \BibitemOpen
  \bibfield  {author} {\bibinfo {author} {\bibfnamefont {V.}~\bibnamefont
  {Skokov}}, \bibinfo {author} {\bibfnamefont {A.~{\relax Yu}.}\ \bibnamefont
  {Illarionov}}, \ and\ \bibinfo {author} {\bibfnamefont {V.}~\bibnamefont
  {Toneev}},\ }\href {\doibase 10.1142/S0217751X09047570} {\bibfield  {journal}
  {\bibinfo  {journal} {Int. J. Mod. Phys.}\ }\textbf {\bibinfo {volume}
  {A24}},\ \bibinfo {pages} {5925} (\bibinfo {year} {2009})},\ \Eprint
  {http://arxiv.org/abs/0907.1396} {arXiv:0907.1396 [nucl-th]} \BibitemShut
  {NoStop}%
\bibitem [{\citenamefont {Gursoy}\ \emph {et~al.}(2014)\citenamefont {Gursoy},
  \citenamefont {Kharzeev},\ and\ \citenamefont {Rajagopal}}]{Gursoy:2014aka}%
  \BibitemOpen
  \bibfield  {author} {\bibinfo {author} {\bibfnamefont {U.}~\bibnamefont
  {Gursoy}}, \bibinfo {author} {\bibfnamefont {D.}~\bibnamefont {Kharzeev}}, \
  and\ \bibinfo {author} {\bibfnamefont {K.}~\bibnamefont {Rajagopal}},\ }\href
  {\doibase 10.1103/PhysRevC.89.054905} {\bibfield  {journal} {\bibinfo
  {journal} {Phys. Rev.}\ }\textbf {\bibinfo {volume} {C89}},\ \bibinfo {pages}
  {054905} (\bibinfo {year} {2014})},\ \Eprint {http://arxiv.org/abs/1401.3805}
  {arXiv:1401.3805 [hep-ph]} \BibitemShut {NoStop}%
\bibitem [{\citenamefont {Thompson}\ and\ \citenamefont
  {Duncan}(1993)}]{Thompson:1993hn}%
  \BibitemOpen
  \bibfield  {author} {\bibinfo {author} {\bibfnamefont {C.}~\bibnamefont
  {Thompson}}\ and\ \bibinfo {author} {\bibfnamefont {R.~C.}\ \bibnamefont
  {Duncan}},\ }\href {\doibase 10.1086/172580} {\bibfield  {journal} {\bibinfo
  {journal} {Astrophys. J.}\ }\textbf {\bibinfo {volume} {408}},\ \bibinfo
  {pages} {194} (\bibinfo {year} {1993})}\BibitemShut {NoStop}%
\bibitem [{\citenamefont {Vachaspati}(1991)}]{Vachaspati:1991nm}%
  \BibitemOpen
  \bibfield  {author} {\bibinfo {author} {\bibfnamefont {T.}~\bibnamefont
  {Vachaspati}},\ }\href {\doibase 10.1016/0370-2693(91)90051-Q} {\bibfield
  {journal} {\bibinfo  {journal} {Phys. Lett.}\ }\textbf {\bibinfo {volume}
  {B265}},\ \bibinfo {pages} {258} (\bibinfo {year} {1991})}\BibitemShut
  {NoStop}%
\bibitem [{\citenamefont {Campanelli}(2013)}]{Campanelli:2013mea}%
  \BibitemOpen
  \bibfield  {author} {\bibinfo {author} {\bibfnamefont {L.}~\bibnamefont
  {Campanelli}},\ }\href {\doibase 10.1103/PhysRevLett.111.061301} {\bibfield
  {journal} {\bibinfo  {journal} {Phys. Rev. Lett.}\ }\textbf {\bibinfo
  {volume} {111}},\ \bibinfo {pages} {061301} (\bibinfo {year} {2013})},\
  \Eprint {http://arxiv.org/abs/1304.6534} {arXiv:1304.6534 [astro-ph.CO]}
  \BibitemShut {NoStop}%
\bibitem [{\citenamefont {de~Forcrand}\ and\ \citenamefont
  {Philipsen}(2007)}]{Forcrand1}%
  \BibitemOpen
  \bibfield  {author} {\bibinfo {author} {\bibfnamefont {P.}~\bibnamefont
  {de~Forcrand}}\ and\ \bibinfo {author} {\bibfnamefont {O.}~\bibnamefont
  {Philipsen}},\ }\href {\doibase 10.1088/1126-6708/2007/01/077} {\bibfield
  {journal} {\bibinfo  {journal} {JHEP}\ }\textbf {\bibinfo {volume} {01}},\
  \bibinfo {pages} {077} (\bibinfo {year} {2007})},\ \Eprint
  {http://arxiv.org/abs/hep-lat/0607017} {arXiv:hep-lat/0607017 [hep-lat]}
  \BibitemShut {NoStop}%
\bibitem [{\citenamefont {de~Forcrand}\ and\ \citenamefont
  {Philipsen}(2008{\natexlab{a}})}]{Forcrand2}%
  \BibitemOpen
  \bibfield  {author} {\bibinfo {author} {\bibfnamefont {P.}~\bibnamefont
  {de~Forcrand}}\ and\ \bibinfo {author} {\bibfnamefont {O.}~\bibnamefont
  {Philipsen}},\ }\href {\doibase 10.1088/1126-6708/2008/11/012} {\bibfield
  {journal} {\bibinfo  {journal} {JHEP}\ }\textbf {\bibinfo {volume} {11}},\
  \bibinfo {pages} {012} (\bibinfo {year} {2008}{\natexlab{a}})},\ \Eprint
  {http://arxiv.org/abs/0808.1096} {arXiv:0808.1096 [hep-lat]} \BibitemShut
  {NoStop}%
\bibitem [{\citenamefont {de~Forcrand}\ and\ \citenamefont
  {Philipsen}(2008{\natexlab{b}})}]{Forcrand3}%
  \BibitemOpen
  \bibfield  {author} {\bibinfo {author} {\bibfnamefont {P.}~\bibnamefont
  {de~Forcrand}}\ and\ \bibinfo {author} {\bibfnamefont {O.}~\bibnamefont
  {Philipsen}},\ }\bibfield  {booktitle} {\emph {\bibinfo {booktitle}
  {{Proceedings, 26th International Symposium on Lattice field theory (Lattice
  2008): Williamsburg, USA, July 14-19, 2008}}},\ }\href {\doibase
  10.22323/1.066.0208} {\bibfield  {journal} {\bibinfo  {journal} {PoS}\
  }\textbf {\bibinfo {volume} {LATTICE2008}},\ \bibinfo {pages} {208} (\bibinfo
  {year} {2008}{\natexlab{b}})},\ \Eprint {http://arxiv.org/abs/0811.3858}
  {arXiv:0811.3858 [hep-lat]} \BibitemShut {NoStop}%
\bibitem [{\citenamefont {Brandt}\ \emph {et~al.}(2016)\citenamefont {Brandt},
  \citenamefont {Bali}, \citenamefont {Endrödi},\ and\ \citenamefont
  {Glässle}}]{Bali_lat}%
  \BibitemOpen
  \bibfield  {author} {\bibinfo {author} {\bibfnamefont {B.~B.}\ \bibnamefont
  {Brandt}}, \bibinfo {author} {\bibfnamefont {G.}~\bibnamefont {Bali}},
  \bibinfo {author} {\bibfnamefont {G.}~\bibnamefont {Endrödi}}, \ and\
  \bibinfo {author} {\bibfnamefont {B.}~\bibnamefont {Glässle}},\ }\bibfield
  {booktitle} {\emph {\bibinfo {booktitle} {{Proceedings, 33rd International
  Symposium on Lattice Field Theory (Lattice 2015): Kobe, Japan, July 14-18,
  2015}}},\ }\href {\doibase 10.22323/1.251.0265} {\bibfield  {journal}
  {\bibinfo  {journal} {PoS}\ }\textbf {\bibinfo {volume} {LATTICE2015}},\
  \bibinfo {pages} {265} (\bibinfo {year} {2016})},\ \Eprint
  {http://arxiv.org/abs/1510.03899} {arXiv:1510.03899 [hep-lat]} \BibitemShut
  {NoStop}%
\bibitem [{\citenamefont {Luschevskaya}\ and\ \citenamefont
  {Larina}(2014)}]{Luschevskaya}%
  \BibitemOpen
  \bibfield  {author} {\bibinfo {author} {\bibfnamefont {E.~V.}\ \bibnamefont
  {Luschevskaya}}\ and\ \bibinfo {author} {\bibfnamefont {O.~V.}\ \bibnamefont
  {Larina}},\ }\href {\doibase 10.1016/j.nuclphysb.2014.04.003} {\bibfield
  {journal} {\bibinfo  {journal} {Nucl. Phys.}\ }\textbf {\bibinfo {volume}
  {B884}},\ \bibinfo {pages} {1} (\bibinfo {year} {2014})},\ \Eprint
  {http://arxiv.org/abs/1203.5699} {arXiv:1203.5699 [hep-lat]} \BibitemShut
  {NoStop}%
\bibitem [{\citenamefont {Aoki}\ \emph
  {et~al.}(2006{\natexlab{a}})\citenamefont {Aoki}, \citenamefont {Endrodi},
  \citenamefont {Fodor}, \citenamefont {Katz},\ and\ \citenamefont
  {Szabo}}]{Aoki}%
  \BibitemOpen
  \bibfield  {author} {\bibinfo {author} {\bibfnamefont {Y.}~\bibnamefont
  {Aoki}}, \bibinfo {author} {\bibfnamefont {G.}~\bibnamefont {Endrodi}},
  \bibinfo {author} {\bibfnamefont {Z.}~\bibnamefont {Fodor}}, \bibinfo
  {author} {\bibfnamefont {S.~D.}\ \bibnamefont {Katz}}, \ and\ \bibinfo
  {author} {\bibfnamefont {K.~K.}\ \bibnamefont {Szabo}},\ }\href {\doibase
  10.1038/nature05120} {\bibfield  {journal} {\bibinfo  {journal} {Nature}\
  }\textbf {\bibinfo {volume} {443}},\ \bibinfo {pages} {675} (\bibinfo {year}
  {2006}{\natexlab{a}})},\ \Eprint {http://arxiv.org/abs/hep-lat/0611014}
  {arXiv:hep-lat/0611014 [hep-lat]} \BibitemShut {NoStop}%
\bibitem [{\citenamefont {Aoki}\ \emph
  {et~al.}(2006{\natexlab{b}})\citenamefont {Aoki}, \citenamefont {Endrodi},
  \citenamefont {Fodor}, \citenamefont {Katz},\ and\ \citenamefont
  {Szabo}}]{Aoki2}%
  \BibitemOpen
  \bibfield  {author} {\bibinfo {author} {\bibfnamefont {Y.}~\bibnamefont
  {Aoki}}, \bibinfo {author} {\bibfnamefont {G.}~\bibnamefont {Endrodi}},
  \bibinfo {author} {\bibfnamefont {Z.}~\bibnamefont {Fodor}}, \bibinfo
  {author} {\bibfnamefont {S.~D.}\ \bibnamefont {Katz}}, \ and\ \bibinfo
  {author} {\bibfnamefont {K.~K.}\ \bibnamefont {Szabo}},\ }\href {\doibase
  10.1038/nature05120} {\bibfield  {journal} {\bibinfo  {journal} {Nature}\
  }\textbf {\bibinfo {volume} {443}},\ \bibinfo {pages} {675} (\bibinfo {year}
  {2006}{\natexlab{b}})},\ \Eprint {http://arxiv.org/abs/hep-lat/0611014}
  {arXiv:hep-lat/0611014 [hep-lat]} \BibitemShut {NoStop}%
\bibitem [{\citenamefont {Adler}\ \emph {et~al.}(2007)\citenamefont {Adler}
  \emph {et~al.}}]{Adler:2006yt}%
  \BibitemOpen
  \bibfield  {author} {\bibinfo {author} {\bibfnamefont {S.~S.}\ \bibnamefont
  {Adler}} \emph {et~al.} (\bibinfo {collaboration} {PHENIX}),\ }\href
  {\doibase 10.1103/PhysRevLett.98.012002} {\bibfield  {journal} {\bibinfo
  {journal} {Phys. Rev. Lett.}\ }\textbf {\bibinfo {volume} {98}},\ \bibinfo
  {pages} {012002} (\bibinfo {year} {2007})},\ \Eprint
  {http://arxiv.org/abs/hep-ex/0609031} {arXiv:hep-ex/0609031 [hep-ex]}
  \BibitemShut {NoStop}%
\bibitem [{\citenamefont {Adare}\ \emph
  {et~al.}(2007{\natexlab{a}})\citenamefont {Adare} \emph
  {et~al.}}]{Adare:2006nq}%
  \BibitemOpen
  \bibfield  {author} {\bibinfo {author} {\bibfnamefont {A.}~\bibnamefont
  {Adare}} \emph {et~al.} (\bibinfo {collaboration} {PHENIX}),\ }\href
  {\doibase 10.1103/PhysRevLett.98.172301} {\bibfield  {journal} {\bibinfo
  {journal} {Phys. Rev. Lett.}\ }\textbf {\bibinfo {volume} {98}},\ \bibinfo
  {pages} {172301} (\bibinfo {year} {2007}{\natexlab{a}})},\ \Eprint
  {http://arxiv.org/abs/nucl-ex/0611018} {arXiv:nucl-ex/0611018 [nucl-ex]}
  \BibitemShut {NoStop}%
\bibitem [{\citenamefont {Adare}\ \emph
  {et~al.}(2007{\natexlab{b}})\citenamefont {Adare} \emph
  {et~al.}}]{Adare:2006ti}%
  \BibitemOpen
  \bibfield  {author} {\bibinfo {author} {\bibfnamefont {A.}~\bibnamefont
  {Adare}} \emph {et~al.} (\bibinfo {collaboration} {PHENIX}),\ }\href
  {\doibase 10.1103/PhysRevLett.98.162301} {\bibfield  {journal} {\bibinfo
  {journal} {Phys. Rev. Lett.}\ }\textbf {\bibinfo {volume} {98}},\ \bibinfo
  {pages} {162301} (\bibinfo {year} {2007}{\natexlab{b}})},\ \Eprint
  {http://arxiv.org/abs/nucl-ex/0608033} {arXiv:nucl-ex/0608033 [nucl-ex]}
  \BibitemShut {NoStop}%
\bibitem [{\citenamefont {Abelev}\ \emph {et~al.}(2007)\citenamefont {Abelev}
  \emph {et~al.}}]{Abelev:2006db}%
  \BibitemOpen
  \bibfield  {author} {\bibinfo {author} {\bibfnamefont {B.~I.}\ \bibnamefont
  {Abelev}} \emph {et~al.} (\bibinfo {collaboration} {STAR}),\ }\href {\doibase
  10.1103/PhysRevLett.106.159902, 10.1103/PhysRevLett.98.192301} {\bibfield
  {journal} {\bibinfo  {journal} {Phys. Rev. Lett.}\ }\textbf {\bibinfo
  {volume} {98}},\ \bibinfo {pages} {192301} (\bibinfo {year} {2007})},\
  \bibinfo {note} {[Erratum: Phys. Rev. Lett.106,159902(2011)]},\ \Eprint
  {http://arxiv.org/abs/nucl-ex/0607012} {arXiv:nucl-ex/0607012 [nucl-ex]}
  \BibitemShut {NoStop}%
\bibitem [{\citenamefont {Aamodt}\ \emph
  {et~al.}(2010{\natexlab{a}})\citenamefont {Aamodt} \emph
  {et~al.}}]{Aamodt:2010pa}%
  \BibitemOpen
  \bibfield  {author} {\bibinfo {author} {\bibfnamefont {K.}~\bibnamefont
  {Aamodt}} \emph {et~al.} (\bibinfo {collaboration} {ALICE}),\ }\href
  {\doibase 10.1103/PhysRevLett.105.252302} {\bibfield  {journal} {\bibinfo
  {journal} {Phys. Rev. Lett.}\ }\textbf {\bibinfo {volume} {105}},\ \bibinfo
  {pages} {252302} (\bibinfo {year} {2010}{\natexlab{a}})},\ \Eprint
  {http://arxiv.org/abs/1011.3914} {arXiv:1011.3914 [nucl-ex]} \BibitemShut
  {NoStop}%
\bibitem [{\citenamefont {Aamodt}\ \emph
  {et~al.}(2010{\natexlab{b}})\citenamefont {Aamodt} \emph
  {et~al.}}]{Aamodt:2010pb}%
  \BibitemOpen
  \bibfield  {author} {\bibinfo {author} {\bibfnamefont {K.}~\bibnamefont
  {Aamodt}} \emph {et~al.} (\bibinfo {collaboration} {ALICE}),\ }\href
  {\doibase 10.1103/PhysRevLett.105.252301} {\bibfield  {journal} {\bibinfo
  {journal} {Phys. Rev. Lett.}\ }\textbf {\bibinfo {volume} {105}},\ \bibinfo
  {pages} {252301} (\bibinfo {year} {2010}{\natexlab{b}})},\ \Eprint
  {http://arxiv.org/abs/1011.3916} {arXiv:1011.3916 [nucl-ex]} \BibitemShut
  {NoStop}%
\bibitem [{\citenamefont {Aamodt}\ \emph {et~al.}(2011)\citenamefont {Aamodt}
  \emph {et~al.}}]{Aamodt:2010jd}%
  \BibitemOpen
  \bibfield  {author} {\bibinfo {author} {\bibfnamefont {K.}~\bibnamefont
  {Aamodt}} \emph {et~al.} (\bibinfo {collaboration} {ALICE}),\ }\href
  {\doibase 10.1016/j.physletb.2010.12.020} {\bibfield  {journal} {\bibinfo
  {journal} {Phys. Lett.}\ }\textbf {\bibinfo {volume} {B696}},\ \bibinfo
  {pages} {30} (\bibinfo {year} {2011})},\ \Eprint
  {http://arxiv.org/abs/1012.1004} {arXiv:1012.1004 [nucl-ex]} \BibitemShut
  {NoStop}%
\bibitem [{\citenamefont {Aad}\ \emph {et~al.}(2010)\citenamefont {Aad} \emph
  {et~al.}}]{Aad:2010bu}%
  \BibitemOpen
  \bibfield  {author} {\bibinfo {author} {\bibfnamefont {G.}~\bibnamefont
  {Aad}} \emph {et~al.} (\bibinfo {collaboration} {ATLAS}),\ }\href {\doibase
  10.1103/PhysRevLett.105.252303} {\bibfield  {journal} {\bibinfo  {journal}
  {Phys. Rev. Lett.}\ }\textbf {\bibinfo {volume} {105}},\ \bibinfo {pages}
  {252303} (\bibinfo {year} {2010})},\ \Eprint {http://arxiv.org/abs/1011.6182}
  {arXiv:1011.6182 [hep-ex]} \BibitemShut {NoStop}%
\bibitem [{\citenamefont {Nambu}\ and\ \citenamefont
  {Jona-Lasinio}(1961{\natexlab{a}})}]{Nambu1}%
  \BibitemOpen
  \bibfield  {author} {\bibinfo {author} {\bibfnamefont {Y.}~\bibnamefont
  {Nambu}}\ and\ \bibinfo {author} {\bibfnamefont {G.}~\bibnamefont
  {Jona-Lasinio}},\ }\href {\doibase 10.1103/PhysRev.124.246} {\bibfield
  {journal} {\bibinfo  {journal} {Phys. Rev.}\ }\textbf {\bibinfo {volume}
  {124}},\ \bibinfo {pages} {246} (\bibinfo {year} {1961}{\natexlab{a}})},\
  \bibinfo {note} {[,141(1961)]}\BibitemShut {NoStop}%
\bibitem [{\citenamefont {Nambu}\ and\ \citenamefont
  {Jona-Lasinio}(1961{\natexlab{b}})}]{Nambu2}%
  \BibitemOpen
  \bibfield  {author} {\bibinfo {author} {\bibfnamefont {Y.}~\bibnamefont
  {Nambu}}\ and\ \bibinfo {author} {\bibfnamefont {G.}~\bibnamefont
  {Jona-Lasinio}},\ }\href {\doibase 10.1103/PhysRev.122.345} {\bibfield
  {journal} {\bibinfo  {journal} {Phys. Rev.}\ }\textbf {\bibinfo {volume}
  {122}},\ \bibinfo {pages} {345} (\bibinfo {year} {1961}{\natexlab{b}})},\
  \bibinfo {note} {[,127(1961)]}\BibitemShut {NoStop}%
\bibitem [{\citenamefont {Klevansky}(1992)}]{Klevansky}%
  \BibitemOpen
  \bibfield  {author} {\bibinfo {author} {\bibfnamefont {S.~P.}\ \bibnamefont
  {Klevansky}},\ }\href {\doibase 10.1103/RevModPhys.64.649} {\bibfield
  {journal} {\bibinfo  {journal} {Rev. Mod. Phys.}\ }\textbf {\bibinfo {volume}
  {64}},\ \bibinfo {pages} {649} (\bibinfo {year} {1992})}\BibitemShut
  {NoStop}%
\bibitem [{\citenamefont {Hatsuda}\ and\ \citenamefont
  {Kunihiro}(1994)}]{Hatsuda1}%
  \BibitemOpen
  \bibfield  {author} {\bibinfo {author} {\bibfnamefont {T.}~\bibnamefont
  {Hatsuda}}\ and\ \bibinfo {author} {\bibfnamefont {T.}~\bibnamefont
  {Kunihiro}},\ }\href {\doibase 10.1016/0370-1573(94)90022-1} {\bibfield
  {journal} {\bibinfo  {journal} {Phys. Rept.}\ }\textbf {\bibinfo {volume}
  {247}},\ \bibinfo {pages} {221} (\bibinfo {year} {1994})},\ \Eprint
  {http://arxiv.org/abs/hep-ph/9401310} {arXiv:hep-ph/9401310 [hep-ph]}
  \BibitemShut {NoStop}%
\bibitem [{\citenamefont {Vogl}\ and\ \citenamefont {Weise}(1991)}]{Vogl}%
  \BibitemOpen
  \bibfield  {author} {\bibinfo {author} {\bibfnamefont {U.}~\bibnamefont
  {Vogl}}\ and\ \bibinfo {author} {\bibfnamefont {W.}~\bibnamefont {Weise}},\
  }\href {\doibase 10.1016/0146-6410(91)90005-9} {\bibfield  {journal}
  {\bibinfo  {journal} {Prog. Part. Nucl. Phys.}\ }\textbf {\bibinfo {volume}
  {27}},\ \bibinfo {pages} {195} (\bibinfo {year} {1991})}\BibitemShut
  {NoStop}%
\bibitem [{\citenamefont {Buballa}(2005)}]{Buballa}%
  \BibitemOpen
  \bibfield  {author} {\bibinfo {author} {\bibfnamefont {M.}~\bibnamefont
  {Buballa}},\ }\href {\doibase 10.1016/j.physrep.2004.11.004} {\bibfield
  {journal} {\bibinfo  {journal} {Phys. Rept.}\ }\textbf {\bibinfo {volume}
  {407}},\ \bibinfo {pages} {205} (\bibinfo {year} {2005})},\ \Eprint
  {http://arxiv.org/abs/hep-ph/0402234} {arXiv:hep-ph/0402234 [hep-ph]}
  \BibitemShut {NoStop}%
\bibitem [{\citenamefont {Klevansky}\ and\ \citenamefont
  {Lemmer}(1989)}]{Klevansky2}%
  \BibitemOpen
  \bibfield  {author} {\bibinfo {author} {\bibfnamefont {S.~P.}\ \bibnamefont
  {Klevansky}}\ and\ \bibinfo {author} {\bibfnamefont {R.~H.}\ \bibnamefont
  {Lemmer}},\ }\href {\doibase 10.1103/PhysRevD.39.3478} {\bibfield  {journal}
  {\bibinfo  {journal} {Phys. Rev.}\ }\textbf {\bibinfo {volume} {D39}},\
  \bibinfo {pages} {3478} (\bibinfo {year} {1989})}\BibitemShut {NoStop}%
\bibitem [{\citenamefont {Mao}(2016)}]{Mao}%
  \BibitemOpen
  \bibfield  {author} {\bibinfo {author} {\bibfnamefont {S.}~\bibnamefont
  {Mao}},\ }\href {\doibase 10.1016/j.physletb.2016.05.018} {\bibfield
  {journal} {\bibinfo  {journal} {Phys. Lett.}\ }\textbf {\bibinfo {volume}
  {B758}},\ \bibinfo {pages} {195} (\bibinfo {year} {2016})},\ \Eprint
  {http://arxiv.org/abs/1602.06503} {arXiv:1602.06503 [hep-ph]} \BibitemShut
  {NoStop}%
\bibitem [{\citenamefont {Fayazbakhsh}\ \emph {et~al.}(2012)\citenamefont
  {Fayazbakhsh}, \citenamefont {Sadeghian},\ and\ \citenamefont
  {Sadooghi}}]{Sadooghi1}%
  \BibitemOpen
  \bibfield  {author} {\bibinfo {author} {\bibfnamefont {S.}~\bibnamefont
  {Fayazbakhsh}}, \bibinfo {author} {\bibfnamefont {S.}~\bibnamefont
  {Sadeghian}}, \ and\ \bibinfo {author} {\bibfnamefont {N.}~\bibnamefont
  {Sadooghi}},\ }\href {\doibase 10.1103/PhysRevD.86.085042} {\bibfield
  {journal} {\bibinfo  {journal} {Phys. Rev.}\ }\textbf {\bibinfo {volume}
  {D86}},\ \bibinfo {pages} {085042} (\bibinfo {year} {2012})},\ \Eprint
  {http://arxiv.org/abs/1206.6051} {arXiv:1206.6051 [hep-ph]} \BibitemShut
  {NoStop}%
\bibitem [{\citenamefont {Ruggieri}\ \emph {et~al.}(2013)\citenamefont
  {Ruggieri}, \citenamefont {Tachibana},\ and\ \citenamefont
  {Greco}}]{Ruggieri}%
  \BibitemOpen
  \bibfield  {author} {\bibinfo {author} {\bibfnamefont {M.}~\bibnamefont
  {Ruggieri}}, \bibinfo {author} {\bibfnamefont {M.}~\bibnamefont {Tachibana}},
  \ and\ \bibinfo {author} {\bibfnamefont {V.}~\bibnamefont {Greco}},\ }\href
  {\doibase 10.1007/JHEP07(2013)165} {\bibfield  {journal} {\bibinfo  {journal}
  {JHEP}\ }\textbf {\bibinfo {volume} {07}},\ \bibinfo {pages} {165} (\bibinfo
  {year} {2013})},\ \Eprint {http://arxiv.org/abs/1305.0137} {arXiv:1305.0137
  [hep-ph]} \BibitemShut {NoStop}%
\bibitem [{\citenamefont {Ruggieri}\ \emph {et~al.}(2014)\citenamefont
  {Ruggieri}, \citenamefont {Oliva}, \citenamefont {Castorina}, \citenamefont
  {Gatto},\ and\ \citenamefont {Greco}}]{Ruggieri2}%
  \BibitemOpen
  \bibfield  {author} {\bibinfo {author} {\bibfnamefont {M.}~\bibnamefont
  {Ruggieri}}, \bibinfo {author} {\bibfnamefont {L.}~\bibnamefont {Oliva}},
  \bibinfo {author} {\bibfnamefont {P.}~\bibnamefont {Castorina}}, \bibinfo
  {author} {\bibfnamefont {R.}~\bibnamefont {Gatto}}, \ and\ \bibinfo {author}
  {\bibfnamefont {V.}~\bibnamefont {Greco}},\ }\href {\doibase
  10.1016/j.physletb.2014.05.073} {\bibfield  {journal} {\bibinfo  {journal}
  {Phys. Lett.}\ }\textbf {\bibinfo {volume} {B734}},\ \bibinfo {pages} {255}
  (\bibinfo {year} {2014})},\ \Eprint {http://arxiv.org/abs/1402.0737}
  {arXiv:1402.0737 [hep-ph]} \BibitemShut {NoStop}%
\bibitem [{\citenamefont {Andersen}\ \emph {et~al.}(2016)\citenamefont
  {Andersen}, \citenamefont {Naylor},\ and\ \citenamefont
  {Tranberg}}]{Andersen}%
  \BibitemOpen
  \bibfield  {author} {\bibinfo {author} {\bibfnamefont {J.~O.}\ \bibnamefont
  {Andersen}}, \bibinfo {author} {\bibfnamefont {W.~R.}\ \bibnamefont
  {Naylor}}, \ and\ \bibinfo {author} {\bibfnamefont {A.}~\bibnamefont
  {Tranberg}},\ }\href {\doibase 10.1103/RevModPhys.88.025001} {\bibfield
  {journal} {\bibinfo  {journal} {Rev. Mod. Phys.}\ }\textbf {\bibinfo {volume}
  {88}},\ \bibinfo {pages} {025001} (\bibinfo {year} {2016})},\ \Eprint
  {http://arxiv.org/abs/1411.7176} {arXiv:1411.7176 [hep-ph]} \BibitemShut
  {NoStop}%
\bibitem [{\citenamefont {Ayala}\ \emph {et~al.}(2016)\citenamefont {Ayala},
  \citenamefont {Dominguez}, \citenamefont {Hernandez}, \citenamefont {Loewe},\
  and\ \citenamefont {Zamora}}]{Ayala2}%
  \BibitemOpen
  \bibfield  {author} {\bibinfo {author} {\bibfnamefont {A.}~\bibnamefont
  {Ayala}}, \bibinfo {author} {\bibfnamefont {C.~A.}\ \bibnamefont
  {Dominguez}}, \bibinfo {author} {\bibfnamefont {L.~A.}\ \bibnamefont
  {Hernandez}}, \bibinfo {author} {\bibfnamefont {M.}~\bibnamefont {Loewe}}, \
  and\ \bibinfo {author} {\bibfnamefont {R.}~\bibnamefont {Zamora}},\ }\href
  {\doibase 10.1016/j.physletb.2016.05.058} {\bibfield  {journal} {\bibinfo
  {journal} {Phys. Lett.}\ }\textbf {\bibinfo {volume} {B759}},\ \bibinfo
  {pages} {99} (\bibinfo {year} {2016})},\ \Eprint
  {http://arxiv.org/abs/1510.09134} {arXiv:1510.09134 [hep-ph]} \BibitemShut
  {NoStop}%
\bibitem [{\citenamefont {Strickland}\ \emph {et~al.}(2012)\citenamefont
  {Strickland}, \citenamefont {Dexheimer},\ and\ \citenamefont
  {Menezes}}]{Strickland}%
  \BibitemOpen
  \bibfield  {author} {\bibinfo {author} {\bibfnamefont {M.}~\bibnamefont
  {Strickland}}, \bibinfo {author} {\bibfnamefont {V.}~\bibnamefont
  {Dexheimer}}, \ and\ \bibinfo {author} {\bibfnamefont {D.~P.}\ \bibnamefont
  {Menezes}},\ }\href {\doibase 10.1103/PhysRevD.86.125032} {\bibfield
  {journal} {\bibinfo  {journal} {Phys. Rev.}\ }\textbf {\bibinfo {volume}
  {D86}},\ \bibinfo {pages} {125032} (\bibinfo {year} {2012})},\ \Eprint
  {http://arxiv.org/abs/1209.3276} {arXiv:1209.3276 [nucl-th]} \BibitemShut
  {NoStop}%
\bibitem [{\citenamefont {Mukherjee}\ \emph {et~al.}(2018)\citenamefont
  {Mukherjee}, \citenamefont {Ghosh}, \citenamefont {Mandal}, \citenamefont
  {Sarkar},\ and\ \citenamefont {Roy}}]{Arghya}%
  \BibitemOpen
  \bibfield  {author} {\bibinfo {author} {\bibfnamefont {A.}~\bibnamefont
  {Mukherjee}}, \bibinfo {author} {\bibfnamefont {S.}~\bibnamefont {Ghosh}},
  \bibinfo {author} {\bibfnamefont {M.}~\bibnamefont {Mandal}}, \bibinfo
  {author} {\bibfnamefont {S.}~\bibnamefont {Sarkar}}, \ and\ \bibinfo {author}
  {\bibfnamefont {P.}~\bibnamefont {Roy}},\ }\href {\doibase
  10.1103/PhysRevD.98.056024} {\bibfield  {journal} {\bibinfo  {journal} {Phys.
  Rev.}\ }\textbf {\bibinfo {volume} {D98}},\ \bibinfo {pages} {056024}
  (\bibinfo {year} {2018})},\ \Eprint {http://arxiv.org/abs/1809.07028}
  {arXiv:1809.07028 [hep-ph]} \BibitemShut {NoStop}%
\bibitem [{\citenamefont {Fayazbakhsh}\ and\ \citenamefont
  {Sadooghi}(2014)}]{Sadooghi}%
  \BibitemOpen
  \bibfield  {author} {\bibinfo {author} {\bibfnamefont {S.}~\bibnamefont
  {Fayazbakhsh}}\ and\ \bibinfo {author} {\bibfnamefont {N.}~\bibnamefont
  {Sadooghi}},\ }\href {\doibase 10.1103/PhysRevD.90.105030} {\bibfield
  {journal} {\bibinfo  {journal} {Phys. Rev.}\ }\textbf {\bibinfo {volume}
  {D90}},\ \bibinfo {pages} {105030} (\bibinfo {year} {2014})},\ \Eprint
  {http://arxiv.org/abs/1408.5457} {arXiv:1408.5457 [hep-ph]} \BibitemShut
  {NoStop}%
\bibitem [{\citenamefont {Chaudhuri}\ \emph {et~al.}(2019)\citenamefont
  {Chaudhuri}, \citenamefont {Ghosh}, \citenamefont {Sarkar},\ and\
  \citenamefont {Roy}}]{Chaudhuri:2019lbw}%
  \BibitemOpen
  \bibfield  {author} {\bibinfo {author} {\bibfnamefont {N.}~\bibnamefont
  {Chaudhuri}}, \bibinfo {author} {\bibfnamefont {S.}~\bibnamefont {Ghosh}},
  \bibinfo {author} {\bibfnamefont {S.}~\bibnamefont {Sarkar}}, \ and\ \bibinfo
  {author} {\bibfnamefont {P.}~\bibnamefont {Roy}},\ }\href {\doibase
  10.1103/PhysRevD.99.116025} {\bibfield  {journal} {\bibinfo  {journal} {Phys.
  Rev.}\ }\textbf {\bibinfo {volume} {D99}},\ \bibinfo {pages} {116025}
  (\bibinfo {year} {2019})},\ \Eprint {http://arxiv.org/abs/1907.03990}
  {arXiv:1907.03990 [nucl-th]} \BibitemShut {NoStop}%
\bibitem [{\citenamefont {Wong}(1995)}]{Wong:1995jf}%
  \BibitemOpen
  \bibfield  {author} {\bibinfo {author} {\bibfnamefont {C.~Y.}\ \bibnamefont
  {Wong}},\ }\href@noop {} {\emph {\bibinfo {title} {{Introduction to
  high-energy heavy ion collisions}}}}\ (\bibinfo {year} {1995})\BibitemShut
  {NoStop}%
\bibitem [{\citenamefont {Arnold}\ \emph {et~al.}(2001)\citenamefont {Arnold},
  \citenamefont {Moore},\ and\ \citenamefont {Yaffe}}]{Arnold:2001ms}%
  \BibitemOpen
  \bibfield  {author} {\bibinfo {author} {\bibfnamefont {P.~B.}\ \bibnamefont
  {Arnold}}, \bibinfo {author} {\bibfnamefont {G.~D.}\ \bibnamefont {Moore}}, \
  and\ \bibinfo {author} {\bibfnamefont {L.~G.}\ \bibnamefont {Yaffe}},\ }\href
  {\doibase 10.1088/1126-6708/2001/12/009} {\bibfield  {journal} {\bibinfo
  {journal} {JHEP}\ }\textbf {\bibinfo {volume} {12}},\ \bibinfo {pages} {009}
  (\bibinfo {year} {2001})},\ \Eprint {http://arxiv.org/abs/hep-ph/0111107}
  {arXiv:hep-ph/0111107 [hep-ph]} \BibitemShut {NoStop}%
\bibitem [{\citenamefont {Aurenche}\ \emph
  {et~al.}(2000{\natexlab{a}})\citenamefont {Aurenche}, \citenamefont {Gelis},\
  and\ \citenamefont {Zaraket}}]{Aurenche:1999tq}%
  \BibitemOpen
  \bibfield  {author} {\bibinfo {author} {\bibfnamefont {P.}~\bibnamefont
  {Aurenche}}, \bibinfo {author} {\bibfnamefont {F.}~\bibnamefont {Gelis}}, \
  and\ \bibinfo {author} {\bibfnamefont {H.}~\bibnamefont {Zaraket}},\ }\href
  {\doibase 10.1103/PhysRevD.61.116001} {\bibfield  {journal} {\bibinfo
  {journal} {Phys. Rev.}\ }\textbf {\bibinfo {volume} {D61}},\ \bibinfo {pages}
  {116001} (\bibinfo {year} {2000}{\natexlab{a}})},\ \Eprint
  {http://arxiv.org/abs/hep-ph/9911367} {arXiv:hep-ph/9911367 [hep-ph]}
  \BibitemShut {NoStop}%
\bibitem [{\citenamefont {Aurenche}\ \emph
  {et~al.}(2000{\natexlab{b}})\citenamefont {Aurenche}, \citenamefont {Gelis},\
  and\ \citenamefont {Zaraket}}]{Aurenche:2000gf}%
  \BibitemOpen
  \bibfield  {author} {\bibinfo {author} {\bibfnamefont {P.}~\bibnamefont
  {Aurenche}}, \bibinfo {author} {\bibfnamefont {F.}~\bibnamefont {Gelis}}, \
  and\ \bibinfo {author} {\bibfnamefont {H.}~\bibnamefont {Zaraket}},\ }\href
  {\doibase 10.1103/PhysRevD.62.096012} {\bibfield  {journal} {\bibinfo
  {journal} {Phys. Rev.}\ }\textbf {\bibinfo {volume} {D62}},\ \bibinfo {pages}
  {096012} (\bibinfo {year} {2000}{\natexlab{b}})},\ \Eprint
  {http://arxiv.org/abs/hep-ph/0003326} {arXiv:hep-ph/0003326 [hep-ph]}
  \BibitemShut {NoStop}%
\bibitem [{\citenamefont {Chatterjee}\ \emph {et~al.}(2010)\citenamefont
  {Chatterjee}, \citenamefont {Bhattacharya},\ and\ \citenamefont
  {Srivastava}}]{Chatterjee:2009rs}%
  \BibitemOpen
  \bibfield  {author} {\bibinfo {author} {\bibfnamefont {R.}~\bibnamefont
  {Chatterjee}}, \bibinfo {author} {\bibfnamefont {L.}~\bibnamefont
  {Bhattacharya}}, \ and\ \bibinfo {author} {\bibfnamefont {D.~K.}\
  \bibnamefont {Srivastava}},\ }\bibfield  {booktitle} {\emph {\bibinfo
  {booktitle} {{QGP Winter School 2008 Jaipur, India, February 1-3, 2008}}},\
  }\href {\doibase 10.1007/978-3-642-02286-9_7} {\bibfield  {journal} {\bibinfo
   {journal} {Lect. Notes Phys.}\ }\textbf {\bibinfo {volume} {785}},\ \bibinfo
  {pages} {219} (\bibinfo {year} {2010})},\ \Eprint
  {http://arxiv.org/abs/0901.3610} {arXiv:0901.3610 [nucl-th]} \BibitemShut
  {NoStop}%
\bibitem [{\citenamefont {Alam}\ \emph {et~al.}(1996)\citenamefont {Alam},
  \citenamefont {Sinha},\ and\ \citenamefont {Raha}}]{Alam:1996fd}%
  \BibitemOpen
  \bibfield  {author} {\bibinfo {author} {\bibfnamefont {J.}~\bibnamefont
  {Alam}}, \bibinfo {author} {\bibfnamefont {B.}~\bibnamefont {Sinha}}, \ and\
  \bibinfo {author} {\bibfnamefont {S.}~\bibnamefont {Raha}},\ }\href {\doibase
  10.1016/0370-1573(95)00084-4} {\bibfield  {journal} {\bibinfo  {journal}
  {Phys. Rept.}\ }\textbf {\bibinfo {volume} {273}},\ \bibinfo {pages} {243}
  (\bibinfo {year} {1996})}\BibitemShut {NoStop}%
\bibitem [{\citenamefont {Alam}\ \emph {et~al.}(2001)\citenamefont {Alam},
  \citenamefont {Sarkar}, \citenamefont {Roy}, \citenamefont {Hatsuda},\ and\
  \citenamefont {Sinha}}]{Alam:1999sc}%
  \BibitemOpen
  \bibfield  {author} {\bibinfo {author} {\bibfnamefont {J.}~\bibnamefont
  {Alam}}, \bibinfo {author} {\bibfnamefont {S.}~\bibnamefont {Sarkar}},
  \bibinfo {author} {\bibfnamefont {P.}~\bibnamefont {Roy}}, \bibinfo {author}
  {\bibfnamefont {T.}~\bibnamefont {Hatsuda}}, \ and\ \bibinfo {author}
  {\bibfnamefont {B.}~\bibnamefont {Sinha}},\ }\href {\doibase
  10.1006/aphy.2000.6091} {\bibfield  {journal} {\bibinfo  {journal} {Annals
  Phys.}\ }\textbf {\bibinfo {volume} {286}},\ \bibinfo {pages} {159} (\bibinfo
  {year} {2001})},\ \Eprint {http://arxiv.org/abs/hep-ph/9909267}
  {arXiv:hep-ph/9909267 [hep-ph]} \BibitemShut {NoStop}%
\bibitem [{\citenamefont {Kajantie}\ \emph {et~al.}(1986)\citenamefont
  {Kajantie}, \citenamefont {Kapusta}, \citenamefont {McLerran},\ and\
  \citenamefont {Mekjian}}]{Kajantie:1986dh}%
  \BibitemOpen
  \bibfield  {author} {\bibinfo {author} {\bibfnamefont {K.}~\bibnamefont
  {Kajantie}}, \bibinfo {author} {\bibfnamefont {J.~I.}\ \bibnamefont
  {Kapusta}}, \bibinfo {author} {\bibfnamefont {L.~D.}\ \bibnamefont
  {McLerran}}, \ and\ \bibinfo {author} {\bibfnamefont {A.}~\bibnamefont
  {Mekjian}},\ }\href {\doibase 10.1103/PhysRevD.34.2746} {\bibfield  {journal}
  {\bibinfo  {journal} {Phys. Rev.}\ }\textbf {\bibinfo {volume} {D34}},\
  \bibinfo {pages} {2746} (\bibinfo {year} {1986})}\BibitemShut {NoStop}%
\bibitem [{\citenamefont {McLerran}\ and\ \citenamefont
  {Toimela}(1985)}]{McLerran:1984ay}%
  \BibitemOpen
  \bibfield  {author} {\bibinfo {author} {\bibfnamefont {L.~D.}\ \bibnamefont
  {McLerran}}\ and\ \bibinfo {author} {\bibfnamefont {T.}~\bibnamefont
  {Toimela}},\ }\href {\doibase 10.1103/PhysRevD.31.545} {\bibfield  {journal}
  {\bibinfo  {journal} {Phys. Rev.}\ }\textbf {\bibinfo {volume} {D31}},\
  \bibinfo {pages} {545} (\bibinfo {year} {1985})}\BibitemShut {NoStop}%
\bibitem [{\citenamefont {Rapp}\ and\ \citenamefont
  {Wambach}(2000)}]{Rapp:1999ej}%
  \BibitemOpen
  \bibfield  {author} {\bibinfo {author} {\bibfnamefont {R.}~\bibnamefont
  {Rapp}}\ and\ \bibinfo {author} {\bibfnamefont {J.}~\bibnamefont {Wambach}},\
  }\href {\doibase 10.1007/0-306-47101-9_1} {\bibfield  {journal} {\bibinfo
  {journal} {Adv. Nucl. Phys.}\ }\textbf {\bibinfo {volume} {25}},\ \bibinfo
  {pages} {1} (\bibinfo {year} {2000})},\ \Eprint
  {http://arxiv.org/abs/hep-ph/9909229} {arXiv:hep-ph/9909229 [hep-ph]}
  \BibitemShut {NoStop}%
\bibitem [{\citenamefont {Weldon}(1990)}]{Weldon:1990iw}%
  \BibitemOpen
  \bibfield  {author} {\bibinfo {author} {\bibfnamefont {H.~A.}\ \bibnamefont
  {Weldon}},\ }\href {\doibase 10.1103/PhysRevD.42.2384} {\bibfield  {journal}
  {\bibinfo  {journal} {Phys. Rev.}\ }\textbf {\bibinfo {volume} {D42}},\
  \bibinfo {pages} {2384} (\bibinfo {year} {1990})}\BibitemShut {NoStop}%
\bibitem [{\citenamefont {Rapp}\ and\ \citenamefont {van
  Hees}(2010)}]{Rapp:2009my}%
  \BibitemOpen
  \bibfield  {author} {\bibinfo {author} {\bibfnamefont {R.}~\bibnamefont
  {Rapp}}\ and\ \bibinfo {author} {\bibfnamefont {H.}~\bibnamefont {van
  Hees}},\ }in\ \href {\doibase 10.1142/9789814293297_0003} {\emph {\bibinfo
  {booktitle} {{Quark-gluon plasma 4}}}}\ (\bibinfo {year} {2010})\ pp.\
  \bibinfo {pages} {111--206},\ \Eprint {http://arxiv.org/abs/0903.1096}
  {arXiv:0903.1096 [hep-ph]} \BibitemShut {NoStop}%
\bibitem [{\citenamefont {Poskanzer}\ and\ \citenamefont
  {Voloshin}(1998)}]{Poskanzer:1998yz}%
  \BibitemOpen
  \bibfield  {author} {\bibinfo {author} {\bibfnamefont {A.~M.}\ \bibnamefont
  {Poskanzer}}\ and\ \bibinfo {author} {\bibfnamefont {S.~A.}\ \bibnamefont
  {Voloshin}},\ }\href {\doibase 10.1103/PhysRevC.58.1671} {\bibfield
  {journal} {\bibinfo  {journal} {Phys. Rev.}\ }\textbf {\bibinfo {volume}
  {C58}},\ \bibinfo {pages} {1671} (\bibinfo {year} {1998})},\ \Eprint
  {http://arxiv.org/abs/nucl-ex/9805001} {arXiv:nucl-ex/9805001 [nucl-ex]}
  \BibitemShut {NoStop}%
\bibitem [{\citenamefont {Voloshin}\ \emph {et~al.}(2010)\citenamefont
  {Voloshin}, \citenamefont {Poskanzer},\ and\ \citenamefont
  {Snellings}}]{Voloshin:2008dg}%
  \BibitemOpen
  \bibfield  {author} {\bibinfo {author} {\bibfnamefont {S.~A.}\ \bibnamefont
  {Voloshin}}, \bibinfo {author} {\bibfnamefont {A.~M.}\ \bibnamefont
  {Poskanzer}}, \ and\ \bibinfo {author} {\bibfnamefont {R.}~\bibnamefont
  {Snellings}},\ }\href {\doibase 10.1007/978-3-642-01539-7_10} {\bibfield
  {journal} {\bibinfo  {journal} {Landolt-Bornstein}\ }\textbf {\bibinfo
  {volume} {23}},\ \bibinfo {pages} {293} (\bibinfo {year} {2010})},\ \Eprint
  {http://arxiv.org/abs/0809.2949} {arXiv:0809.2949 [nucl-ex]} \BibitemShut
  {NoStop}%
\bibitem [{\citenamefont {Ollitrault}(1992)}]{Ollitrault:1992bk}%
  \BibitemOpen
  \bibfield  {author} {\bibinfo {author} {\bibfnamefont {J.-Y.}\ \bibnamefont
  {Ollitrault}},\ }\href {\doibase 10.1103/PhysRevD.46.229} {\bibfield
  {journal} {\bibinfo  {journal} {Phys. Rev.}\ }\textbf {\bibinfo {volume}
  {D46}},\ \bibinfo {pages} {229} (\bibinfo {year} {1992})}\BibitemShut
  {NoStop}%
\bibitem [{\citenamefont {Kolb}\ and\ \citenamefont
  {Heinz}(2003)}]{Kolb:2003dz}%
  \BibitemOpen
  \bibfield  {author} {\bibinfo {author} {\bibfnamefont {P.~F.}\ \bibnamefont
  {Kolb}}\ and\ \bibinfo {author} {\bibfnamefont {U.~W.}\ \bibnamefont
  {Heinz}},\ }\href@noop {} {\ ,\ \bibinfo {pages} {634} (\bibinfo {year}
  {2003})},\ \Eprint {http://arxiv.org/abs/nucl-th/0305084}
  {arXiv:nucl-th/0305084 [nucl-th]} \BibitemShut {NoStop}%
\bibitem [{\citenamefont {Adams}\ \emph {et~al.}(2005)\citenamefont {Adams}
  \emph {et~al.}}]{Adams:2005dq}%
  \BibitemOpen
  \bibfield  {author} {\bibinfo {author} {\bibfnamefont {J.}~\bibnamefont
  {Adams}} \emph {et~al.} (\bibinfo {collaboration} {STAR}),\ }\href {\doibase
  10.1016/j.nuclphysa.2005.03.085} {\bibfield  {journal} {\bibinfo  {journal}
  {Nucl. Phys.}\ }\textbf {\bibinfo {volume} {A757}},\ \bibinfo {pages} {102}
  (\bibinfo {year} {2005})},\ \Eprint {http://arxiv.org/abs/nucl-ex/0501009}
  {arXiv:nucl-ex/0501009 [nucl-ex]} \BibitemShut {NoStop}%
\bibitem [{\citenamefont {Matsui}\ and\ \citenamefont
  {Satz}(1986)}]{Matsui:1986dk}%
  \BibitemOpen
  \bibfield  {author} {\bibinfo {author} {\bibfnamefont {T.}~\bibnamefont
  {Matsui}}\ and\ \bibinfo {author} {\bibfnamefont {H.}~\bibnamefont {Satz}},\
  }\href {\doibase 10.1016/0370-2693(86)91404-8} {\bibfield  {journal}
  {\bibinfo  {journal} {Phys. Lett.}\ }\textbf {\bibinfo {volume} {B178}},\
  \bibinfo {pages} {416} (\bibinfo {year} {1986})}\BibitemShut {NoStop}%
\bibitem [{\citenamefont {Wang}\ and\ \citenamefont
  {Gyulassy}(1992)}]{Wang:1991xy}%
  \BibitemOpen
  \bibfield  {author} {\bibinfo {author} {\bibfnamefont {X.-N.}\ \bibnamefont
  {Wang}}\ and\ \bibinfo {author} {\bibfnamefont {M.}~\bibnamefont
  {Gyulassy}},\ }\href {\doibase 10.1103/PhysRevLett.68.1480} {\bibfield
  {journal} {\bibinfo  {journal} {Phys. Rev. Lett.}\ }\textbf {\bibinfo
  {volume} {68}},\ \bibinfo {pages} {1480} (\bibinfo {year}
  {1992})}\BibitemShut {NoStop}%
\bibitem [{\citenamefont {Tuchin}(2013{\natexlab{a}})}]{Tuchin:2012mf}%
  \BibitemOpen
  \bibfield  {author} {\bibinfo {author} {\bibfnamefont {K.}~\bibnamefont
  {Tuchin}},\ }\href {\doibase 10.1103/PhysRevC.87.024912} {\bibfield
  {journal} {\bibinfo  {journal} {Phys. Rev.}\ }\textbf {\bibinfo {volume}
  {C87}},\ \bibinfo {pages} {024912} (\bibinfo {year} {2013}{\natexlab{a}})},\
  \Eprint {http://arxiv.org/abs/1206.0485} {arXiv:1206.0485 [hep-ph]}
  \BibitemShut {NoStop}%
\bibitem [{\citenamefont {Tuchin}(2013{\natexlab{b}})}]{Tuchin:2013bda}%
  \BibitemOpen
  \bibfield  {author} {\bibinfo {author} {\bibfnamefont {K.}~\bibnamefont
  {Tuchin}},\ }\href {\doibase 10.1103/PhysRevC.88.024910} {\bibfield
  {journal} {\bibinfo  {journal} {Phys. Rev.}\ }\textbf {\bibinfo {volume}
  {C88}},\ \bibinfo {pages} {024910} (\bibinfo {year} {2013}{\natexlab{b}})},\
  \Eprint {http://arxiv.org/abs/1305.0545} {arXiv:1305.0545 [nucl-th]}
  \BibitemShut {NoStop}%
\bibitem [{\citenamefont {Sadooghi}\ and\ \citenamefont
  {Taghinavaz}(2017)}]{Sadooghi:2016jyf}%
  \BibitemOpen
  \bibfield  {author} {\bibinfo {author} {\bibfnamefont {N.}~\bibnamefont
  {Sadooghi}}\ and\ \bibinfo {author} {\bibfnamefont {F.}~\bibnamefont
  {Taghinavaz}},\ }\href {\doibase 10.1016/j.aop.2016.11.008} {\bibfield
  {journal} {\bibinfo  {journal} {Annals Phys.}\ }\textbf {\bibinfo {volume}
  {376}},\ \bibinfo {pages} {218} (\bibinfo {year} {2017})},\ \Eprint
  {http://arxiv.org/abs/1601.04887} {arXiv:1601.04887 [hep-ph]} \BibitemShut
  {NoStop}%
\bibitem [{\citenamefont {Mamo}(2013)}]{Mamo:2013efa}%
  \BibitemOpen
  \bibfield  {author} {\bibinfo {author} {\bibfnamefont {K.~A.}\ \bibnamefont
  {Mamo}},\ }\href {\doibase 10.1007/JHEP08(2013)083} {\bibfield  {journal}
  {\bibinfo  {journal} {JHEP}\ }\textbf {\bibinfo {volume} {08}},\ \bibinfo
  {pages} {083} (\bibinfo {year} {2013})},\ \Eprint
  {http://arxiv.org/abs/1210.7428} {arXiv:1210.7428 [hep-th]} \BibitemShut
  {NoStop}%
\bibitem [{\citenamefont {Bandyopadhyay}\ \emph {et~al.}(2016)\citenamefont
  {Bandyopadhyay}, \citenamefont {Islam},\ and\ \citenamefont
  {Mustafa}}]{Bandyopadhyay:2016fyd}%
  \BibitemOpen
  \bibfield  {author} {\bibinfo {author} {\bibfnamefont {A.}~\bibnamefont
  {Bandyopadhyay}}, \bibinfo {author} {\bibfnamefont {C.~A.}\ \bibnamefont
  {Islam}}, \ and\ \bibinfo {author} {\bibfnamefont {M.~G.}\ \bibnamefont
  {Mustafa}},\ }\href {\doibase 10.1103/PhysRevD.94.114034} {\bibfield
  {journal} {\bibinfo  {journal} {Phys. Rev.}\ }\textbf {\bibinfo {volume}
  {D94}},\ \bibinfo {pages} {114034} (\bibinfo {year} {2016})},\ \Eprint
  {http://arxiv.org/abs/1602.06769} {arXiv:1602.06769 [hep-ph]} \BibitemShut
  {NoStop}%
\bibitem [{\citenamefont {Bandyopadhyay}\ and\ \citenamefont
  {Mallik}(2017)}]{Bandyopadhyay:2017raf}%
  \BibitemOpen
  \bibfield  {author} {\bibinfo {author} {\bibfnamefont {A.}~\bibnamefont
  {Bandyopadhyay}}\ and\ \bibinfo {author} {\bibfnamefont {S.}~\bibnamefont
  {Mallik}},\ }\href {\doibase 10.1103/PhysRevD.95.074019} {\bibfield
  {journal} {\bibinfo  {journal} {Phys. Rev.}\ }\textbf {\bibinfo {volume}
  {D95}},\ \bibinfo {pages} {074019} (\bibinfo {year} {2017})},\ \Eprint
  {http://arxiv.org/abs/1704.01364} {arXiv:1704.01364 [hep-ph]} \BibitemShut
  {NoStop}%
\bibitem [{\citenamefont {Ghosh}\ and\ \citenamefont
  {Chandra}(2018)}]{Ghosh:2018xhh}%
  \BibitemOpen
  \bibfield  {author} {\bibinfo {author} {\bibfnamefont {S.}~\bibnamefont
  {Ghosh}}\ and\ \bibinfo {author} {\bibfnamefont {V.}~\bibnamefont
  {Chandra}},\ }\href {\doibase 10.1103/PhysRevD.98.076006} {\bibfield
  {journal} {\bibinfo  {journal} {Phys. Rev.}\ }\textbf {\bibinfo {volume}
  {D98}},\ \bibinfo {pages} {076006} (\bibinfo {year} {2018})},\ \Eprint
  {http://arxiv.org/abs/1808.05176} {arXiv:1808.05176 [hep-ph]} \BibitemShut
  {NoStop}%
\bibitem [{\citenamefont {Islam}\ \emph {et~al.}(2019)\citenamefont {Islam},
  \citenamefont {Bandyopadhyay}, \citenamefont {Roy},\ and\ \citenamefont
  {Sarkar}}]{Islam:2018sog}%
  \BibitemOpen
  \bibfield  {author} {\bibinfo {author} {\bibfnamefont {C.~A.}\ \bibnamefont
  {Islam}}, \bibinfo {author} {\bibfnamefont {A.}~\bibnamefont
  {Bandyopadhyay}}, \bibinfo {author} {\bibfnamefont {P.~K.}\ \bibnamefont
  {Roy}}, \ and\ \bibinfo {author} {\bibfnamefont {S.}~\bibnamefont {Sarkar}},\
  }\href {\doibase 10.1103/PhysRevD.99.094028} {\bibfield  {journal} {\bibinfo
  {journal} {Phys. Rev.}\ }\textbf {\bibinfo {volume} {D99}},\ \bibinfo {pages}
  {094028} (\bibinfo {year} {2019})},\ \Eprint
  {http://arxiv.org/abs/1812.10380} {arXiv:1812.10380 [hep-ph]} \BibitemShut
  {NoStop}%
\bibitem [{\citenamefont {Das}\ \emph {et~al.}(2019)\citenamefont {Das},
  \citenamefont {Haque}, \citenamefont {Mustafa},\ and\ \citenamefont
  {Roy}}]{Das:2019nzv}%
  \BibitemOpen
  \bibfield  {author} {\bibinfo {author} {\bibfnamefont {A.}~\bibnamefont
  {Das}}, \bibinfo {author} {\bibfnamefont {N.}~\bibnamefont {Haque}}, \bibinfo
  {author} {\bibfnamefont {M.~G.}\ \bibnamefont {Mustafa}}, \ and\ \bibinfo
  {author} {\bibfnamefont {P.~K.}\ \bibnamefont {Roy}},\ }\href {\doibase
  10.1103/PhysRevD.99.094022} {\bibfield  {journal} {\bibinfo  {journal} {Phys.
  Rev.}\ }\textbf {\bibinfo {volume} {D99}},\ \bibinfo {pages} {094022}
  (\bibinfo {year} {2019})},\ \Eprint {http://arxiv.org/abs/1903.03528}
  {arXiv:1903.03528 [hep-ph]} \BibitemShut {NoStop}%
\bibitem [{\citenamefont {Mallik}\ and\ \citenamefont
  {Sarkar}(2016)}]{Mallik:2016anp}%
  \BibitemOpen
  \bibfield  {author} {\bibinfo {author} {\bibfnamefont {S.}~\bibnamefont
  {Mallik}}\ and\ \bibinfo {author} {\bibfnamefont {S.}~\bibnamefont
  {Sarkar}},\ }\href {\doibase 10.1017/9781316535585} {\emph {\bibinfo {title}
  {{Hadrons at Finite Temperature}}}}\ (\bibinfo  {publisher} {Cambridge
  University Press},\ \bibinfo {address} {Cambridge},\ \bibinfo {year}
  {2016})\BibitemShut {NoStop}%
\bibitem [{\citenamefont {Greiner}\ \emph {et~al.}(2011)\citenamefont
  {Greiner}, \citenamefont {Haque}, \citenamefont {Mustafa},\ and\
  \citenamefont {Thoma}}]{Greiner:2010zg}%
  \BibitemOpen
  \bibfield  {author} {\bibinfo {author} {\bibfnamefont {C.}~\bibnamefont
  {Greiner}}, \bibinfo {author} {\bibfnamefont {N.}~\bibnamefont {Haque}},
  \bibinfo {author} {\bibfnamefont {M.~G.}\ \bibnamefont {Mustafa}}, \ and\
  \bibinfo {author} {\bibfnamefont {M.~H.}\ \bibnamefont {Thoma}},\ }\href
  {\doibase 10.1103/PhysRevC.83.014908} {\bibfield  {journal} {\bibinfo
  {journal} {Phys. Rev.}\ }\textbf {\bibinfo {volume} {C83}},\ \bibinfo {pages}
  {014908} (\bibinfo {year} {2011})},\ \Eprint {http://arxiv.org/abs/1010.2169}
  {arXiv:1010.2169 [hep-ph]} \BibitemShut {NoStop}%
\bibitem [{\citenamefont {Schwartz}(2014)}]{Schwartz:2013pla}%
  \BibitemOpen
  \bibfield  {author} {\bibinfo {author} {\bibfnamefont {M.~D.}\ \bibnamefont
  {Schwartz}},\ }\href@noop {} {\emph {\bibinfo {title} {{Quantum Field Theory
  and the Standard Model}}}}\ (\bibinfo  {publisher} {Cambridge University
  Press},\ \bibinfo {year} {2014})\BibitemShut {NoStop}%
\bibitem [{\citenamefont {Peskin}\ and\ \citenamefont
  {Schroeder}(1995)}]{Peskin:1995ev}%
  \BibitemOpen
  \bibfield  {author} {\bibinfo {author} {\bibfnamefont {M.~E.}\ \bibnamefont
  {Peskin}}\ and\ \bibinfo {author} {\bibfnamefont {D.~V.}\ \bibnamefont
  {Schroeder}},\ }\href@noop {} {\emph {\bibinfo {title} {{An Introduction to
  quantum field theory}}}}\ (\bibinfo  {publisher} {Addison-Wesley},\ \bibinfo
  {address} {Reading, USA},\ \bibinfo {year} {1995})\BibitemShut {NoStop}%
\bibitem [{\citenamefont {Halzen}\ and\ \citenamefont
  {Martin}(1984)}]{Halzen:1984mc}%
  \BibitemOpen
  \bibfield  {author} {\bibinfo {author} {\bibfnamefont {F.}~\bibnamefont
  {Halzen}}\ and\ \bibinfo {author} {\bibfnamefont {A.~D.}\ \bibnamefont
  {Martin}},\ }\href@noop {} {\emph {\bibinfo {title} {{QUARKS AND LEPTONS: AN
  INTRODUCTORY COURSE IN MODERN PARTICLE PHYSICS}}}}\ (\bibinfo {year}
  {1984})\BibitemShut {NoStop}%
\bibitem [{\citenamefont {Ghosh}\ \emph {et~al.}(2017)\citenamefont {Ghosh},
  \citenamefont {Mukherjee}, \citenamefont {Mandal}, \citenamefont {Sarkar},\
  and\ \citenamefont {Roy}}]{Ghosh:2017rjo}%
  \BibitemOpen
  \bibfield  {author} {\bibinfo {author} {\bibfnamefont {S.}~\bibnamefont
  {Ghosh}}, \bibinfo {author} {\bibfnamefont {A.}~\bibnamefont {Mukherjee}},
  \bibinfo {author} {\bibfnamefont {M.}~\bibnamefont {Mandal}}, \bibinfo
  {author} {\bibfnamefont {S.}~\bibnamefont {Sarkar}}, \ and\ \bibinfo {author}
  {\bibfnamefont {P.}~\bibnamefont {Roy}},\ }\href {\doibase
  10.1103/PhysRevD.96.116020} {\bibfield  {journal} {\bibinfo  {journal} {Phys.
  Rev.}\ }\textbf {\bibinfo {volume} {D96}},\ \bibinfo {pages} {116020}
  (\bibinfo {year} {2017})},\ \Eprint {http://arxiv.org/abs/1704.05319}
  {arXiv:1704.05319 [hep-ph]} \BibitemShut {NoStop}%
\bibitem [{\citenamefont {Ghosh}\ \emph {et~al.}(2019)\citenamefont {Ghosh},
  \citenamefont {Mukherjee}, \citenamefont {Roy},\ and\ \citenamefont
  {Sarkar}}]{Ghosh:2019fet}%
  \BibitemOpen
  \bibfield  {author} {\bibinfo {author} {\bibfnamefont {S.}~\bibnamefont
  {Ghosh}}, \bibinfo {author} {\bibfnamefont {A.}~\bibnamefont {Mukherjee}},
  \bibinfo {author} {\bibfnamefont {P.}~\bibnamefont {Roy}}, \ and\ \bibinfo
  {author} {\bibfnamefont {S.}~\bibnamefont {Sarkar}},\ }\href {\doibase
  10.1103/PhysRevD.99.096004} {\bibfield  {journal} {\bibinfo  {journal} {Phys.
  Rev.}\ }\textbf {\bibinfo {volume} {D99}},\ \bibinfo {pages} {096004}
  (\bibinfo {year} {2019})},\ \Eprint {http://arxiv.org/abs/1901.02290}
  {arXiv:1901.02290 [hep-ph]} \BibitemShut {NoStop}%
\bibitem [{\citenamefont {Zhang}\ \emph {et~al.}(2016)\citenamefont {Zhang},
  \citenamefont {Fu},\ and\ \citenamefont {Liu}}]{Zhang}%
  \BibitemOpen
  \bibfield  {author} {\bibinfo {author} {\bibfnamefont {R.}~\bibnamefont
  {Zhang}}, \bibinfo {author} {\bibfnamefont {W.-j.}\ \bibnamefont {Fu}}, \
  and\ \bibinfo {author} {\bibfnamefont {Y.-x.}\ \bibnamefont {Liu}},\ }\href
  {\doibase 10.1140/epjc/s10052-016-4123-8} {\bibfield  {journal} {\bibinfo
  {journal} {Eur. Phys. J.}\ }\textbf {\bibinfo {volume} {C76}},\ \bibinfo
  {pages} {307} (\bibinfo {year} {2016})},\ \Eprint
  {http://arxiv.org/abs/1604.08888} {arXiv:1604.08888 [hep-ph]} \BibitemShut
  {NoStop}%
\bibitem [{\citenamefont {Shovkovy}\ and\ \citenamefont
  {Huang}(2003)}]{Shovkovy:2003uu}%
  \BibitemOpen
  \bibfield  {author} {\bibinfo {author} {\bibfnamefont {I.}~\bibnamefont
  {Shovkovy}}\ and\ \bibinfo {author} {\bibfnamefont {M.}~\bibnamefont
  {Huang}},\ }\href {\doibase 10.1016/S0370-2693(03)00748-2} {\bibfield
  {journal} {\bibinfo  {journal} {Phys.\ Lett.\ B}\ }\textbf {\bibinfo {volume}
  {564}},\ \bibinfo {pages} {205} (\bibinfo {year} {2003})},\ \Eprint
  {http://arxiv.org/abs/hep-ph/0302142} {arXiv:hep-ph/0302142} \BibitemShut
  {NoStop}%
\bibitem [{\citenamefont {Roberts}\ and\ \citenamefont
  {Schmidt}(2000)}]{Roberts:2000aa}%
  \BibitemOpen
  \bibfield  {author} {\bibinfo {author} {\bibfnamefont {C.~D.}\ \bibnamefont
  {Roberts}}\ and\ \bibinfo {author} {\bibfnamefont {S.~M.}\ \bibnamefont
  {Schmidt}},\ }\href {\doibase 10.1016/S0146-6410(00)90011-5} {\bibfield
  {journal} {\bibinfo  {journal} {Prog. Part. Nucl. Phys.}\ }\textbf {\bibinfo
  {volume} {45}},\ \bibinfo {pages} {S1} (\bibinfo {year} {2000})},\ \Eprint
  {http://arxiv.org/abs/nucl-th/0005064} {arXiv:nucl-th/0005064 [nucl-th]}
  \BibitemShut {NoStop}%
\bibitem [{\citenamefont {Roberts}(2008)}]{Roberts:2007ji}%
  \BibitemOpen
  \bibfield  {author} {\bibinfo {author} {\bibfnamefont {C.~D.}\ \bibnamefont
  {Roberts}},\ }\bibfield  {booktitle} {\emph {\bibinfo {booktitle} {{Quarks in
  hadrons and nuclei. Proceedings, International Workshop on Nuclear Physics,
  29th Course, Erice, Italy, September 16-24, 2007}}},\ }\href {\doibase
  10.1016/j.ppnp.2007.12.034} {\bibfield  {journal} {\bibinfo  {journal} {Prog.
  Part. Nucl. Phys.}\ }\textbf {\bibinfo {volume} {61}},\ \bibinfo {pages} {50}
  (\bibinfo {year} {2008})},\ \Eprint {http://arxiv.org/abs/0712.0633}
  {arXiv:0712.0633 [nucl-th]} \BibitemShut {NoStop}%
\bibitem [{\citenamefont {Bhagwat}\ \emph {et~al.}(2003)\citenamefont
  {Bhagwat}, \citenamefont {Pichowsky}, \citenamefont {Roberts},\ and\
  \citenamefont {Tandy}}]{Bhagwat:2003vw}%
  \BibitemOpen
  \bibfield  {author} {\bibinfo {author} {\bibfnamefont {M.}~\bibnamefont
  {Bhagwat}}, \bibinfo {author} {\bibfnamefont {M.}~\bibnamefont {Pichowsky}},
  \bibinfo {author} {\bibfnamefont {C.}~\bibnamefont {Roberts}}, \ and\
  \bibinfo {author} {\bibfnamefont {P.}~\bibnamefont {Tandy}},\ }\href
  {\doibase 10.1103/PhysRevC.68.015203} {\bibfield  {journal} {\bibinfo
  {journal} {Phys.\ Rev.\ C}\ }\textbf {\bibinfo {volume} {68}},\ \bibinfo
  {pages} {015203} (\bibinfo {year} {2003})},\ \Eprint
  {http://arxiv.org/abs/nucl-th/0304003} {arXiv:nucl-th/0304003} \BibitemShut
  {NoStop}%
\bibitem [{\citenamefont {Siringo}(2016)}]{Siringo:2016jrc}%
  \BibitemOpen
  \bibfield  {author} {\bibinfo {author} {\bibfnamefont {F.}~\bibnamefont
  {Siringo}},\ }\href {\doibase 10.1103/PhysRevD.94.114036} {\bibfield
  {journal} {\bibinfo  {journal} {Phys. Rev.}\ }\textbf {\bibinfo {volume}
  {D94}},\ \bibinfo {pages} {114036} (\bibinfo {year} {2016})},\ \Eprint
  {http://arxiv.org/abs/1605.07357} {arXiv:1605.07357 [hep-ph]} \BibitemShut
  {NoStop}%
\bibitem [{\citenamefont {Liu}\ \emph {et~al.}(2015)\citenamefont {Liu},
  \citenamefont {Shuryak},\ and\ \citenamefont {Zahed}}]{Liu:2015jsa}%
  \BibitemOpen
  \bibfield  {author} {\bibinfo {author} {\bibfnamefont {Y.}~\bibnamefont
  {Liu}}, \bibinfo {author} {\bibfnamefont {E.}~\bibnamefont {Shuryak}}, \ and\
  \bibinfo {author} {\bibfnamefont {I.}~\bibnamefont {Zahed}},\ }\href
  {\doibase 10.1103/PhysRevD.92.085007} {\bibfield  {journal} {\bibinfo
  {journal} {Phys.\ Rev.\ D}\ }\textbf {\bibinfo {volume} {92}},\ \bibinfo
  {pages} {085007} (\bibinfo {year} {2015})},\ \Eprint
  {http://arxiv.org/abs/1503.09148} {arXiv:1503.09148 [hep-ph]} \BibitemShut
  {NoStop}%
\bibitem [{\citenamefont {Shuryak}(2018)}]{Shuryak:2018fjr}%
  \BibitemOpen
  \bibfield  {author} {\bibinfo {author} {\bibfnamefont {E.}~\bibnamefont
  {Shuryak}},\ }\href@noop {} {\  (\bibinfo {year} {2018})},\ \Eprint
  {http://arxiv.org/abs/1812.01509} {arXiv:1812.01509 [hep-ph]} \BibitemShut
  {NoStop}%
\bibitem [{\citenamefont {Bowman}\ \emph {et~al.}(2003)\citenamefont {Bowman},
  \citenamefont {Heller}, \citenamefont {Leinweber},\ and\ \citenamefont
  {Williams}}]{Bowman:2002kn}%
  \BibitemOpen
  \bibfield  {author} {\bibinfo {author} {\bibfnamefont {P.~O.}\ \bibnamefont
  {Bowman}}, \bibinfo {author} {\bibfnamefont {U.~M.}\ \bibnamefont {Heller}},
  \bibinfo {author} {\bibfnamefont {D.~B.}\ \bibnamefont {Leinweber}}, \ and\
  \bibinfo {author} {\bibfnamefont {A.~G.}\ \bibnamefont {Williams}},\
  }\bibfield  {booktitle} {\emph {\bibinfo {booktitle} {{Lattice field theory.
  Proceedings: 20th International Symposium, Lattice 2002, Cambridge, USA, Jun
  24-29, 2002}}},\ }\href {\doibase 10.1016/S0920-5632(03)01533-0} {\bibfield
  {journal} {\bibinfo  {journal} {Nucl. Phys. Proc. Suppl.}\ }\textbf {\bibinfo
  {volume} {119}},\ \bibinfo {pages} {323} (\bibinfo {year} {2003})},\ \Eprint
  {http://arxiv.org/abs/hep-lat/0209129} {arXiv:hep-lat/0209129 [hep-lat]}
  \BibitemShut {NoStop}%
\bibitem [{\citenamefont {Bowman}\ \emph {et~al.}(2005)\citenamefont {Bowman},
  \citenamefont {Heller}, \citenamefont {Leinweber}, \citenamefont
  {Parappilly}, \citenamefont {Williams},\ and\ \citenamefont
  {Zhang}}]{Bowman:2005vx}%
  \BibitemOpen
  \bibfield  {author} {\bibinfo {author} {\bibfnamefont {P.~O.}\ \bibnamefont
  {Bowman}}, \bibinfo {author} {\bibfnamefont {U.~M.}\ \bibnamefont {Heller}},
  \bibinfo {author} {\bibfnamefont {D.~B.}\ \bibnamefont {Leinweber}}, \bibinfo
  {author} {\bibfnamefont {M.~B.}\ \bibnamefont {Parappilly}}, \bibinfo
  {author} {\bibfnamefont {A.~G.}\ \bibnamefont {Williams}}, \ and\ \bibinfo
  {author} {\bibfnamefont {J.-b.}\ \bibnamefont {Zhang}},\ }\href {\doibase
  10.1103/PhysRevD.71.054507} {\bibfield  {journal} {\bibinfo  {journal} {Phys.
  Rev.}\ }\textbf {\bibinfo {volume} {D71}},\ \bibinfo {pages} {054507}
  (\bibinfo {year} {2005})},\ \Eprint {http://arxiv.org/abs/hep-lat/0501019}
  {arXiv:hep-lat/0501019 [hep-lat]} \BibitemShut {NoStop}%
\bibitem [{\citenamefont {Ebert}\ and\ \citenamefont
  {Vshivtsev}(1998)}]{Ebert:1998gx}%
  \BibitemOpen
  \bibfield  {author} {\bibinfo {author} {\bibfnamefont {D.}~\bibnamefont
  {Ebert}}\ and\ \bibinfo {author} {\bibfnamefont {A.~S.}\ \bibnamefont
  {Vshivtsev}},\ }\href@noop {} {\  (\bibinfo {year} {1998})},\ \Eprint
  {http://arxiv.org/abs/hep-ph/9806421} {arXiv:hep-ph/9806421 [hep-ph]}
  \BibitemShut {NoStop}%
\bibitem [{\citenamefont {Inagaki}\ \emph {et~al.}(2004)\citenamefont
  {Inagaki}, \citenamefont {Kimura},\ and\ \citenamefont
  {Murata}}]{Inagaki:2004ih}%
  \BibitemOpen
  \bibfield  {author} {\bibinfo {author} {\bibfnamefont {T.}~\bibnamefont
  {Inagaki}}, \bibinfo {author} {\bibfnamefont {D.}~\bibnamefont {Kimura}}, \
  and\ \bibinfo {author} {\bibfnamefont {T.}~\bibnamefont {Murata}},\
  }\bibfield  {booktitle} {\emph {\bibinfo {booktitle} {{Finite density QCD.
  Proceedings, International Workshop, Nara, Japan, July 10-12, 2003}}},\
  }\href {\doibase 10.1143/PTPS.153.321} {\bibfield  {journal} {\bibinfo
  {journal} {Prog. Theor. Phys. Suppl.}\ }\textbf {\bibinfo {volume} {153}},\
  \bibinfo {pages} {321} (\bibinfo {year} {2004})},\ \Eprint
  {http://arxiv.org/abs/hep-ph/0404219} {arXiv:hep-ph/0404219 [hep-ph]}
  \BibitemShut {NoStop}%
\bibitem [{\citenamefont {Landau}\ and\ \citenamefont
  {Lifshitz}(1980)}]{Landau:1980mil}%
  \BibitemOpen
  \bibfield  {author} {\bibinfo {author} {\bibfnamefont {L.~D.}\ \bibnamefont
  {Landau}}\ and\ \bibinfo {author} {\bibfnamefont {E.~M.}\ \bibnamefont
  {Lifshitz}},\ }\href@noop {} {\emph {\bibinfo {title} {{Statistical Physics,
  Part 1}}}},\ \bibinfo {series} {Course of Theoretical Physics}, Vol.~\bibinfo
  {volume} {5}\ (\bibinfo  {publisher} {Butterworth-Heinemann},\ \bibinfo
  {address} {Oxford},\ \bibinfo {year} {1980})\BibitemShut {NoStop}%
\end{thebibliography}%

\end{document}